\documentclass[10pt]{article}
\usepackage{verbatim}
\usepackage{makeidx}
\usepackage{amssymb}
\usepackage{slashed}
\usepackage{graphicx}
\usepackage{epstopdf}

\newcommand{\tl}{\tilde{\l}}
\newcommand{\lra}{\leftrightarrow}

\def\half{{1\over 2}}
\def\ben{\begin{equation}}
\def\een{\end{equation}}
\def\bea{\begin{eqnarray}}
\def\eea{\end{eqnarray}}
\newcommand{\be}{\begin{equation}}
\newcommand{\ee}{\end{equation}}
\newcommand{\beq}{\begin{equation}}
\newcommand{\eeq}{\end{equation}}
\newcommand{\ba}{\begin{eqnarray}}
\newcommand{\ea}{\end{eqnarray}}
\newcommand{\nn}{\nonumber}
\newcommand{\reef}[1]{(\ref{#1})}

\def \bv{{\bf v}}

\def \bu {\bf u}

\newcommand{\bbox}{\lower.2ex\hbox{$\Box$}}

\def\diag{\mathop{\rm diag}\nolimits}
\newsavebox{\uuunit}
\sbox{\uuunit}
    {\setlength{\unitlength}{0.825em}
     \begin{picture}(0.6,0.7)
        \thinlines
        \put(0,0){\line(1,0){0.5}}
        \put(0.15,0){\line(0,1){0.7}}
        \put(0.35,0){\line(0,1){0.8}}
       \multiput(0.3,0.8)(-0.04,-0.02){12}{\rule{0.5pt}{0.5pt}}
     \end {picture}}

\begin{document}


\def\del{{\partial}}
\def\vev#1{\left\langle #1 \right\rangle}
\def\cn{{\cal N}}
\def\co{{\cal O}}

\def\IC{{\mathbb C}}
\def\IR{{\mathbb R}}
\def\IZ{{\mathbb Z}}
\def\RP{{\bf RP}}
\def\CP{{\bf CP}}
\def\Poincare{{Poincar{\'e} }}
\def\tp{{\tilde \Phi}}

\def\TL{\hfil$\displaystyle{##}$}
\def\TR{$\displaystyle{{}##}$\hfil}
\def\TC{\hfil$\displaystyle{##}$\hfil}
\def\TT{\hbox{##}}
\def\HLINE{\noalign{\vskip1\jot}\hline\noalign{\vskip1\jot}}
\def\seqalign#1#2{\vcenter{\openup1\jot
  \halign{\strut #1\cr #2 \cr}}}
\def\lbldef#1#2{\expandafter\gdef\csname #1\endcsname {#2}}
\def\eqn#1#2{\lbldef{#1}{(\ref{#1})}%
\begin{equation} #2 \label{#1} \end{equation}}
\def\eqalign#1{\vcenter{\openup1\jot
    \halign{\strut\span\TL & \span\TR\cr #1 \cr
   }}}
\def\eno#1{(\ref{#1})}
\def\href#1#2{#2}
\def\half{{1 \over 2}}
\def\lab{\label}

\def\ads{{\it AdS}}
\def\adsp{{\it AdS}$_{p+2}$}
\def\cft{{\it CFT}}

\newcommand{\ber}{\begin{eqnarray}}
\newcommand{\eer}{\end{eqnarray}}

\newcommand{\beqar}{\begin{eqnarray}}
\newcommand{\cN}{{\cal N}}
\newcommand{\cO}{{\cal O}}
\newcommand{\cA}{{\cal A}}
\newcommand{\cT}{{\cal T}}
\newcommand{\cF}{{\cal F}}
\newcommand{\cC}{{\cal C}}
\newcommand{\cR}{{\cal R}}
\newcommand{\cW}{{\cal W}}
\newcommand{\eeqar}{\end{eqnarray}}
\newcommand{\lm}{\lambda}
\newcommand{\eps}{\epsilon}


\newcommand{\nonu}{\nonumber}
\newcommand{\oh}{\displaystyle{\frac{1}{2}}}
\newcommand{\dsl}
  {\kern.06em\hbox{\raise.15ex\hbox{$/$}\kern-.56em\hbox{$\partial$}}}
\newcommand{\id}{i\!\!\not\!\partial}
\newcommand{\as}{\not\!\! A}
\newcommand{\ps}{\not\! p}
\newcommand{\ks}{\not\! k}
\newcommand{\D}{{\cal{D}}}
\newcommand{\dv}{d^2x}
\newcommand{\Z}{{\cal Z}}
\newcommand{\N}{{\cal N}}
\newcommand{\Dsl}{\not\!\! D}
\newcommand{\Bsl}{\not\!\! B}
\newcommand{\Psl}{\not\!\! P}
\newcommand{\eeqarr}{\end{eqnarray}}
\newcommand{\ZZ}{{\rm \kern 0.275em Z \kern -0.92em Z}\;}


\tolerance =10000

\def\be{\begin{equation}}
\def\ee{\end{equation}}

\def\bea{\begin{eqnarray}}
\def\eea{\end{eqnarray}}

\def\ba{\begin{array}}
\def\ea{\end{array}}
\def\bd{\begin{displaymath}}
\def\ed{\end{displaymath}}

\def\eq{\begin{equation}}
\def\eqe{\end{equation}}
\def\eqa{\begin{eqnarray}}
\def\eqae{\end{eqnarray}}
\def\ena{\end{eqnarray}}

\def\ap{{\alpha^{\prime}}}
\def\aap{{a^{\prime}}}
\def\eg{{\it e.g.~}}
\def\ie{{\it i.e.~}}
\def\nn{\nonumber}

\def\tz{\tilde{z}}
\def\ra{\rangle}
\def\la{\langle}
\def\ud{{\mathrm{d}}}
\def\unit{1 \hskip-.3em \raise2pt\hbox{$ \scriptstyle |$ } }

\def\a{\alpha}
\def\b{\beta}
\def\g{\gamma}
\def\c{\gamma}
\def\d{\delta}
\def\e{\epsilon}           
\def\f{\phi}               
\def\vf{\varphi}  \def\tvf{\tilde{\varphi}}
\def\g{\gamma}
\def\h{\eta}
\def\j{\psi}
\def\k{\kappa}                    
\def\l{\lambda}
\def\m{\mu}
\def\n{\nu}
\def\o{\omega}  \def\w{\omega}
\def\p{\pi}                
\def\th{\theta}                  
\def\r{\rho}                                     
\def\s{\sigma}                                   
\def\t{\tau}
\def\u{\upsilon}
\def\x{\xi}
\def\z{\zeta}
\def\D{\Delta}
\def\F{\Phi}
\def\G{\Gamma}
\def\J{\Psi}
\def\L{\Lambda}
\def\O{\Omega}  \def\W{\Omega}
\def\P{\Pi}
\def\Q{\Theta}
\def\Th{\Theta}
\def\S{\Sigma}
\def\X{\Xi}
\def\del{\partial}

\def\ca{{\cal A}}
\def\cb{{\cal B}}
\def\cc{{\cal C}}
\def\cd{{\cal D}}
\def\ce{{\cal E}}
\def\cf{{\cal F}}
\def\cg{{\cal G}}
\def\ch{{\cal H}}
\def\ci{{\cal I}}
\def\cj{{\cal J}}
\def\ck{{\cal K}}
\def\cl{{\cal L}}
\def\cm{{\cal M}}
\def\cn{{\cal N}}
\def\co{{\cal O}}
\def\cp{{\cal P}}
\def\cq{{\cal Q}}
\def\car{{\cal R}}
\def\cs{{\cal S}}
\def\ct{{\cal T}}
\def\cu{{\cal U}}
\def\cv{{\cal V}}
\def\cw{{\cal W}}
\def\cx{{\cal X}}
\def\cy{{\cal Y}}
\def\cz{{\cal Z}}
\def\half{{1 \over 2}}

\def\Bf#1{\mbox{\boldmath $#1$}}       
\def\Sf#1{\hbox{\sf #1}}               

\def\bop#1{\setbox0=\hbox{$#1M$}\mkern1.5mu
        \vbox{\hrule height0pt depth.04\ht0
        \hbox{\vrule width.04\ht0 height.9\ht0 \kern.9\ht0
        \vrule width.04\ht0}\hrule height.04\ht0}\mkern1.5mu}
\def\Box{{\mathpalette\bop{}}}                        
\def\pa{\partial}                              
\def\de{\nabla}                                       
\def\dell{\bigtriangledown} 
\def\su{\sum}                                         
\def\pr{\prod}                                        
\def\iff{\leftrightarrow}                      
\def\conj{{\hbox{\large *}}} 
\def\lconj{{\hbox{\footnotesize *}}}          
\def\dg{\sp\dagger} 
\def\ddg{\sp\ddagger} 
\def\>{\rangle} 

\def\<{\langle} 
\def\Dsl{D \hskip-.6em \raise1pt\hbox{$ / $ } }
\def\to{\rightarrow}
\def\tf{\tilde{\f}}

\def\pa{\partial}
\def\del{\nabla}
\def\delbar{\bar{\nabla}}
\def\+{\oplus}
\def\sg{supergravity}
\def\tQ{\tilde{Q}}
\def\tA{\tilde{A}}
\def\wi{Ward identity}
\def\wis{Ward identities}

\def\da{{\dot\alpha}}
\def\db{{\dot\beta}}
\def\dg{{\dot\gamma}}
\def\dd{{\dot\delta}}
\def\bu{\bar{u}}
\def\bv{\bar{v}}

\begin{titlepage}

\begin{flushright}
MIT-CTP-3943 \\
ROM2F/2008/13 \\
CERN-TH-2008-096
\end{flushright}
\vspace{1cm}

\begin{center}
{\Large\bf
  Generating Tree Amplitudes in
 ${\cal N} =4$ SYM and  ${\cal N} = 8$ SG}  \\
\vspace{1cm}
{\bf Massimo Bianchi${}^{a,b}$, Henriette Elvang${}^{d}$ and
Daniel Z. Freedman${}^{c\,d\,e}$} \\
\vspace{0.7cm}
{${}^{a}${\it Physics Department, Theory Unit, CERN}\\
         {\it CH1211, Geneva 23, Switzerland}}\\[2mm]
{${}^{b}${\it Dipartimento di Fisica, }\\
         {\it Universit\'a di Roma ``Tor Vergata''}\\
         {\it I.N.F.N. Sezione di Roma ``Tor Vergata''}\\
         {\it Via della Ricerca Scientifica, 00133 Roma, Italy}}\\[2mm]
{${}^{c}${\it Department of Mathematics}}\\
{${}^{d}${\it Center for Theoretical Physics}\\
         {\it Massachusetts Institute of Technology}\\
         {\it 77 Massachusetts Avenue}\\
         {\it Cambridge, MA 02139, USA}}\\[2mm]
{${}^{e}${\it Department of Applied Mathematics and Theoretical Physics}\\{\it Centre for
Mathematical Sciences}\\{\it Cambridge University}\\{\it Cambridge CB3 0WA, United
Kingdom}}\\[5mm] {\small \tt Massimo.Bianchi@roma2.infn.it, elvang@lns.mit.edu,
 dzf@math.mit.edu}
\end{center}
\vskip .3truecm

\begin{abstract}
We study $n$-point tree amplitudes of $\cn =4$ super Yang-Mills theory and $\cn =8$ supergravity for
general configurations of external particles of the two theories. We construct generating functions for $n$-point MHV and NMHV amplitudes with general external states. Amplitudes derived from
them obey SUSY Ward identities, and the generating functions characterize and count amplitudes in the MHV and NMHV sectors. The MHV generating function provides an efficient
way to perform the intermediate state helicity sums required to obtain loop amplitudes
from trees. The NMHV generating functions rely on the MHV-vertex expansion obtained from recursion relations associated with a
3-line shift of external momenta involving a reference spinor $|X]$. The recursion
relations remain valid for a subset of $\cn =8$ supergravity amplitudes which do not vanish asymptotically for all  $|X]$.
The MHV-vertex expansion of
the $n$-graviton NMHV
amplitude for $n=5,6,...,11$ is independent of $|X]$ and exhibits the
asymptotic behavior $z^{n-12}$.  This presages difficulties for $n > 12.$
Generating  functions show how the
symmetries of supergravity can be implemented in the quadratic map
between supergravity and gauge theory embodied in the KLT and other
similar relations between amplitudes in the two theories.
\end{abstract}

\end{titlepage}

\tableofcontents

\newpage

\setcounter{equation}{0}
\section{Introduction}
Recent calculations and conjectures \cite{bdr, bcdjkr} on the
possible ultraviolet finiteness of $\cN =8$ supergravity theory
motivate a search for simplifications of the difficult perturbative calculations
needed for further progress.\footnote{There are also earlier relevant calculations \cite{earlier} as well as more recent work \cite{VanBjB,unexp}. Additional references are given in  \cite{bdr,bcdjkr,VanBjB}.}
Three important techniques used in those calculations are the following:
\begin{enumerate}
\item[i.] The integrands of loop diagrams are constructed from tree
amplitudes using generalized unitarity cuts. Even when external lines are
gravitons, the unitarity sum includes processes involving all possible states
of the supergravity theory. New information on these tree amplitudes
can be helpful at the loop level.
\item[ii.]  On-shell tree amplitudes in gauge theory and gravity are best expressed
using the spinor helicity formalism and are most easily obtained from the modern form
of recursion relations \cite{bcf,bcfw,qmc,CSgr} which relate $n$-point amplitudes to
 those for smaller values of $n$. The simplest expressions appear in the MHV
sectors of each theory, but perturbative calculations have reached the point
 where NMHV amplitudes are required. These have been studied for external gluons and
gravitons, but there is less information on amplitudes involving other particles
 of the theory.
\item[iii.] Relatively complicated supergravity trees are constructed from
the simpler tree
amplitudes of $\cN =4$ super-Yang-Mills theory using the quadratic
relation between gravity and  gauge theory embodied in the KLT
relations \cite{klt}.
Implicit in this relation is a map between two
copies of the gauge theory and supergravity which we denote by
\be
\lab{map}
\left[\cN = 4 ~\mathrm{SYM}\right]_L \otimes \left[\cN = 4~
\mathrm{SYM}\right]_R
~\leftrightarrow ~ \left[\cN = 8 ~\mathrm{SG}\right] \, .
\ee
There are 16
distinct particle states in each $\cN =4$ SYM factor and 256
states in $\cN =8$ SG.
\end{enumerate}

This paper is motivated by all three issues above.  We focus on the construction
of $n$-point MHV and NMHV tree amplitudes in $\cn = 4$ SYM and $\cn =8$ \sg\ with
general external states.  Toward this end we develop and generalize to \sg\ the
generating function for MHV amplitudes in gauge theory discovered in \cite{nair}
and further developed and extended to NMHV amplitudes in \cite{ggk}.
The generating functions encode the
external state dependence in a compact way and furnish precise
characterizations of the MHV and NMHV sectors. To entice the reader we
pose three questions to which the formalism gives simple answers. The
MHV sector of $\cn = 4$ SYM consists of the $n$-gluon amplitude
$A_n(--++\ldots +)$ with two negative helicity gluons plus all amplitudes
related by SUSY transformations. Would the reader have
guessed that this sector contains the 8-point amplitude with 8 +ve
helicity gluini?
In \sg\ the MHV sector consists of all amplitudes related by SUSY to the
$n$-graviton amplitude $M_n(--++\ldots +)$ with two negative helicity
gravitons.  Would the reader have guessed that there are 186 distinct
processes,\footnote{For $n < 16$ the number of processes is
  smaller.}
each with a different set of particles, in this sector? And would
the reader have  anticipated the external state dependence of
$n$-point MHV amplitudes has a simple  direct relationship to the
properties of $n$-point CFT correlators?

Generating functions provide useful answers to a number of questions,
and they appear to have practical applications. For
example, the unitarity sums over intermediate states required to obtain
1-loop Feynman integrands from the product of tree amplitudes in both
gauge theory and \sg\ can be
done quite efficiently using the generating function.

The generating function for $n$-point amplitudes in gauge theory is an $SU(4)$
invariant function $F_n(p_i,\h_{ia})$ of the momenta and of $4n$ Grassmann variables $\h_{ia}$. Here
$i=1,\ldots ,n$ refers to the momentum $p_i$ of each external particle and $a =
1,2,3,4$ is the $SU(4)$ flavor index. The generalization to gravity is
straightforward in the MHV sector, in which the generating function is an
$SU(8)$ invariant function $\O(p_i,\h_{iA})$ of $8n$ Grassmann
variables $\h_{iA}$ where $A=1,\dots,8$ is an $SU(8)$ index.
It is very simple to
calculate any MHV amplitude from the generating function by applying
Grassmann derivatives specific to the external states. All symmetry transformations can be implemented at
the level of the generating function as operations involving the
$\h_{ia}$ and $\h_{iA}$ variables, and one can show that all
amplitudes automatically satisfy SUSY Ward identities.

The NMHV sector of gauge theory (or respectively, \sg) consists of
all amplitudes linked by SUSY \wis\  to the $n$-gluon amplitude
$A_n(---++\ldots +)$ (or respectively, the $n$-graviton
$M_n(---+\ldots +)$) with 3 negative helicity particles. The
construction of a generating function is formally straightforward
in the NMHV sector, but its justification is more subtle. There is a
different generating function for each diagram in the MHV-vertex
expansion of an amplitude. The MHV-vertex expansion was first
obtained in ($\cn = 0$) gauge theory in \cite{csw} and extended to
gravity in \cite{greatdane}.   The contribution of each diagram
depends on the choice of an arbitrary reference spinor $|X]$, but
the full amplitude, which is the sum of all diagrams, should be
independent of $|X]$.

The simplest justification of the
expansion comes from the recursion relation associated with a complex
shift of the spinors $|1],\,|2],\,|3]$ of the negative helicity lines
\cite{risager}. The required shift is
\be \lab{shift3}
|m_i] \to |\hat{m}_i] = |m_i]  + z \<m_j\,m_k\> |X]\,,
\ee
where $m_i, m_j, m_k$ are the cyclic permutations of the momentum labels for a choice of three of the external lines. For pure gluon or graviton amplitudes, these will be the three negative helicity particles.
The recursion relation, and therefore the diagrammatic expansion, is
valid if the continued amplitude vanishes as $z \to \infty.$ This
desired property was proven  for gluon amplitudes $A_n(---++\ldots +)$ in \cite{csw,risager}, but was observed in numerical calculation of the graviton amplitudes $M_n(---+\ldots +)$
only for $n=6,7$ in \cite{greatdane}.  It is also known for simpler shifts of two external momenta that the large
$z$ falloff is slower for amplitudes in which some gluons, or gravitons,
are replaced by lower spin particles of the supermultiplet.   For these reasons we must be
cautious in our applications of the MHV-vertex expansion.

If an amplitude vanishes as $z \to \infty$ for all choices of $|X]$, Cauchy's theorem ensures that the
sum of MHV-vertex diagrams is independent of $|X]$.  For all NMHV amplitudes in $\cn = 4$ SYM
we show that there is always a choice of 3 lines to shift such that the contribution of each diagram
falls at least as fast as $1/z$. We have verified $|X]$-independence of the expansion numerically for a large number of 6-point NMHV amplitudes.
Thus we detect no problems, and the generating function appears to be valid for the whole NMHV sector of the $\cn = 4$ theory.

In gravity the situation is more problematic.  For graviton amplitudes $M_n(---+\ldots +)$ we show that the falloff as $z \to \infty$ depends on the number of external legs $n$. Specifically, we have verified numerically for $n=5,\dots,11$ that
\bea
  \lab{Mnz}
  M_n(\hat{1}^-\, \hat{2}^- \,\hat{3}^-\, 4^+ \ldots n^+) &\to& \frac{1}{z^{12-n}}
  ~~~~\mathrm{as}~~~~~z \to \infty \, .
\eea
This means that the MHV-vertex decomposition of the $n$-graviton NHMV amplitude can be expected to fail for $n \ge 12$. Indeed for $n=12$ we find that the sum of 1533 MHV-vertex diagrams fails to be independent of $|X]$.

The evaluation and summation of diagrams is more complicated for  general external states in  \sg\, so our analysis is limited to 6-point NMHV processes.  There are 151 such processes, each with
several functionally independent amplitudes obtained by inequivalent assignments of $SU(8)$
indices to the external particles.  For each amplitude there are up to 21 non-vanishing diagrams.
Most 6-point amplitudes have the same good
properties as those of gauge theory; they vanish under large shifts, and they are constructed
correctly  using the MHV-vertex expansion with  diagrams  obtained from the  generating function.

The large $z$ behavior of individual diagrams for any amplitude can be determined analytically.
The result depends on which set of 3 lines are shifted.  Our analysis shows that there are processes
for which even the best shift contains diagrams with either $O(1)$ or $O(z)$ behavior at large $z$.
Numerical evaluation can then test whether the sum of diagrams depends of $|X]$. This would indicate
that the undesired large $z$ behavior persists in the full amplitude, and we have found that it does for a number\footnote{Of the total of 151 6-point NMHV processes, we estimate that about half will include amplitudes with asymptotic $O(1)$ behavior.}  of examples. In these cases we recalculate the amplitude using the KLT formula
which provides a correct evaluation of any $\cn =8$ amplitude as a sum of products of $\cn = 4$ SYM amplitudes. The result from KLT can be continued to complex momenta by shifting spinors and the large $z$ behavior is then extracted. By this method\footnote{We have automated the process by writing a Mathematica code which evaluates the KLT expansion as well as the MHV-vertex decomposition for any 6-point NMHV amplitude of the $\cn = 8$ theory. }
 we have explored about 20
amplitudes whose best shifts give asymptotic $O(1)$ behavior. We call  these cases ``bad'' amplitudes, as opposed to ``good'' amplitudes which vanish asymptotically for one or more 3-line shifts.  The large $z$ limit of these ``bad" amplitudes is a ratio of polynomials in the reference spinor $|X]$.  The amplitude does not vanish asymptotically for all $|X]$, but it does vanish when $|X]$ is
chosen to be a root of the numerator polynomial. The recursion relation becomes valid for these special values of $|X]$, and the sum of MHV-vertex diagrams then agrees with the KLT evaluation.
In this way we have developed  a good interpretation, and justification, of the generating function even for ``bad" amplitudes.

Finally we must mention that our analysis locates two ``very bad'' 6-point NMHV amplitudes whose KLT evaluations show linear growth in $z$ as $z \to \infty$.  There are no values of $|X]$ which validate  the MHV-vertex decomposition, so the generating function is not useful. However, the SUSY
Ward identities can be used to express each ``very bad" amplitude as a sum of two other amplitudes which are constructible via the MHV-vertex expansion and generating function.

Complications with the large $z$ behavior in supergravity suggest that it may be difficult to apply the generating function to intermediate state helicity  sums involving NMHV amplitudes. It is important to explore this question, but it is  beyond the scope of the present paper.

Let's return to the map \reef{map} because another focus of this paper
concerns how the $\cn = 8$ supersymmetry and global $SU(8)$ symmetry
of \sg\ are implemented in the tensor product of gauge theory states.
One question of concern is how the $SU(4)_L\otimes
SU(4)_R$ flavor symmetry of the product of gauge theory factors is
promoted to the  $SU(8)$ global symmetry of supergravity.  The
derivation of the  KLT relations from string theory does not
really settle this question, since $SU(8)$ only emerges as an
accidental symmetry in the $\a' \to 0$ limit.

To investigate such questions we write the detailed algebra of the SUSY charges
and the annihilation and creation operators of the gauge and \sg\ theories
and provide a detailed version of the map \reef{map} which is compatible with
these symmetry operations. In the map, any $SU(8)$ index $A,B,\ldots
\in 1,\dots,8$ on the \sg\ side splits into $a,b\ldots \in 1,\dots,4$  in the
left ($L$) factor of the gauge theory and $r,s\ldots \in 5,\dots,8$ in the
right ($R$) factor.
Although not manifested in this split, $SU(8)$ transformations can
be formally implemented on the gauge theory side of the map of
states. We
 take the attitude that the implementation of $SU(8)$ is better tested
 on amplitudes, for example through the
KLT relations, which read for $n=4$,
\bea \lab{klt4}
M_4(1,2,3,4) &=& - s_{12}\,  A_4(1,2,3,4)_L\, A_4(1,2,4,3)_R \, .
\eea

To apply these to a \sg\ process, one places the images of the
supergravity operators under the map \reef{map} into the gauge factors
on the right side of the relations. We will discuss one example,
although the notation is not fully described until section \ref{s:susy}.
Consider the
scattering amplitude $\<b^-(1)\, b^-_{AB}(2) \,  b^{CD}_+(3)\, b_+(4)\>$ of two gravitons, $b^-$ and $b_+$, and two graviphotons, $b^-_{AB}$ and $b^{CD}_+$, with helicities as indicated.
The gauge theory images of these operators involve  gluons $B^-$,
 gluinos $F^-_a$, and scalars $B_{ab}$, and
 the images of the graviphotons depend on whether the $SU(8)$ indices lie in
the range  $ a,b,.. \in 1,\dots,4$ or $r,s,.. \in 5,\dots,8$. In other words, the helicity-1 particles can decompose either as $0 \otimes 1$ or as $\frac{1}{2} \otimes \frac{1}{2}$.
Using the KLT result \reef{klt4} leads to the formulas
\bea
\nonumber
\big\<b^-(1)\,  b^-_{ab}(2) \,  b^{cd}_+(3)\, b_+(4)\big\>&=&
-s_{12}\, \<B^-(1) \, B^-_{ab}(2)\,  B^{cd}(3)\, B_+(4)\big\>_L\;
\big\<B^-(1)\,  B^-(2)\, B_+(4)\, B_+(3)\>_R \, , \\[1mm]
\nonumber
\big\<b^-(1)\,  b^-_{ar}(2) \,  b^{cs}_+(3)\, b_+(4)\big\>&=&
~s_{12}\, \<B^-(1) \, F^-_{a}(2)\,  F_+^{c}(3)\, B_+(4)\big\>_L\;
\big\<B^-(1)\,  F^-_r(2)\, B_+(4)\, F_+^s(3)\>_R\,.
\eea
The \sg\ amplitude is proportional to the antisymmetric $SU(8)$ tensor
$\d^{CD}_{AB}$, so the product of two bosonic amplitudes in the
first expression must equal (to within a sign)
the product of fermion amplitudes in the second. This agreement is not
a miracle. It must work because the KLT relations are derived from the
low energy limit of superstring theory. Nevertheless we are happy to
see the sometimes intricate way it does work in this and several
other examples we have studied.

The generating function enables us to
 go beyond examples and give a simple argument that all supergravity
 symmetries are consistent with the map \reef{map}. In the MHV sector the \sg\
 generating function factors into the product of gauge theory generating
 functions as
\be
\lab{fact1} \O_n(p_i,\h_{iA}) ~\propto~
F_n(p_i,\h_{ia})_L\,F_n(p_i,\h_{ir})_R\,.
\ee
Symmetry
transformations of supergravity, written in terms of the $\h_{iA}$
variables, automatically work correctly when the $\h_{iA}$ split
into $\h_{ia}$ and $\h_{ir}$, and the transformations applied to the
 product of gauge theory
generating functions on the right side of \reef{fact1}. The situation
is somewhat more complicated, but very similar in the NMHV sector,
where factorization occurs at the level of diagrams.

The plan of the paper is as follows. In section \ref{s:susy} we
discuss the algebra of supercharges and the annihilation operators
in gauge theory and supergravity and then the operator map. We
also discuss the derivation of SUSY Ward identities and their
application in the MHV sector. In section \ref{s:genfunct}  we derive the generating
functions for the MHV sectors of gauge theory and gravity.  An
application to the intermediate state helicity sums is presented
in section \ref{inthelsum}. The connection between state dependence of MHV amplitudes and CFT correlators is discussed in section \ref{s:sfcft}.
We turn our attention to NMHV amplitudes
in section \ref{nmhvampl}. We first discuss recursion relations,
especially those derived from \reef{shift3} which lead to the
MHV-vertex expansion. Using this we derive the NMHV generating
function for gauge theory and discuss its properties. Then we
define the NMHV generating function for gravity and discuss the key
properties of independence of $|X]$ and behavior as $z \to
\infty$. A discussion section concludes the main text.
Our conventions are summarized in appendix \ref{app:conv}.
In appendix \ref{app:N1susy}, we derive the solution of the SUSY \wis\ for 6-point NMHV $\cn =1$ amplitudes.


\setcounter{equation}{0}
\section{SUSY Ward identities and the operator map}
\lab{s:susy}

In section \ref{s:algebras} we set up our notation and present
the $\cn = 4$ and $\cn = 8$ SUSY transformation rules for
annihilation operators of the bosons and fermions of the gauge and
\sg\ theories we are concerned with.
Further information about our conventions is given in
appendix \ref{app:conv}.  In section \ref{s:map} we present the detailed
correspondence between the $16\times 16$ products of pairs of
gauge theory annihilators and the 256 annihilation operators in
\sg , and in section \ref{s:su8} we show how $SU(8)$ transformations can be
implemented formally in the product space. We
discuss SUSY \wis\ for on-shell amplitudes in $\cn =4$ SYM and $\cn
= 8$ supergravity in section \ref{s:swi}. We show by example how to solve the \wis\ in the MHV
sectors of the two theories.

\subsection{Transformation rules of annihilation operators}
\lab{s:algebras}

We focus on annihilation operators because
we adopt the common convention that all particle momenta in an
$n$-point process are viewed as outgoing. An amplitude, such as
the $n$-gluon MHV amplitude, can therefore be viewed as a string
of annihilation operators acting to the left on the
``out'' vacuum.
Thus if $B_+(i)$ and $B^-(i)$ are annihilation operators for gluons of
energy-momentum $p^\m_i$ and helicity $\pm$, we can represent the
color-ordered amplitude as
\be \label{amp}
A_n(1^-,2^-,3^+,\dots,n^+)
\,=\,\big\<B^-(1)B^-(2)B_+(3),\ldots, B_+(n)\big\> \, .
\ee In general the amplitudes are regarded as functions
of complex null energy-momentum vectors $p^\m_i$ which may be
continued to the physical region. If the energy-momentum $p^\m$ is
physical, i.e.~a positive real null vector, then the operator
$B_+(i)$ (or $B^-(i)$) describes a particle in the final state of a physical
process. If $p^\m$ is real, but negative null, then the operator
corresponds to the anti-particle of opposite helicity in the
initial state, which carries physical momentum $-p^\m$.

The bosons and fermions of $\cn = 4$ SYM  theory are described by
the following annihilators, which are listed in order of
descending helicity:
\be \lab{n4ops}
 B_+(p)\, , \hspace{5mm}
 F^a_+(p)\, , \hspace{5mm}
 B^{ab}(p) = {1\over 2}\, \a_4 \,\e^{abcd}B_{cd}(p)\, ,\hspace{5mm}
 F_a^-(p)\, ,\hspace{5mm}
 B^-(p) \, .
\ee
The scalar particles are complex, and satisfy the
indicated $SU(4)$ self-duality condition with $\a_4 = \pm 1$.
The gauge group of the
theory is $SU(N)$ with all particles in the adjoint
representation. Notation for the ``color'' degree of freedom is
omitted, and we consider only ``color-ordered'' amplitudes.

The global symmetry group is $SU(4)$, and we use upper and lower
indices $a, b, = 1,2,3,4$ to distinguish the two inequivalent
conjugate four-dimensional representations. To achieve an $SU(4)$
covariant notation, we separate the left and right chiral
components of the $\cn = 4$ supercharges and write them as $Q^a_\a$
and $\tilde{Q}^{\dot{\a}}_a$ respectively. We then define $Q^a
\equiv - \e^\a\,Q^a_\a$ and
$\tilde{Q}_a = \tilde{\e}_{\dot{\a}} \tilde{Q}^{\dot{\a}}_a$,
where  $\e^\a, \tilde{\e}_{\dot{\a}}$ is the anti-commuting parameter
of SUSY transformations. (See appendix \ref{app:conv} for details.)
Note that $(\tQ_a)^\dagger\,=\,Q^a$.

We now state the independent commutation rules for the operators $Q^a$
and $\tilde{Q}_a$ with the various annihilators:
\bea
\begin{array}{rcl}
\big[\tQ_a,B_+(p)\big] &=& 0\, ,\\[2mm]
\big[\tQ_a,F_+^b(p)\big] &=& \<\epsilon\, p\>\,\d^b_a\,B_+(p)\, ,\\[2mm]
\big[\tQ_a,B^{bc}(p)\big]
&=&\<\epsilon\,p\> \,\big(\d^b_a\,F^c_+(p)- \d^c_a\,F^b_+(p)\big)\, ,\\[2mm]
\big[\tQ_a,B_{bc}(p)\big] &=&\<\epsilon\, p\>\,\a_4 \,
\e_{abcd}\,F_+^d(p)\, ,\\[2mm]
\big[\tQ_a,F_b^-(p)\big] &=&\<\epsilon\, p\>\, B_{ab}(p)\, ,\\[2mm]
\big[\tQ_a,B^-(p)\big] &=& -\<\epsilon\,p\>\,F_a^-(p) \, ,
\end{array}
\hspace{1mm}
\begin{array}{rcl}
[Q^a,B_+(p)] &=& [p\,\epsilon]\,F^a_+(p)\, , \\[2mm]
\big[Q^a,F^b_+(p)\big] &=& [p\,\epsilon]\, B^{ab}(p)\, , \\[2mm]
\big[Q^a,B^{bc}(p)\big]
&=&[p\,\epsilon]\,\a_4 \,\e^{abcd}\,F_d^-(p)\, , \\[2mm]
\big[Q^a,B_{bc}(p)\big]
&=&[p\,\epsilon]\,\big(\d^{a}_b\,F_c^-(p)- \d^{a}_c\,F_b^-(p)\big)\, ,\\[2mm]
\big[Q^a,F_b^-(p)\big]
&=&-  [p\,\epsilon]\,\d^{a}_b\,B^-(p)\, , \\[2mm]
\big[Q^a,B^-(p)\big] &=& 0\, .
\end{array}
\lab{n4tQ}
\eea
Note that $\tQ_a$  raises the helicity of all operators and involves
the spinor angle bracket $\<\e\,p\>$ in which $|p\>
\leftrightarrow \l_{p}^{\dot{\a}}$ is the dotted spinor for a
particle of momentum $p^\m$. Similarly, $Q^a$
lowers the helicity and spinor square brackets $[p\, \e]$ appear.
Commutators with $B^{bc}(p)$ and $B_{bc}(p)$ are related by self-duality.
The $Q^a$ and $\tQ_a$ operators generate independent Ward
identities for $n$-point amplitudes. We will primarily be
concerned with those for $\tQ_a$.

For distinct SUSY parameters $\epsilon_1,\,\tilde{\epsilon}_1$ and
$\epsilon_2,\,\tilde{\epsilon}_2$, we define $Q^a_i = -
\epsilon_i^\a Q^a_\a$ and $\tQ_{ia} =
\tilde{\epsilon}_{i\dot\a}\tilde{Q}^{\dot{\a}}_{ia}$. For any
operator $\co$ above, the SUSY algebra reads
\bea
\nonumber
\big[ [Q_1^a,\tQ_{2b}], \co\big]
&=&
\big[Q_1^a,[\tQ_{2b},\co]\big] \,-\,
\big[\tQ_{2b},[Q_1^a,\co]\big]\,
=\, \<\epsilon_2\,p\>[p\,\epsilon_1]\, \d^a_b \, \co \, ,\\[1mm]
\lab{n4comm}
\big[[Q_1^a,Q_2^b], \co\big]
&=&
\big[Q_1^a,[Q_2^b,\co]\big] \,-\,
\big[Q_2^b,[Q_1^a,\co]\big]\,=\,0 \, ,\\[1mm]
\nonumber
\big[[\tQ_{1a},\tQ_{2b}], \co\big]
&=&
\big[\tQ_{1a},[\tQ_{2b},\co]\big] \,-\,
\big[\tQ_{2b},[\tQ_{1a},\co]\big]=0\,.
\eea

Next we proceed in a similar fashion to discuss the transformation
rules of $\cn = 8$ \sg, in which the annihilation operators for
the 128 bosons and 128 fermions are
\bea
\nonumber
&b_+(p)\, , \hspace{7mm}
f^A_+(p)\, , \hspace{7mm}
b^{AB}_+(p)\, , \hspace{7mm}
f^{ABC}_+(p)\, , \\[2mm]
\lab{n8ops}
&
b^{ABCD}(p)\, =\, \frac{1}{4!}\,\a_8 \,\e^{ABCDEFGH}\, b_{EFGH}(p)\, ,
\\[2mm] \nonumber &
f_{ABC}^-(p)\,, \hspace{7mm}
b_{AB}^-(p)\, , \hspace{7mm}
f_A^-(p)\,, \hspace{7mm}
b^-(p)\, .
\eea
The 70 scalars satisfy an $SU(8)$ self-duality condition with $\a_8 =
\pm 1$. The notation is redundant, since the information on particle
type and helicity is determined by the number and position of the
global symmetry indices.

There are chiral spinor supercharges $Q^A_\a$ and
$\tQ^{\dot{\a}}_A$ which transform in the inequivalent 8 and
$\bar{8}$ representations. We contract these charges with a SUSY
Grassmann parameter and define  $Q^A \equiv - \e^\a\,Q^A_\a$ and
$\tilde{Q}_A \equiv \tilde{\e}_{\dot{\a}} \tilde{Q}^{\dot{\a}}_A$.
It is then straightforward to write  $SU(8)$ covariant commutators
with annihilation operators:

\begin{eqnarray}
 \begin{array}{rcl}
 \scriptstyle{\big[\tQ_A, ~ b_+\big]} &=& \scriptstyle{0}\, ,\\[2mm]
 \scriptstyle{\big[\tQ_A,~ f^B_+\big]} &=& \scriptstyle{\<\epsilon\, p\>\, \d_A^{B} \, b_+} \, ,\\[2mm]
 \scriptstyle{\big[\tQ_A,~ b^{BC}_+\big]} &=&
   \scriptstyle{\< \epsilon\,  p\> \big(\d_A^B\, f^C_+-\d_A^{C}\, f^B_+\big)} \,  ,\\[2mm]
 \scriptstyle{\big[\tQ_A,~ f^{BCD}_+\big]} &=&
  \scriptstyle{\<\epsilon\,  p\>\big(\d_A^{B}\,b^{CD}_++\d_A^{C}\,b^{DB}_+
    +\d_A^{D}\,b^{BC}_+\big) \, ,} \\[2mm]
 \scriptstyle{\big[\tQ_A,~ b^{BCDE}\big]}
   &=& \scriptstyle{\<\epsilon\,  p\> \big(\d_A^{B}\,f^{CDE}_+-
  \d_A^{C}\,f^{DEB}_+} \\[1mm]
  && ~~~~~~~  \scriptstyle{+\d_A^{D}\,f^{EBC}_+
  -\d_A^{E}\,f^{BCD}_+\big)\,  , }\\[2mm]
 \scriptstyle{\big[\tQ_A,~ b_{BCDE}\big]} &=&
   \scriptstyle{\<\epsilon\,  p\> \, \frac16 \, \a_8 \,
   \e_{ABCDEFGH}\, f^{FGH}_+ \, ,}\\[2mm]
 \scriptstyle{\big[\tQ_A,~ f_{BCD}^-\big]} &=&
     \scriptstyle{\< \epsilon\,  p\> \,b_{ABCD}  \, , }\\[2mm]
 \scriptstyle{\big[\tQ_A,~ b_{BC}^-\big]} &=&
  \scriptstyle{ - \<\epsilon\,  p \>\, f_{ABC}^- \, ,}\\[2mm]
  \scriptstyle{\big[\tQ_A,~ f_B^-\big]} &=&
    \scriptstyle{\< \epsilon\,  p\>  \,b_{AB}^- \, ,}\\[2mm]
  \scriptstyle{\big[\tQ_A, ~ b^-\big]} &=&
  \scriptstyle{-\<\epsilon\,  p\> \, f_A^- \, ,}
\end{array}
\begin{array}{rcl}
  \scriptstyle{\big[Q^A, ~ b_+\big]} &=&
  \scriptstyle{\left[p\,  \epsilon \right] f^A_+ \, ,}\\[2mm]
 \scriptstyle{\big[Q^A,~ f^B_+\big]} &=&
 \scriptstyle{\left[p\,  \epsilon \right]\, b^{AB}_+ \, ,}\\[2mm]
 \scriptstyle{\big[Q^A,~ b^{BC}_+\big]} &=&
    \scriptstyle{\left[p\,  \epsilon \right]\, f^{ABC}_+ \, ,}\\[2mm]
 \scriptstyle{\big[Q^A,~ f^{BCD}_+\big]} &=&
  \scriptstyle{\left[p\, \epsilon \right]\, b^{ABCD}  \, ,}\\[2mm]
 \scriptstyle{\big[Q^A,~ b^{BCDE}\big]}
   &=& \scriptstyle{\left[p\,  \epsilon \right] \,
   \frac16\, \a_8 \,\e^{ABCDEFGH} \, f_{FGH}^-  \, ,}\\[2mm]
 \scriptstyle{\big[Q^A,~ b_{BCDE}\big]} &=&
   \scriptstyle{\left[p\, \epsilon \right] \big(\d^A_{B}\, f_{CDE}^- -
    \d^A_{C}\, f_{DEB}^-}\\[1mm]
   && ~~~~~~~ \scriptstyle{+\d^A_{D}\, f_{EBC}^- -\d^A_{E}\,
   f_{BCD}^-\big) \, ,}\\[2mm]
 \scriptstyle{\big[Q^A,~ f_{BCD}^-\big]} &=&
  \scriptstyle{-\left[p\, \epsilon \right]
  \big(\d^A_{B}\,b_{CD}^- +\d^A_{C}\,b_{DB}^-
    +\d^A_{D}\,b_{BC}^- \big)  \, ,} \\[2mm]
  \scriptstyle{\big[Q^A,~ b_{BC}^-\big]} &=&
   \scriptstyle{\left[p\,  \epsilon  \right] \big(\d^A_{B}\, f_C^-
   -\d^A_{C}\,f_B^-\big)  \, ,}\\[2mm]
  \scriptstyle{\big[Q^A,~ f_B^-\big]} &=&
  \scriptstyle{-\left[p\, \epsilon  \right]\d^A_{B} \,b^- \, ,}\\[2mm]
 \scriptstyle{\big[Q^A, ~ b^-\big]} &=& \scriptstyle{0 \, .}
 \lab{n8susy}
 \end{array}
\end{eqnarray}
The supersymmetry generators satisfy \reef{n4comm} for any operator $\co$ above.

Supercharge commutators with creation operators can be
obtained as the adjoints of the relations given in \reef{n4tQ} and
\reef{n8susy}.  Phases in these commutators have been fixed to be
compatible with crossing.  Crossing  symmetry relates an S-matrix
element containing a particle  with physical (positive null)
momentum in the initial state to the amplitude containing its
anti-particle with opposite helicity and unphysical (negative
null) momentum in the final state. Thus  the SUSY transformation
of any creator $a(p,\pm)^*$ must agree with that of the
annihilator  $a(-p, \mp)$ multiplied by the conventional
\cite{aderetal} crossing phase $(-)^{s-\l}$ of helicity amplitudes
(which has the value $-1$ only for negative helicity fermions).
Note that spinors for negative null momenta satisfy
$|-p\>=-|p\>,~~ |-p]=|p]$.


\subsection{The operator map }
\lab{s:map}

The precise operator map between $(\cn =8) \leftrightarrow (\cn =
4)_L \otimes (\cn =4)_R$ is presented in Table \ref{tab:map}.
Operators in the $R$ gauge theory  are dressed with tildes whereas
the operators of the $L$ factor are undecorated. The entries in
the map are determined, up to signs, by matching the helicity and
global symmetry properties of \sg\ operators with products of
gauge theory operators.  Unfixed signs are then determined by
compatibility with the scalar self-duality conditions and
especially by the consistent action of the supercharges of the
$\cn =8$ and $\cn =4$ theories.

To discuss the implementation of the SUSY commutators we denote a
generic annihilation operator  by $a$ in \sg\ and by $A$ and $\tA$
in  the $L$ and $R$ copies of the gauge theory.  The image  of any
$a$ under the map \reef{map} is a specific
 product $A\otimes \tA$.  A supercharge
component $Q^a$ from the first $SU(4)$ sector acts non-trivially
only on $A$, while $Q^r$ from the second sector acts non-trivially
only on $\tA$. Thus we have the scheme
\bea
  \nonumber
  a(p) ~&\leftrightarrow&~ A(p)\otimes \tA(p) \, ,\\[1mm]
  \lab{wigen}
  \left[Q^a,a\right] ~&\leftrightarrow&~ [Q^a,A\otimes\tA]
  \equiv [Q^a,A]\otimes \tA \, ,\\[1mm]
  \nonumber
  \left[Q^r,a\right] ~&\leftrightarrow&~ [Q^r,A\otimes\tA]\,
  \equiv A \otimes [Q^r,\tA]\,,
  \eea
with similar definitions of the action of $\tQ_a$ and $\tQ_r$.  We
then require that the left and right sides of \reef{wigen} still
map properly when the transformation rules of
section \ref{s:algebras} are used.  This determines the signs of
entries in Table 1.

Here are two examples, interesting because the two sectors mix. The first example is
\bea
\left[Q^a\,,\, b^{br}_+(p)\right] &=& [p\,\epsilon]  \,f^{abr}_+(p) \, ,\\
\left[Q^a\,,\,F^b_+(p)\otimes \tilde{F}^r_+(p)\right] &=&
[p\,\epsilon] \, B(p)^{ab}\otimes \tilde{F}^r_+(p)\, .
\eea
This is
compatible with the supersymmetry algebras because the right sides are images under
the map \reef{84map}. The other example is
\bea
\left[\tQ_r\,,\, b^{abcs}(p)\right] &=&-\<\epsilon\,p\>\,\d^s_r\,f^{abc}_+(p)\, ,\\
\left[\tQ_r\,,\,F_d^-(p)\otimes \tilde{F}^s_+(p)\right]
&=&-\<\epsilon\,p\>\, \d^s_r\, F_d^-(p)\tilde{B}_+(p)\,. \eea After
multiplication of the second equation by $\a_4\,\e^{abcd}$, we see
that the map works properly. We have checked explicitly that all
entries in the map are consistent with the transformation rules.

There is a choice of the scalar self-duality phases $\a_8$, $\a_4$, and
$\tilde{\a}_4$ in the $\cn=8$ \sg\ theory and in the two $N=4$ SYM
factors. It turns out that consistency of the map with the
commutator algebras requires that
\bea
\lab{mapcon}
  \a_4 \, \tilde{\a}_4 = \a_8 \, .
\eea
We leave $\a_4$, $\tilde{\a}_4$, and $\a_8$ arbitrary
in the map in Table \ref{tab:map}, but in applications below we will often set
$\a_4=\tilde{\a}_4=\a_8=1$.

\begin{table}[t]
\begin{eqnarray}
  \begin{array}{rcl|rcl}
  \hline\\[-3mm]
  b_+ &=& B_+ \, \tilde{B}_+ & ~~~~~
  b^- &=& B^- \, \tilde{B}^- \\[1mm]
  \hline\\[-3mm]
  f^{a}_+ &=& F^{a}_+ \, \tilde{B}_+ &
  f_a^- &=& F_a^- \, \tilde{B}^- \\[1mm]
  f^{r}_+ &=& B_+ \, \tilde{F}^{r}_+ &
  f_r^- &=& B^- \, \tilde{F}_r^- \\[1mm]
  \hline\\[-3mm]
  b^{ab}_+ &=& B^{ab} \, \tilde{B}_+ &
  b_{ab}^- &=& B_{ab} \, \tilde{B}^- \\[1mm]
  b^{ar}_+ &=& F^{a}_+ \, \tilde{F}^{r}_+ &
  b_{ar}^- &=& - F_{a}^- \, \tilde{F}_r^- \\[1mm]
  b^{rs}_+ &=& B_+ \, \tilde{B}^{rs} &
  b_{rs}^- &=& B^- \, \tilde{B}_{rs} \\[1mm]
  \hline\\[-3mm]
  f^{abc}_+&=& \a_4\, \eps^{abcd} \, F_d^- \, \tilde{B}_+ ~~&
  f_{abc}^- &=& -\a_4\, \eps_{abcd} \, F^{d}_+ \, \tilde{B}^- \\[1mm]
  f^{abr}_+ &=& \, B^{ab} \, \tilde{F}^{r}_+ ~~&
  f_{abr}^- &=& \, B_{ab} \, \tilde{F}_r^- \\[1mm]
  f^{ars}_+ &=& \, F^{a}_+ \, \tilde{B}^{rs} ~~&
  f_{ars}^- &=& \, F_{a}^- \, \tilde{B}_{rs} \\[1mm]
  f^{rst}_+ &=& \, \tilde{\a}_4\, \eps^{rstu} \, B_+ \, \tilde{F}_u^- &
  f_{rst}^-
  &=& \, -\tilde{\a}_4\, \eps_{rstu} \, B^- \, \tilde{F}^{u}_+
  \\[1mm]
  \hline\\[-3mm]
  b^{abcd} &=& \a_4\, \eps^{abcd} \, B^- \, \tilde{B}_+ ~~&
  b_{abcd} &=& \a_4\, \eps_{abcd} \, B_+ \, \tilde{B}^- \\[1mm]
  b^{abcr} &=& \a_4\, \eps^{abcd} \, F_d^- \, \tilde{F}^{r}_+ ~~&
  b_{abcr} &=& \a_4\, \eps_{abcd} \, F^{d}_+ \, \tilde{F}_r^- \\[1mm]
  b^{abrs} &=& \, B^{ab} \, \tilde{B}^{rs} ~~&
  b_{abrs} &=& \, B_{ab} \, \tilde{B}_{rs} \\[1mm]
  b^{arst} &=& \tilde{\a}_4\, \eps^{rstu} \, F^{a}_+ \, \tilde{F}_u^- ~~&
  b_{arst}
  &=& \tilde{\a}_4\, \eps_{rstu} \, F_a^- \, \tilde{F}^{u}_+
  \\[1mm]
  b^{rstu} &=& \tilde{\a}_4\, \eps^{rstu} \, B_+ \, \tilde{B}^- ~~&
  b_{rstu} &=& \tilde{\a}_4\, \eps_{rstu} \, B^- \, \tilde{B}_+
  \\[1mm]
  \hline
  \end{array}
\label{84map}
\end{eqnarray}
\caption{Operator map for annihilators of $(\cn =8)
\leftrightarrow (\cn = 4)_L \otimes (\cn =4)_R$. Indices $a,b,c,d=
(1,2,3,4)$ and $r,s,t,u=(5,6,7,8)$ refer  to the splitting of
$SU(8)$ into the two separate $SU(4)$ factors.} \lab{tab:map}
\end{table}

\subsection{$SU(8)$ symmetry and the operator map}
\lab{s:su8}

The generators of the fundamental representation of $SU(8)$ are
the set of $63~8\times 8$ traceless matrices:
\be \lab{su8}
(T^A_B)^C{}_D\,=\, \d^A_D\,\d^C_B - \frac18 \d^A_B\,\d^C_D \, ,
\ee
in which  $A,B$ denote the Lie algebra element, and $C,D$ are row and
column indices.  The commutators are
\be \lab{com8}
\left[T^A_B\,,\,T^C_D\right]\,=\,\d^A_D\,T^C_B\,-\,\d_B^C\, T^A_D\, .
\ee

The algebra decomposes with respect to the subgroup $SU(4)_L\otimes
SU(4)_R\otimes U(1)$.  We use indices ($a, b,\dots =
1,2,3,4$ and $r, s,\dots = 5,6,7,8$). After a minor rearrangement of
the basis of \reef{su8}, we obtain a set of 63 matrices whose
non-vanishing elements are
\bea
\nonumber
&&\hspace{1.2cm}
(T^a_b)^c{}_d ~=~  \d^a_d\,\d^c_b - \frac14 \d^a_b\,\d^c_d\, ,
 \hspace{1cm}
(T^r_s)^t{}_u ~=~ \d^r_u\,\d^t_s - \frac14 \d^r_s\,\d^t_u\, ,\\[2mm]
&&
(T)^c{}_d ~=~\d^c_d\, ,
\hspace{8mm}
(T)^t{}_u~=~-\d^t_u\, ,
\hspace{8mm}
(T^a_s)^t{}_d~=~\d^a_d\d^t_s\, ,
\hspace{8mm}
(T^r_b)^a{}_s  = \d^r_s \d^a_b \, .
 \lab{dec}
\eea
We now define the action of the corresponding Hilbert space
operators on the states of the operator map.  The generators
$T^a_b$ and $T^r_s$ have the usual matrix action of $SU(4)$,
defined in \reef{dec}, on gauge theory operators. Nothing special
is required.  Examples make things clear:
\bea
\nonumber
\big[T^a_b\,,\, b^{ct}_+\big]&=&\d^c_b\, b^{at}_+ \, ,\hspace{1.25cm}
\big[T^a_b\,,\, F^{c}_+\otimes\tilde{F}^{t}_+\big]
~=~ \d^c_b\, F^{a}_+\otimes\tilde{F}^{t}_+\, ,\\[1mm]
\big[T^a_b\,,\, f^-_{c}\big]&=&-\d^a_c\,f^-_b\, ,
~~~~~~~~
\big[T^a_b,F^-_c\otimes \tilde{B}^-\big]
~=~ -\d^a_c\,F^-_b\otimes
\tilde{B}^-\,.
\lab{unmix}
\eea

The remaining generators are more subtle, but very simple. They have
no well defined action on single operators of the gauge theory,
but we {\it define}  their action on tensor products of gauge
theory operators to match the appropriate supergravity states. The
generator $T$ is diagonal on all states.   Thus, for example, \be
\lab{exD} \big[T\,,\,f^{+abc}\big]~=~3\,f^{+abc}\, , ~~~~~~~
\big[T\,,\,\a_4\e^{abcd}F^-_d\tilde{B}^+\big]~\equiv~ 3 \,
\a_4\e^{abcd}F^-_d\tilde{B}^+\,. \ee For the mixed generators
$T^a_s,~T^r_b$ the definitions require changes from boson to
fermion operators in each gauge theory factor. Hence 
\be
\lab{exmix} \big[T^r_b\,,\,f^-_{cds} \big]~=~ -\d^r_s\,f^-_{cdb}\,
, ~~~~~~~~ \big[T^r_b\,,\,B_{cd}\otimes\tilde{F}^-_s\big]
~=~\d^r_s\,\a_4\e_{cdbe}F^{+e}\otimes\tilde{B}^-\, 
\ee 
The
consistency test for any claimed realization of $SU(8)$ is that
the commutation relations \reef{com8} are satisfied.  But this is
certainly true here, by explicit construction, since our
definitions simply track the conventional implementation of
$SU(8)$ in \sg.

This implementation of $SU(8)$ in the operator map is correct but
formal. The acid test is that  \sg\ amplitudes constructed from
gauge theory transform correctly.  This requires the kind of
non-miracle discussed in the introduction. The dynamical parts of
products of very different gauge theory amplitudes must agree, and
so must their group theory factors. To show that this non-miracle
happens, we will use SUSY \wis.

 \subsection{SUSY Ward identities for on-shell amplitudes}
\lab{s:swi}

 To begin the discussion, we use the generic notation of \cite{gris}.
 An annihilation operator of $\cn = 4$ SYM or $\cn = 8$ \sg\ is denoted either by
 $\a_i$ or $\b_i$.  The subscript $i$ indicates particle momentum, while helicity
 and global symmetry indices are suppressed.
 For a pair of supercharges $Q^a,~\tQ_a$ of $\cn = 4$ SYM with fixed $SU(4)$ index,
 an $\a$ operator is defined as one for which $[Q^a, \a]$  is
 non-vanishing, and a $\b$ operator is one
for which $[\tQ_a,\b]$ is non-vanishing.  It is clear that $ [Q^a,
\a]= [p\,\e]\, \b$ and $[\tQ_a,\b]=\<\e\, p\>\,  \a$.  The division into
$\a$- and $\b$-operators depends on the index choice $a$. For example,
the $\a,~\b$ operators for the supercharge pair $Q^1,~\tQ_1$ are
\bea
\a~\mathrm{operators:}&& B_+(p)\, , ~~F^b_+(p) \, ,~~
B^{bc}(p)\, ,~~B_{1b}(p)\,, ~~F_1^-(p)\, . \\[1mm]
\b~\mathrm{operators:}&& F^1_+(p)\, ,~~B^{1b}(p)\, ,~~B_{bc}\,
,~~F_b^-(p)\, ,~~B^-(p)\, , \eea
 where $b,c \ne 1$. The definition of $\a,~\b$ operators in $\cn = 8$ \sg\ is entirely analogous.

 The basic Ward identities are simply the statements that, since supercharges annihilate  the vacuum,
\bea \lab{wif}
 0 &=& \big\<\big[\tQ_a, \b_1\b_2\ldots \b_n\,\a_{n+1}\a_{n+2}\dots\a_{n+m}\big]\big\> \, , \\[1mm] \nonumber
  0 &=& \big\<\big[Q^a, \b_1\b_2\ldots \b_n\,\a_{n+1}\a_{n+2}\dots\a_{n+m}\big]\big\>\, .
\eea
By adding and subtracting terms, we convert \reef{wif} into a sum
of commutators $[\tQ_a,\b_i]$ or $[Q^a,\a_j]$. We can then rewrite
\reef{wif} as
\bea
\lab{wiq}
 0\,&=&\,\sum_{i=1}^n\<\epsilon \, i\> \,
\<\b_1\dots\a_i\ldots\b_n\,\a_{n+1}\ldots\a_{n+m}\> \, , \\
\lab{witq}
0 &=&
\sum_{j=n+1}^{n+m}[j\,\epsilon]\, \<\b_1\ldots\b_n\,\a_{n+1}\dots\b_j\ldots\a_{n+m}\>\,.
 \eea
Since the spinors have two components, the analytic and
anti-analytic expressions each contain two independent constraints
on the amplitudes. To obtain a useful identity one must start
with a string of operators in \reef{wif} which contains an odd
number of fermion annihilators. Then the individual amplitudes
which appear in the constraints contain an even number of
fermions. Otherwise they vanish trivially.  The ordering of operators is relevant
in gauge theory because amplitudes are color ordered, but it has no significance in supergravity.

Let's consider the two cases in which the initial string of
operators in \reef{wif} contains only one or two $\a$
operators, respectively. Then the $Q^a$ \wis\ read
\bea \lab{onea}
  \left[ (n+1)\,\epsilon\right]\<\b_1\ldots\b_n\b_{n+1}\>&=&0 \, , \\[1.5mm]
\lab{twoa}
 \left[ (n+1)\,\epsilon\right]\<\b_1\ldots\b_n\a_{n+1}\b_{n+2}\>\,+\,\left[ (n+2)\,\epsilon\right]\<\b_1\ldots\b_n\b_{n+1}\a_{n+2}\>&=&0\,.
\eea
 We now exploit the freedom to choose the two-component spinor $\e_\a$.
 We can choose it so that $ [ (n+1)\,\e] \ne 0$.
 Then \reef{onea} tells us that any amplitude which contains only
 $\b$ operators must vanish.  To exploit the information in
 \reef{twoa} we choose, in turn, $|\e ] \sim |n+2]$ and then $|\e]\sim|n+1]$.
 We learn that any amplitude with $n$ $\beta$ operators and one $\a$ operator must vanish.
 By similar arguments, we can use the $\tQ_a$ Ward identity to
 show that any amplitude containing at most one $\b$ operator must vanish.
 These statements comprise the well known helicity conservation rules for $n$-point functions.
 For amplitudes containing only gluons, they read $A_n(1^+,2^+,\ldots,n^+)=0$ and
\mbox{$A_n(1^+,\ldots,(n-1)^+,n^-)=0$.}

Relations between different amplitudes are obtained when the
initial string contains $k\ge3~\b$ operators plus $n-k\ge1~\a$
operators. The case of exactly three $\b$ operators is
particularly easy to analyze and very useful. The analytic Ward
identity reads \be\lab{mhv}
 \<\e\,1\>\,\<\a_1\b_2\b_3\a_4\ldots\a_n\>\,+\,\<\e\,2\>\,
 \<\b_1\a_2\b_3\a_4\ldots\a_n\>\,+\,\<\e\,3\>\,\<\b_1\b_2\a_3\a_4\ldots\a_n\>\,\,=\,0\,.
 \ee
 As stated above this equation contains two independent relations
 among the three amplitudes involved. By choosing $|\e\>=|2\>$ and then $|\e\>=|1\>$, we obtain
 \bea \lab{mhva}
 \<\a_1\b_2\b_3\a_4\ldots\a_n\>
 &=& -\frac{\<2\,3\>}{\<2\,1\>}\<\b_1\b_2\a_3\a_4\ldots\a_n\>\, ,\\
 \<\b_1\a_2\b_3\a_4\ldots\a_n\>&=&-\frac{\<1\,3\>}
 {\<1\,2\>}\<\b_1\b_2\a_3\a_4\ldots\a_n\>\,.
\eea

An example of these relations is  the
$\big\<[ \tQ_a,B^-(1)B^-(2)F^b_+(3) B_+(4)\ldots B_+(n)]\big\>=0$ Ward identity
in gauge theory.  The two constraints above then become
\bea
\lab{gGg}
\big\<F_a^-(1) B^-(2) F^b_+(3) B_+(4)\ldots B_+(n)\big\> &=&
\d^b_a\frac{\<2\,3\>}{\<2\,1\>}
\big\<B^-(1)B^-(2)B_+(3)\ldots B_+(n)\big\>\, , \\
\lab{Ggg}
 \big\<B^-(1)F_a^-(2) F^b_+(3) B_+(4)\ldots B_+(n)\big\> &=&
 \d^b_a\frac{\<1\,3\>}{\<1\,2\>}
 \big\<B^-(1)B^-(2)B_+(3)\ldots B_+(n)\big\> \, .
\eea
Thus an amplitude containing a pair of opposite
helicity gluinos is related to the well known MHV $n$-gluon
amplitude. For this reason the set of amplitudes with two $\b$
operators and $n-2$ $\a$ operators is called the MHV sector of the
theory. Note that the gluinos can be placed in any positions by
change in the placement of the three initial $\b$ operators.

As another example  of an MHV Ward identity in the gauge theory,
consider
$$
\big\<  [\tQ_a, B^-(1)  F_b^-(2) B^{cd}(3)\, B_+(4)\dots B_+(n) ] \big\> = 0 \, ,
$$
and use \reef{Ggg} to simplify
the $\< \e\, 3 \>$ terms. With $|\e\>\sim |1\>$ or $\sim |2\>$ we get
two \wis:
\bea
  \lab{GSS}
  \big\<  B^-(1)\,  B_{ab}(2) \, B^{cd}(3) \, B_+(4)\dots B_+(n)  \big\>
  &=&
  2\d_{ab}^{cd}\, \frac{\<13\>^2}{\<12\>^2}  \,
  \big\<  B^-(1)\,  B^-(2) \, B_+(3)\dots B_+(n)  \big\> \, ,\\[1mm]
  \lab{ggS}
  \big\<  F_a^-(1)\,  F_b^-(2) \, B^{cd}(3) \, B_+(4)\dots B_+(n)   \big\>
  &=&
  2 \d_{ab}^{cd}\, \frac{\<13\>\<23\>}{\<12\>^2} \,
  \big\<  B^-(1)\,  B^-(2) \,  B_+(3)\dots B_+(n)  \big\> \, . ~~~~~~~
\eea
Anti-symmetrizers are defined as $\d_{a_1\dots
a_n}^{b_1\dots b_n} = \frac{1}{n!}(\d_{a_1}^{b_1} \cdots
\d_{a_n}^{b_n} \pm \mathrm{perms})$. It is also easy to derive, see \cite{dixon},
\bea
  \lab{YMid}
  \big\<  B_+(1) \ldots   B^-(i) \ldots B^-(j) \ldots B_+(n)  \big\>
  &=&
  \frac{\<i\, j\>^4}{\<1\,2\>^4}\,
  \big\<  B^-(1)\,  B^-(2) \, B_+(3)\dots B_+(n)  \big\> \, .
\eea

Let's examine the anti-analytic $\big\<[Q^b,B^-(1)\,F_a^-(2)
B_+(3)\ldots B_+(n)]\big\>=0$ Ward identity, which gives the
relation
\be
  \lab{nonal}
  [2\,\e] \d^b_a
  \big\<B^-(1)B^-(2)B_+(3)\ldots B_+(n)\big\> + \sum_{j=3}^{n} [j \,\e] \big\<B^-(1)F_a^-(2) B_+(3)\ldots F_+^b(j)\ldots B_+(n)\big\> ~=~0\,.
\ee
If we use the previous result \reef{Ggg} and its extension to
the case where $F_+^a(j)$ appears, then \reef{nonal} reduces to
\be
\lab{moco}
\bigg(\sum_{j=2}^{n} \<1\,j\>[j\,\e]\bigg)
\d^b_a \big\<B^-(1)B^-(2)B_+(3)\ldots B_+(n)\big\> = 0\,.
\ee
But the sum of products of spinor brackets vanishes because of
momentum conservation, so the anti-analytic Ward identity
\reef{nonal} is satisfied, after information from the analytic
\wis\ is used. This is a general feature of the MHV sector, but it
is not true in the NMHV sector and beyond.

Ward identities for amplitudes in the MHV sector of $\cn = 8$ \sg\
can be obtained in a similar fashion. With appropriate choices of
the $\a,\b$ operators (and of $|\e\>$) one can derive the useful
results:
\bea
\lab{n8wi}
\big\<b^-(1) f_A^-(2) \,b_+(3)\ldots f^B_+(k)\ldots b_+(n)\big\>
&=&\d^B_A\,\frac{\<1 k\>}{\<1 2\>}\big\<b^-(1)b^-(2) b_+(3) \ldots b_+(n)\big\>\, ,\\
\nonumber
 \big\<b^-(1) b_{AB}^-(2)\, b_+(3)\ldots b^{CD}_+(k)\ldots b_+(n)\big\>
 &=& 2 \, \d^{CD}_{AB}\,\frac{\<1 k\>^2}{\<1 2\>^2}\big\<b^-(1)b^-(2) b_+(3) \ldots b_+(n)\big\>\, ,\\
\nonumber
\big\<b^-(1) f_{ABC} ^-(2)\,b_+(3)\ldots f^{DEF}_+(k)\ldots b_+(n)\big\>
&=&3! \, \d^{DEF}_{ABC}\,\frac{\<1 k\>^3}{\<1 2\>^3}\big\<b^-(1)b^-(2) b_+(3) \ldots b_+(n)\big\>\, ,\\
\nonumber
 \big\<b^-(1) b_{ABCD}(2)\, b_+(3)\ldots b^{EFGH}(k)\ldots b_+(n)\big\>
 &=& 4! \, \d^{EFGH}_{ABCD}\,\frac{\<1 k\>^4}{\<1 2\>^4}\big\<b^-(1)b^-(2) b_+(3) \ldots b_+(n)\big\>\, .
\eea

 It is easy to write generic \wis\ in the NMHV sector, but much
 harder to extract useful information from them.
 From $\big\<[\tQ_a\,,\, \b_1\b_2\b_3\b_4\a_5\dots \a_n]\big\>\,=\,0$,  one derives
 \be \lab{nmwi}
 \<\e\, 1\>\, f_1+
 \<\e\, 2\>\, f_2+
 \<\e\, 3\>\, f_3+
 \<\e\, 4\>\, f_4~=~0\,,
 \ee
 with
 $f_1=\<\a_1\b_2\b_3\b_4 \a_5\dots\a_n\>$,
 $f_2=\<\b_1\a_2\b_3\b_4 \a_5\ldots\a_n\>$,
 $f_3=\<\b_1\b_2\a_3\b_4 \a_5\ldots\a_n\>$,
 and
  $f_4=\<\b_1\b_2\b_3\a_4 \a_5 \ldots\a_n\>$.
 By choice of $\e$ one can derive two independent relations among
the four amplitudes. Given one set of amplitudes $f_i$ which
satisfy \reef{nmwi}, then one may use the Schouten identity to show
that another one is given by $f_1 +
\<2\,3\>f_0,~~f_2+\<3\,1\>f_0,~~f_3+\<1\,2\>f_0,~~f_4$, where
$f_0$ is an arbitrary function.  Thus additional information is
required to specify the amplitudes \cite{gris}.  The solution for
$\cn = 1$ 6-point functions in \cite{gris} is rederived by
spinor-helicity methods in appendix \ref{app:N1susy}. It could be
very useful to develop techniques to solve the NMHV \wis,
particularly for extended SUSY.

\vspace{3mm}

Many of the properties we have illustrated above in the examples are neatly encoded in generating functions for MHV amplitudes. This is our next subject.

\setcounter{equation}{0}
\section{Generating functions for MHV amplitudes}
\label{s:genfunct}
In section \ref{s:swi} we showed that SUSY \wis\
are quite simple in the MHV sectors of $\cn = 4$ SYM and $\cn =8$
\sg, indeed amenable to step-by-step solution. Nevertheless, a
systematic method of solution for the entire MHV sector is awkward
at best.  Nor do we yet know a simple way to determine whether
particular amplitudes, such as $\<B^-(1)\,F_+^a(2)\,F^b_+(3)\,
F^c_+(4) \,F^d_+(5)\,B_+(6)\dots B_+(n)\>$ or the 8-gluino
amplitude  mentioned in the Introduction are within the MHV
sector. The remarkable generating function derived for the gauge
theory by Nair \cite{nair}, and further developed by Georgiou,
Glover, and Khoze \cite{ggk}, provides very simple answers to these
questions.  In this section we explain and elucidate new
properties of this generating function and generalize it to the
MHV sector of $ \cn =8$ \sg.  Then we show that it embodies a
clear explanation of the compatibility of $\cn = 8$ SUSY and
$SU(8)$ global symmetry with the map \reef{map}.

\subsection{Gauge theory}
\lab{s:gt}

Suppose that we are interested in the full sector of MHV $n$-point
functions  in the gauge theory. Following Nair, we introduce a set
of $4n$ anti-commuting variables $\eta_{i a}$ in which
$i=1,\ldots,n$ indicates particle momentum, and $a=1,2,3,4$ is the
$SU(4)$ index.
The generating function depends on the $\h_{i a}$ and the
(commuting) spinors $\tilde{\l}^{\dot{\a}}_i \leftrightarrow |i\>$
which encode particle momenta. The generating function is
\be
\lab{n4gen} F_n =\Big(\prod_{i=1}^{n}\<i\,
(i+1)\>\Big)^{-1}\d^{(8)}\Big(\sum_{i=1}^{n}|i\>\eta_{ia}\Big)\,,
\ee and the 8-dimensional $\d$-function can be expressed as
\be\lab{del8}
 \d^{(8)}\Big(\sum_{i=1}^{n}|i\>\eta_{ia}\Big)\,=\,
 \frac{1}{16}\prod_{a=1}^{4} \sum_{i,j=1}^{n}\<i\,j\>\,\eta_{ia}\,\eta_{ja}\,.
\ee
It is a sum of $ \big(\frac{1}{2}n(n-1)\big)^4$ terms, each involving a product of 8
distinct $\eta_{ia}$;  it is invariant under $SU(4)$
transformations of the $\eta_{ai}$ and under cyclic permutations
of the momentum labels $i$.

The coefficient of each distinct product of 8 $\eta_{ai}$  is an
MHV amplitude when interpreted by means of the prescription of
\cite{ggk}. We restate this prescription in terms of products  of
derivatives. Each annihilation operator of the gauge theory is associated with a
differential operator\footnote{The product structure of $F_n$
suggests that we use upper indices for all fields, thus
$(F^-)^{abc}(i) \leftrightarrow
\frac{\pa^3}{\pa\h_{ia}\pa\h_{ib}\pa\h_{ic}}$. The lower index
field is then defined as the dual,
i.e.~$(F^-)^{abc}(i)=\e^{abcd}F_d^-(i)$. This definition leads to
the $-$ sign in the derivative $D_{ia}$.  Similar remarks apply to
the negative chirality fields in supergravity and the associated
differential operators. See \reef{newder} below.} as follows: 
\bea
  \nonumber
  \hspace{-5mm}
  &&B_+(i)\leftrightarrow 1,~~~
  F^a_+(i)\leftrightarrow D_i^a= \frac{\pa}{\pa\h_{ia}}\,,~~~
  B^{ab}(i)\leftrightarrow D_i^{ab}=\frac{\pa^2}{\pa\h_{ia}\pa\h_{ib}}\,,~~~
  B_{ab}(i)\leftrightarrow D_{iab} = \half \e_{abcd}D_i^{cd}\, ,\\[1mm]
  \hspace{-5mm}
  &&
  F_a^-(i)\leftrightarrow D_{ia}
   =-\frac16 \e_{abcd}\frac{\pa^3}{\pa\h_{ib}\pa\h_{ic}\pa\h_{id}}\,,~~~~
  B^-(i)\leftrightarrow D_i
   = \frac{1}{24}\e_{abcd}
    \frac{\pa^4}{\pa\h_{ia}\pa\h_{ib}\pa\h_{ic}\pa\h_{id}}\,.
\lab{n4ders} 
\eea 
Any desired MHV amplitude is obtained by
applying an  8th order differential operator  composed as  the
product of appropriate factors from \reef{n4ders}. For example,
the $n$-gluon  Parke-Taylor \cite{pt} amplitude is given by
\be
  \lab{nglu}
  A_n(1^-,2^-,3^+,\dots , n^+) =
  D_1\, D_2\, F_n = \frac{\<1\, 2\>^4}{ \prod_{i=1}^{n}\<i\, (i+1)\>}\,.
\ee
We can use this to write the generating function in the alternate form
\be \lab{n4alt}
F_n =\frac{A_n(1^-,2^-,3^+,\dots , n^+)}{\<1\, 2\>^4}\d^{(8)}\Big(\sum_{i=1}^{n}|i\>\eta_{ia}\Big)\,,
\ee
which is useful to compare  with extensions discussed below.

Any  product of 8 derivatives produces an amplitude in the  MHV
sector of the gauge theory. Since the maximum order of any individual
operator is 4,  each 8th order differential operator is associated with a partition of
the integer 8 with maximum summand $n_{{\rm max}} \le 4$.  Each
partition corresponds to a particular set of particles in an
$n$-point MHV amplitude. There are 15 such partitions, which
correspond to the 15 types of MHV amplitude listed in (5.4) of
\cite{ggk}. For example, the $\<B^-(1)\,F_+^a(2)\,F^b_+(3)\,
F^c_+(4) \,F^d_+(5)\,B_+(6)\dots B_+(n)\>$ amplitude mentioned in
the first paragraph of this section corresponds to the partition
$8=4+1+1+1+1$, and  the 8-gluino amplitude of the Introduction is
$8=1+1+1+1+1+1+1+1$.

How do we know that all amplitudes obtained by differentiation  of
$F_n$ agree with those produced by explicit stepwise solution of
the \wis?  To answer this question we show below that the
amplitudes obtained from $F_n$ satisfy the SUSY \wis, and we
already observed above that the $n$-gluon Parke-Taylor amplitude
is correctly produced. The solution of the MHV  \wis\ is unique
under these conditions, so the favorable conclusion is valid.

We define supercharges
\be \lab{schg} \tQ_a \,= \,\sum_{i=1}^{n}
|i\> \, \h_{ia} \, ,~~~~~~
Q^a\, = \,\sum_{i=1}^{n} [\,i\,| \,
\frac{\pa}{\pa\h_{ia}} \, ,
\ee
which act by multiplication and
differentiation in the space of  functions of the $\h$'s.  Their
anticommutator is \be \lab{anti} \{Q^a\,,\,\tQ_b\} \;=\;\d^a_b \,
\sum_{i=1}^{n} |\,i\>\,[i\,| \,=\,0\,. \ee The fact that it
vanishes due to momentum conservation should not be a surprise,
since \reef{anti} corresponds exactly to the basic SUSY
anticommutator $\{Q^{a\a}\,,\,\tQ_b^{\db}\} = \d^a_b\,P^{\db\a}$
which also vanishes when the operator $P^{\dot{\b}\a}$  is applied
to an amplitude.

Consider the spinor contraction $\<\e\, \tQ_a\> =
\sum_{i}\<\e \, i\> \h_{ia}$ of the supercharge $\tQ_a$ in
\reef{schg} with the para\-meter $\e$. The set of commutators of this
operator with the differential operators of \reef{n4ders} is
isomorphic to the commutator algebra of \reef{n4tQ}.  For example,
\bea
  \nonumber
  \left[\<\e\, \tQ_a\>\,,\,1\right] &=&0\, ,\\
  \left[ \<\e\, \tQ_a\>\,,\,D_i^b\right]&=& \<\e \,i\>\, \d^b_a\, 1\, ,\\
  \nonumber
  \left[\<\e\, \tQ_a\>\,,\, D_i^{bc}\right]&=&
  \<\e \,i\> \, \big(\d^b_a\,D_i^c- \d^c_a\,D_i^b \big)\, ,
  ~~~~~\mathrm{etc}.
\eea
Thus the correspondence \reef{n4ders} between particle annihilators and
differential operators respects SUSY.

It may seem that there is at most a half-truth here since the
commutators of $[Q^a\,\e] = \sum_{i} [i\,\e]\pa/\pa\,\h_{ia}$ with
all  operators of \reef{n4ders} {\it vanish} rather than mirror
the structure of \reef{n4tQ}. This apparent paradox requires more
thought.  It may be related to the fact that the $Q^a$ Ward
identities are automatically satisfied in the MHV sector and are thus
suggestive of a type of $1/2$-BPS property which we discuss further in section \ref{s:disc}.

The SUSY \wis\ hold formally in the form
\be
  \lab{fwi}
  \tQ_aF_n \,=\, 0 \, ,
  ~~~~~~~~~~~ Q^a F_n \,=\,0\,,
\ee the first because we are multiplying $\d^{(8)}$ by its own
argument and the second  by momentum conservation.  We view these
formal \wis\ as the analogue of the statement \reef{wif}. The
concrete \wis\ of section \ref{s:swi} are obtained from products of
the form
\bea
  \lab{fwi2}
   D^{(9)} \, \big( \<\e\, \tQ_a\>   F_n  \big) ~=~ 0\, ,
\eea
where $D^{(9)}$ is a
product of operators from the correspondence \reef{n4ders}  of
total order 9. Similarly explicit \wis\ of the
supercharge $Q^a$ are obtained from products of the form
\be
\lab{fwi3} \left[\e\, Q^a\right] D^{(7)}F_n ~=~0\,.
\ee
This
expression is a sum of 8th order derivatives. There are two
possibilities depending on the $SU(4)$ indices of the product
operator $D^{(7)}$. Either each individual term vanishes due to
$SU(4)$ symmetry, or there are three non-vanishing
terms\footnote{In some cases  more than three terms appear, but all
except three vanish when definite values are assigned to the
$SU(4)$ indices.} which constitute an explicit $Q^a$ \wi
~relating three amplitudes.  These comments about $SU(4)$ symmetry
also apply to the $\tQ_a$ \wis\ of \reef{fwi2}.

\subsection{Practicalities}

As an example will show, the computation of spin factors from
$D^{(8)} \d^{(8)}\Big(\sum_{i=1}^{n}|i\>\eta_{ia}\Big),$ reduces to a
simple  Wick contraction algorithm of the basic operators $\pa^a_i
\equiv \pa/\pa \h_{ia}$.  The elementary contraction is
\be
\lab{cont}
\hat{\pa}^a_i\ldots\hat{\pa}^b_j \,=\,\pm
\d^{ab}\<i\,j\>\,.
\ee
The $\dots$ indicates other operators
between those which are contracted and the sign depends on whether
the number of these operators is even or odd. Suppose we want to
obtain the amplitude of \reef{ggS} for the specific index values
$a=c=1,~b=d=2$.  From the correspondence \reef{n4ders}  we see
that we must compute
\bea
\nonumber
D^{(8)} \d^{(8)}\Big(\sum_{i=1}^{n}|i\>\eta_{ia}\Big)&=&-\pa^2_1\pa^3_1\pa^4_1 \, \pa^1_2\pa^3_2\pa^4_2 \, \pa^1_3\pa^2_3 \frac{1}{16}\prod_{a=1}^{4} \sum_{i,j=1}^{n}\<i\,j\>\,\eta_{ia}\,\eta_{ja}\\ 
&=& \<1\,2\>^2\<2\,3\>\<1\,3\>\,.
\lab{exffs}
\eea
The spin factor obtained by
explicit action on $\d^{(8)}$ is more easily found by pairwise
Wick contraction of the operators in the string $D^{(8)}$.    When
the spin factor in \reef{exffs} is multiplied by the dynamical
prefactor in the generating function \reef{n4gen}, one finds
exactly the amplitude produced by explicit solution of the \wis\
in \reef{ggS}.


\subsection{Gravity}
\lab{s:mhvGF}

The good news now is that it is a very  straightforward matter to
write down a generating function for the MHV sector of $\cn = 8$
\sg.  To describe $n$-point functions one now needs $8n$
anti-commuting variables $\h_{iA}$ in which $A$ is an $SU(8)$
index.   The generating function is then
 \bea  
 \lab{n8genMHV}
  &&
  \O_n ~=~  \frac{M_n(1^-,2^-,3^+,\ldots n^+)}{\<1\,2\>^8} \,
  \d^{(16)}\Big(\sum_{i=1}^{n}|i\>\eta_{iA}\Big)\, ,\\ \nonumber
 && \mathrm{with}~~~~
 \d^{(16)}\Big(\sum_{i=1}^{n}|i\>\eta_{iA}\Big)~=~ \frac{1}{256}\prod_{A=1}^{8} \sum_{i,j=1}^{n}\<i\,j\>\,\eta_{iA}\,\eta_{jA}\,.
\eea 
The quantity $M_n(1^-,2^-,3^+,\ldots, n^+)$ is the
 $n$-graviton MHV amplitude which can be written using the KLT
relations \cite{klt} or one of the several specific forms
available for MHV amplitudes \cite{bgk,unexp,ef}.
Although it is not obvious, the quantity $M_n(1^-,2^-,3^+,\ldots
n^+)/\<1\,2\>^8$  is invariant under  the exchange
$i\leftrightarrow j$ of {\it any} pair of lines. This property
actually follows from the SUSY \wis\ \cite{dixon}. Thus the
formula \reef{n8genMHV} is entirely Bose symmetric. It is also
$SU(8)$ invariant. It is a sum of products of 16 distinct
$\eta$'s.

To use the generating function $\O_n$ we define a new set of differential operators:
\bea
 \nonumber
  \hspace{-4mm}
  &&\cd_i^A = \frac{\pa}{\pa\h_{iA}}\,,~~~~
  \cd_i^{AB}=\frac{\pa^2}{\pa\h_{iA}\pa\h_{iB}}\,,~~~~
  \cd_i^{ABC} =
  \frac{ \pa^3}{\pa\h_{iA}\pa\h_{iB}\pa\h_{iC}}\,,~~~~
  \cd_i^{ABCD} =
  \frac{ \pa^4}{\pa\h_{iA}\pa\h_{iB}\pa\h_{iC}\pa\h_{iD}}\,,\\
 \nonumber
 \hspace{-4mm}
 &&
 \cd_{iABC}=\,-\,\frac{1}{5!} \e_{ABCDEFGH}
 \frac{\pa^5}{\pa\h_{iD}\cdots\pa\h_{iH}}\,,~~~~~~
 \cd_{iAB} = \frac{1}{6!}\e_{ABCDEFGH}
 \frac{\pa^6}{\pa\h_{iC}\cdots\pa\h_{iH }}\,,~~\\
 \hspace{-4mm}
 &&
  \cd_{iA} =\,-\, \frac{1}{7!}\e_{ABCDEFGH}
  \frac{\pa^7}{\pa\h_{iB}\cdots\pa\h_{iH }}\, ,~~~~~~
  \cd_{i} = \frac{1}{8!}\e_{ABCDEFGH}
  \frac{\pa^8}{\pa\h_{iA}\cdots\pa\h_{iH }}\,.
  \lab{newder}
\eea

As in the case of gauge theory, the  fields of supergravity are
associated with these operators as follows: \bea
  \hspace{-4mm}
  \nonumber
  &&
  b_+(i)\leftrightarrow 1\,,~~~~~
  f^A_+(i) \leftrightarrow \cd_i^A\,,~~~~~
  b^{AB}_+(i)\leftrightarrow\cd_i^{AB}\,,~~~~~
  f^{ABC}_+(i)\leftrightarrow\cd_i^{ABC}\,,~~~~~
  b^{ABCD}(i)\leftrightarrow\cd_i^{ABCD}\,, \\[2mm]
  \hspace{-4mm}
  &&
  f_{ABC}^-(i)\leftrightarrow\cd_{iABC}\,,~~~~~
  b_{AB}^-(i)\leftrightarrow\cd_{iAB}\,,~~~~~
  f_{A}^-(i)\leftrightarrow\cd_{iA}\,,~~~~~
  b^-(i)\leftrightarrow\cd_i\, .
  \lab{fops}
\eea

To obtain the MHV amplitude for a particular set of external
lines one simply applies a 16th order differential operator which
is the product of the corresponding operators from \reef{fops}. As a typical example, we write
\bea
  \nonumber
  \big\<b^-(1)\,b_{AB}(2)\,b^{CD}(3)\,b_+(4)\ldots b_+(n)\big\>
  &=& \cd_1 \,\cd_{2AB} \, \cd_3^{CD} \, \O_n\\[1mm]
  &=& 2 \, \d^{CD}_{AB}\,\frac{\<1\,3\>^2}{\<1\,2\>^2} \,
  M_n(1^-, 2^-, 3^+ ,  \dots ,n^+)\,,
  \lab{typex}
\eea
which agrees with \reef{n8wi}.

It is significant that the state dependent spinor factors obtained
from $\d^{(16)}$ involve only analytic spinor brackets $\<i\,j\>$,
although complete \sg\ amplitudes also involve anti-analytic
spinor brackets $[i\,j].$

It should be clear that any product of  the derivatives in
\reef{newder} of order 16 produces an amplitude in the MHV sector,
and that we can associate a partition of 16 with $n_{{\rm max}}\le
8$ with each distinct product. There are 186 such partitions, each
of which corresponds to an $n$-point amplitude for a particular
set of external fields. For example the amplitude in \reef{typex}
corresponds to the partition $16=8+6+2$.

It is also clear from the preceding  discussion in gauge theory
that the amplitudes generated in this way satisfy the SUSY \wis\
for $\cn = 8$ \sg.  Since these \wis\ have a unique solution in
the MHV sector, the amplitudes so constructed are correct. Each of
the 186 MHV amplitudes is the product of the $n$-graviton
amplitude times a state-dependent spin factor which is a homogeneous
function  with $k \le 8$ angle bracket factors $\<i\,j\>$ in the
numerator  and  $\<1\,2\>^k$ in the denominator.

We now put readers on notice that the punch line of our argument
concerning the realization of $SU(8)$ global symmetry in the map
from gauge theory to \sg\ is near, at least for MHV amplitudes.
This follows  from the simple factorization properties of the
generating function $\O_n$ and the differential operators in
\reef{newder}.  To exhibit these properties we split the set of
$8n~\h_{iA}$ into two subsets, namely a subset $\eta_{i a}$ in which
$A$ is restricted to index values $A \to a = 1,2,3,4$ and a subset
$\eta_{i r}$ in which $A \to r = 5,6,7,8$. Remarkably, and very
simply, the supergravity  generating function $\O_n(\eta_{i A})$
factors as \be \lab{n8fac} \O_n \,=\,
\frac{M_n(1^-,2^-,3^+,\ldots , n^+)}{\<1\,2\>^8} \,
\d^{(8)}\Big(\sum_{i=1}^{n}|i\>\, \eta_{ia}\Big)\,
\d^{(8)}\Big(\sum_{j=1}^{n}|j\>\, \eta_{j r}\Big) \, . \ee

Remarkably and equally simply, the differential operators
factorize precisely in accordance with the map \reef{84map},
including all signs.  As an example,  we write the map of
graviphoton operators with mixed $SU(4)$ indices to illustrate how
the $-$ sign in the negative helicity sector arises: \bea
 b^{ar}_+(i) &\lra& \frac{\pa^2}{\pa\h_{ia}\pa\h_{ir}}
  ~=~\frac{\pa}{\pa\h_{ia}}\, \frac{\pa}{\pa\h_{ir}}
  ~\lra~ F^a_+(i) \,  \tilde{F}^r_+(i) \, ,\\[3mm]
  \nonumber
  b_{ar}^-(i) &\lra&  \frac{1}{6!} \,  {6 \choose 3} \, \eps_{arbcdstu}
  \frac{\pa^6}{\pa\h_{ib}\pa\h_{ic}\pa\h_{id}\pa\h_{is}  \pa\h_{it}\pa\h_{iu}} \\[2mm]
  &\lra&
  - \bigg( - \frac{1}{3!} \eps_{abcd} \frac{\pa^3}{\pa\h_{ib}\pa\h_{ic}\pa\h_{id}} \bigg)
   \bigg( -\frac{1}{3!} \eps_{rstu} \frac{\pa^3}{\pa\h_{is}\pa\h_{it}\pa\h_{iu}}  \bigg)
   ~\lra~ - F_a^-(i) \, \tilde{F}_r^-(i)\, .~~~~~
\eea
We have checked that all $\cn = 8$ supergravity operators factor correctly.
This also implies that the differential operators \reef{fops} satisfy the $\cn =8$ supersymmetry algebra.

\vspace{2mm}
The factorized structure ensures many desiderata, namely
\begin{enumerate}
\item[a.]  Supergravity amplitudes satisfy $\cn = 8$ supersymmetry
Ward identities, and they are $SU(8)$ covariant.
\item[b.] The spin dependence of $\cn = 8$ supergravity amplitudes
for all helicity states factorizes into products of gauge theory spin factors.
This works for MHV amplitudes because the spin factors obtained by applying differential
operators to the product of $\d^{(8)}$-functions  in \reef{n8fac}
are the \emph{same} for all permutations in  formulas such  as the KLT
formula or the formula \reef{bbgk} below, which relate the graviton amplitude $M_n$  to products of
two $n$-gluon amplitudes $A_n$.
\item[c.]  $\cn = 8$ supersymmetry
and $SU(8)$ global symmetry can indeed be implemented in the map
\reef{map}.
\end{enumerate}

These statements have been  checked in a number of examples. We discuss some in the next section.

\subsection{Tests of the operator map}
\lab{s:exWIs}

We now discuss the construction of two examples of MHV amplitudes in $\cn =8$ \sg\
from the map \reef{map} using the operator correspondence in Table 1. We need an explicit formula  which relates the $n$-graviton amplitude to products of $n$-gluon amplitudes. The KLT formula is available and will be used in the NMHV sector.  However, for MHV amplitudes, there is a simpler choice, namely the form recently derived \cite{ef} by  rearrangement of the BGK formula \cite{bgk,unexp}. It reads
\bea
  &&M_n(1^-, 2^-, 3^+,\dots ,n^+)
  ~=~
  \sum_{\mathcal{P}(i_4,\dots, i_n)}
  \frac{\<1 \, 2 \> \< i_3 \, i_4 \>}{\<1 \, i_3 \> \<2\, i_4 \>} \,
  s_{1 i_n} \left( \prod_{s=4}^{n-1} \beta_s  \right)
  A_n(1^-, 2^-, i_3^+,\dots ,i_n^+)^2
  \, , ~~~~~~~~~~~
  \lab{bbgk} \\[1mm]
 &&
 \lab{beta}
  \beta_s ~=~ -\frac{\< i_s\, i_{s+1} \>}{\< 2\, i_{s+1} \>}
    \; \< 2 | \, i_3 + i_4 + \dots + i_{s-1} | i_s ] \, .
\eea To apply \reef{bbgk}, one chooses one distinguished positive
helicity line $i_3$ and then sums over permutations of the
remaining $n-3$ such lines. This formula embodies the
identifications $b^- \lra B^-\otimes \tilde{B}^-$ and $b_+ \lra B_+
\otimes\tilde{B}_+$ in the operator map of of Table 1.

As the first example,  we consider the two gravitino MHV amplitude
$\big\< b^-(1)\, f_A^-(2) \,b_+(3)\ldots f^B_+(k)\ldots b_+(n)\big\> $
which was obtained in the first line of \reef{n8wi} by solving the relevant $\cN = 8$ SUSY \wi . For a non-zero result, the $SU(8)$ indices must be chosen in the same $SU(4)$ factor, $A,B \to a,b$.  For each permutation in \reef{bbgk} we make use Table 1 to decompose $f_a^- \lra F_a^-\otimes \tilde{B}^-$ and write
\bea
  \nonumber
  && \hspace{1cm}
  \big\< B^-\, F_a^-(2)\, B_+(3)\ldots F^b_+(k)\ldots B_+(n)\big\>_L\,
  \big\< \tilde{B}^-\, \tilde{B}^-(2)\, \tilde{B}_+(3)
    \ldots \tilde{B}_+(k)\ldots \tilde{B}_+(n)\big\>_R \\
  \nonumber
  && \hspace{1cm} =
  \d^b_a\, \frac{\<1\,k\>}{\<1\,2\>}\<B^-(1)B^-(2)B_+(3)\ldots B_+(n)\>_L\,
  \big\< \tilde{B}^-\, \tilde{B}^-(2)\, \tilde{B}_+(3)
    \ldots \tilde{B}_+(k)\ldots \tilde{B}_+(n)\big\>_R  \,.
\eea
The spin factor can be obtained either from the  $\cN =4$ \wi , see \reef{Ggg}, or from the gauge theory generating function. The spin factor $\<1\,k\>/\<1\,2\>$ is common to all permutations in \reef{bbgk} and may be extracted as an overall factor. The result via the map \reef{map} therefore agrees with the \sg\ formula in \reef{n8wi}.

The next example is the two scalar MHV amplitude given in
the fourth  line of \reef{n8wi}. There are three distinct
decompositions  of the $SU(8)$ indices into distinct
$SU(4)$ sectors, and we consider each in turn.  It is interesting
to note how products of rather different gauge theory amplitudes
conspire to produce the common spin factor required by \sg.

Choose first all group indices in one $SU(4)$, say $SU(4)_L$,  so
that $b_{abcd}\lra B^-\otimes \tilde{B}_+$. Then  (with momentum
labels implicit by order) we have \bea
  \nonumber
  \big\<  b^-\,  b_{abcd}^- \, b^{efgh}_+ \,  b_+ \ldots b_+  \big\>
  &\to&
  \a_4^2 \, \eps_{abcd}\, \eps^{efgh}
  \big\< B^- \, B^+ \, B^- \, B_+ \ldots B_+ \big\>_L
  \big\< \tilde{B}^- \, \tilde{B}^- \, \tilde{B}_+ \tilde{B}_+\ldots \tilde{B}_+\big\>_R
  \\ \nonumber
  &=& 4! \, \d_{abcd}^{efgh}\, \frac{\<13\>^4}{\<12\>^4}
   \big\< B^- \, B^- \, B_+ \, B_+ \ldots B_+ \big\>_L
  \big\< \tilde{B}^- \, \tilde{B}^- \, \tilde{B}_+\, \tilde{B}_+\ldots \tilde{B}_+\big\>_R
  \, .
\eea In the second line we have used the gluon MHV identity
\reef{YMid} to obtain the spin factor $\<1\,3\>^4/\<1\,2\>^4$
(which is common to all  permutations in the formula \reef{bbgk}).
The identity $\eps_{abcd} \eps^{efgh}  = 4! \, \d_{abcd}^{efgh}$
is also used. The result agrees  perfectly with  \reef{n8wi}.

Next split the $SU(8)$ indices such that one leg of each scalar
lies  in the $SU(4)_R$ and the rest in $SU(4)_L$. Reducing the
4-index antisymmetrizer in \reef{n8wi} this way gives
$\d_{abcr}^{efgs} = 3! \, \d_{r}^{s} \, \d_{abc}^{efg}$. The
operator map in Table 1 tells us
\bea
  \nonumber
  &&\big\<  b^-\,  b_{abcr} \, b^{efgs} \,  b_+ \ldots b_+   \big\>
  \to
 (-1)\,\a_4^2 \, \eps_{abcd}\, \eps^{efgh}
  \big\< B^- \, F^d_+ \, F_h^- \, B_+ \ldots B_+\big\>_L
  \big\< \tilde{B}^- \, \tilde{F}_r^- \, \tilde{F}^s_+ \, \tilde{B}_+\ldots \tilde{B}_+ \big\>_R\\[1mm]
  \nonumber
  && \hspace{8mm} =~
    - \eps_{abcd}\, \eps^{efgh}
  \bigg(-\d_{d}^{h} \,\frac{\<12\>}{\<13\>} \bigg)
  \,\big\< B^- \, B_+ \, B^- \,  B_+ \ldots B_+\big\>_L\,
  \bigg(\d_r^s \,\frac{\<13\>}{\<12\>} \bigg)
  \big\< \tilde{B}^- \, \tilde{B}^- \, \tilde{B}_+ \, \tilde{B}_+\ldots \tilde{B}_+\big\>_R ~~~~~
  \\[1mm] \nonumber
  && \hspace{8mm} =~
  3! \, \d_r^s \, \d_{abc}^{efg}\, \frac{\<13\>^4}{\<12\>^4}\,
  \big\< B^- \, B^- \, B_+ \, B_+ \ldots B_+ \big\>_L
  \big\< \tilde{B}^- \, \tilde{B}^- \, \tilde{B}_+ \tilde{B}_+\ldots \tilde{B}_+\big\>_R \, .
\eea
The minus sign $(-1)$ in the first line comes from conscientiously
moving $\tilde{F}_r^-$ past $F_h^-$ when separating the operators
into the $L$ and $R$ gauge theory amplitudes. In the second line we used the
gluino \wi\ \reef{Ggg}. In the last line, \reef{YMid} again gives the correct overall spin factor.  Observe how gauge theory results (either from \wis\ or the generating function) combine to produce the \sg\ amplitude which agrees with \reef{n8wi}.

The third distinct split of the scalar $SU(8)$ indices places two
of the four indices in $SU(4)_L$ and the other two in $SU(4)_R$.
The antisymmetrizer splits as $\d_{abrs}^{cdtu} = (2!)^2 \,
\d_{ab}^{cd} \, \d_{rs}^{tu}$. Table 1 tells us that  the gauge
theory amplitudes needed in \reef{bbgk} involve two scalars and
$n-2$ gluons. This amplitude  is given in \reef{GSS} and contains
the spin factor $\<1\,3\>^2/\<1\,2\>^2$.  In the product of the
two gauge theory amplitudes this factor is squared exactly as
needed to agree with \reef{n8wi}.

Several other examples of MHV amplitudes in \sg\  have been studied
using the map in Table \ref{tab:map} to identify the appropriate
gauge theory amplitudes. In every case the application of
\reef{bbgk} produces the same result as  a straightforward
calculation using the \sg\ generating function.

\setcounter{equation}{0}
\section{An application: intermediate state helicity sums}
\label{inthelsum}

So far the generating function has been shown to be a useful
bookkeeper for the spin dependence of MHV amplitudes in gauge
theory and \sg.  In this section we outline a further application,
namely to sums over intermediate helicity states needed when the
product of MHV trees occurs in a unitarity cut of a 1-loop
amplitude.

 First we use the generating function to reproduce the intermediate
state sum in a 2-particle cut in gauge theory, as discussed in
section 5  of \cite{bddk}.  Figure \ref{fig:MHVcut}(a)\footnote{This
reproduces figure 4 of \cite{bddk}.} indicates the 2-particle cut of a 1-loop amplitude containing  MHV amplitudes to the left and right.
Each amplitude contains one negative helicity gluon, on line $i^-$ in
the left factor and line $j^-$ on the right, plus arbitrary
numbers of positive helicity gluons denoted as lines $m$ on the
left and $n$ on the right. The intermediate state is a pair of
particles of momenta $l_1, l_2$.  Conservation laws allow these to
be either a  gluon pair, a gluino pair, or a pair of scalars. In
the approach of \cite{bddk}, one must solve the Ward identities to
find the amplitudes and sum their contributions weighted by the
multiplicities, 1-4-6-4-1, of the states in the $\cn =4$ gauge
theory. This is not difficult, nor is the resulting binomial sum
and spinor algebra which is required to obtain the final answer.
However, we find it interesting to put the generating function to
work on the problem.

We are interested only in the helicity sum  so we drop the
dynamical prefactors in the generating function \reef{n4gen}.  The
core situation is then governed by the product
\bea
\lab{core}
&&D_{1}D_{2} \, \d^{(8)}(I) \d^{(8)}(J) \, ,
\eea
where
\bea
I &=& |l_1\>\h_{1a} + |-l_2\>\h_{2a} + |i\>\h_{ia}
+\sum_m |m\>\h_{ma} \, ,\\
J &=&  |-l_1\>\h_{1a} + |l_2\>\h_{2a} + |j\>\h_{ja}
+\sum_n |n\>\h_{na} \, ,\\
D_{l} &=& \prod_{a=1}^4\frac{\pa}{\pa\h_{la}},~~~~l=1,2 \, .
\eea
We
see that $I$ and $J$ are the arguments of the $\d$-functions in
the generating functions for the left and right amplitudes
respectively. The derivatives $D_{1}D_{2}$ act on the Grassmann
variables $\h_{1a}$ and $\h_{2a}$ in both $I$ and $J$.  They
reproduce the intermediate state sum in a very compact fashion,
automatically keeping track of phases and multiplicities.  Each
intermediate state comes from a particular split of the individual
derivatives in $D_{1}D_{2}$ so that some factors act on $I$ and
the rest on $J$.

To see this first note that, because of the outgoing line
convention, the particles on the two ends of an internal line have
opposite helicity. One term of the spin sum is the case where a
positive helicity gluino $F^{b}_+$, with $SU(4)$ index $b$, is emitted
from the left on line $l_1$ and absorbed as a negative helicity gluino
$F^-_{b}$ on the right. This case corresponds to the split of the operator
$D_{1}$ with $\pa/\pa\h_{1b}$ acting on $I$ and the third
order $D_{1b}$ from the list in \reef{n4ders} acting on $J$. After
the 4th order $D_i$ is applied to describe the emission of the -ve
helicity external gluon,  we must apply 3 further derivatives to
$\d^{(8)}(I)$ to have a non-vanishing result. Thus the derivative
$D_{2}$ is forced to split with the third order $D_{2b}$ applied
to $I$ and the first order $D_2^b$ applied to $J$. The negative
sign associated with the fermion loop comes from anti-commutation
of derivatives. The multiplicity factor 4 for gluinos comes from
the sum over the 4 choices of the index $b$.  This description is
unnecessarily tedious. In practice all of the bookkeeping is done
automatically (while the physicist sips his tea).

\begin{figure}[t!]
    \begin{center}
            \raisebox{0.9cm}{\includegraphics[width=5.3cm]{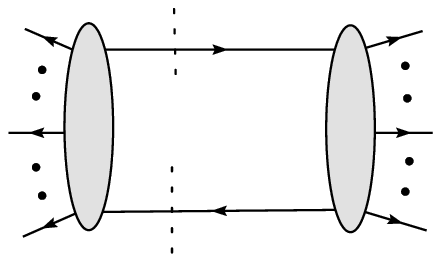}}
        \hspace{1.5cm}
         \includegraphics[width=5.7cm]{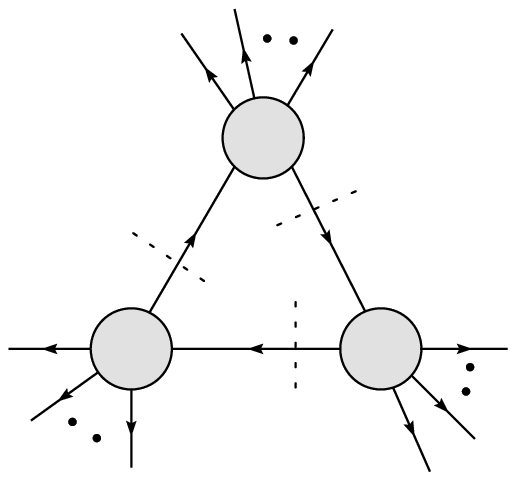}
      \begin{picture}(0,0)(0,0)
      \put(-375,69.5){$i^-$}
      \put(-210,69){$j^-$}
      \put(-292,106){$l_1$}
      \put(-292,30){$l_2$}
      \put(-338,68){$I$}
      \put(-248,68){$J$}
      \put(-93,153){$i^-$}
      \put(-11,5){$j^-$}
      \put(-170,12){$k^-$}
      \put(-54,75){$l_1$}
      \put(-90,28){$l_2$}
      \put(-115,82){$l_3$}
      \put(-84,105.5){$I$}
      \put(-48,37.5){$J$}
      \put(-129,37.5){$K$}
     \put(-297,-5){(a)}
     \put(-90,-5){(b)}
    \end{picture}
    \end{center}
\caption{Intermediate spin sums.}
\lab{fig:MHVcut}
\end{figure}

Let's now proceed to the full calculation; we must compute
\be \lab{aut1}
D_{1}D_{2}\big(D_i\d^{(8)}(I)\,D_j\d^{(8)}(J)\big) \, .
\ee
The computation is simpler in the order indicated. We write\footnote{We use the continuation $|-l\> = -|l\>$ of spinors for negative null momenta.}
\bea
\nonumber
D_i\d^{(8)}(I)&=&
D_i \prod_{a=1}^4
\big( - \<l_1\,l_2\> \h_{1a}\h_{2a}\,+\, \<l_1\, i\>\h_{1a}\h_{ia} \,
-\, \<l_2\, i\>\h_{2a}\h_{ia} +\ldots\big)\\
&=&\prod_{a=1}^4
\big( - \<i\,l_1\>\h_{1a}\,+\,\<i\,l_2\>\h_{2a}+\ldots\big).
\lab{aut2}
\eea
The omitted terms $\ldots$ involve the Grassmann variables
$\h_{ma}$. They can be dropped since no derivatives
$\pa/\pa\h_{ma}$ will be applied. Hence
\bea
\lab{aut3}
D_i\d^{(8)}(I)\,D_j\d^{(8)}(J)
&=&\prod_{a=1}^4\,\prod_{b=1}^4X_a Y_b\,=\, (X_1Y_1)(X_2Y_2)(X_3Y_3)(X_4Y_4)\, , \\[1mm] \nonumber
X_a &=& -\<i\,l_1\>\h_{1a}\,+\,\<i\,l_2\>\h_{2a}\, , \\ \nonumber
Y_a &=& \; ~\<j\,l_1\>\h_{1a}\,-\,\<j\,l_2\>\h_{2a} \, . \eea Each
product simplifies by the Schouten identity, viz 
\bea \nonumber
X_aY_a &=&
\big(\<j\,l_1\>\<i\,l_2\>\,-\,\<j\,l_2\>\<i\,l_1\>\big)\h_{1a}\h_{2a}~~~~~~~{\rm (no~sum)}\\
&=& - \<i\,j\>\<l_1\,l_2\>\, \h_{1a}\h_{2a} \, .
\lab{aut4}
\eea
Finally we obtain
\be
\lab{aut5}
D_{1}D_{2}\big(D_i\d^{(8)}(I)\,D_j\d^{(8)}(J)\big)~=~\<i\,j\>^4\<l_1\,l_2\>^4,
\ee
which is equivalent to (5.6) of \cite{bddk}.  We did this
calculation in gauge theory to facilitate comparison with
\cite{bddk}, but it is just as easy  in \sg. The final result
there is  $\<i\,j\>^8\<l_1\,l_2\>^8$.

It is no more difficult to handle the spin sum for the 3-particle
cut shown in figure \ref{fig:MHVcut}(b),  which is related to the \sg\ calculation
discussed in section 4B of \cite{unexp}. The external states involve
one negative helicity graviton  on each sub-amplitude. The core
involves a product of three generating $\d$-functions to which
operators $D_1 D_2 D_3$  which effect the automatic spin sum are
applied: \bea \lab{3sum}
&& D_1 D_2 D_3 \big[\d^{(16)}(I) \d^{(16)}(J)  \d^{(16)}(K)\big]
\, ,\\[3mm] \nonumber
I &=& |l_1\>\h_{1a} + |-l_3\>\h_{3a} + |i\>\h_{ia}
+\sum_m |m\>\h_{ma} \, ,\\ \nonumber
J &=&  |-l_1\>\h_{1b} + |l_2\>\h_{2a} + |j\>\h_{ja}
+\sum_n |n\>\h_{na}\, ,\\ \nonumber
K&=&  |-l_2\>\h_{1b} + |l_3\>\h_{3a} + |k\>\h_{ka}
+\sum_p |p\>\h_{na}\, .
\eea
The differential operators are now all eighth order, given by
the last entry in the list \reef{newder}. Derivatives  $D_i, D_j, D_k$
for the external gravitons require only simple calculations
similar to \reef{aut2}  which give
\bea
\nonumber
D_i\d^{(8)}(I)&=&
\prod_{a=1}^8\big(\<i\, l_1\>\h_{1a} - \<i \,l_3\>\h_{3a}\big)
\,\equiv\,\prod_{a=1}^8 X_a \, ,\\
\lab{3sum1}
D_j\d^{(8)}(J)&=& \prod_{b=1}^8\big(- \<j\,l_1\>\h_{1b}
+\<j\, l_2\>\h_{2b}\big)
\,\equiv\,\prod_{b=1}^8 Y_b\, ,\\ \nonumber
D_k\d^{(8)}(K)&=& \prod_{c=1}^8\big( -\<k \,l_2\>\h_{2c}
+\<k \,l_3\>\h_{3c}\big)
\,\equiv\,\prod_{c=1}^8 Z_c \, .
\eea Next we assemble the product
\be
\lab{asse}
\prod_{a=1}^8
(X_aY_aZ_a) =  \prod_{a=1}^8 \big[\<i\,l_1\>\<j\,l_2\>\<k l_3\> -
\<i\,l_3\>\<j\,l_1\>\<k l_2\>\big]\h_{1a}\h_{2a}\h_{3a} \, .
\ee
Finally we apply  $D_1D_2D_3$ which trivially gives
the result
\be
\lab{atlast}
\big[\<i\,l_1\>\<j\,l_2\>\<k l_3\> - \<i\,l_3\>\<j\,l_1\>\<k l_2\>\big]^8
\ee
 and agrees\footnote{The denominator in (4.23) is included in the
 prefactors omitted in our calculation.} with (4.23) of \cite{unexp}.

We have applied the generating function to situations which are
fairly straightforward in their original form in \cite{bddk} and
\cite{unexp}. However, we wanted to strut our stuff
in the hope that the technique will be useful
in more complex situations where intermediate spin sums are
required.\footnote{Pilot calculations of 1- and 2-loop helicity sums
involving NMHV tree amplitudes in $\cn =4$ SYM indicate that the generating function method is applicable.}

\setcounter{equation}{0}
\section{Spin factors and CFT correlators}
\lab{s:sfcft}

There is a spectacular analogy  between the spin factors for MHV
diagrams and holomorphic correlators in conformal field theory on
the complex plane. Suppose we are interested in the spin factor
for a general MHV $n$-point  amplitude in \sg\ which we denote by
$\<\phi_1 \phi_2 \ldots \phi_n\>$.  Each of the operators $\phi_1$
has an "$\h$-count"  $r_i \in 0,1,\dots, 8$ and a specific
assignment of $SU(8)$ indices which we omit in the notation. Of
course $\sum_{i} r_i\,=\,16$ for an MHV process as we have
emphasized in section \ref{s:mhvGF}, but this constraint will not play a major
role.  The conformal analogy we now develop is equally valid in
the gauge theory.

The spin factor is defined in this section as 
\be \lab{sfcft}
\<\phi_1 \phi_2 \ldots \phi_n\>\,=\, 
\cd^{(r_1)}_1\ldots \cd^{(r_n)}_n
\d^{(16)}\big(\sum_{i}|i\>\h_{\h_{iA}}\big)\,, 
\ee 
in which
$\cd_i^{(r_i)}$ is a differential operator of order $r_i$ which
carries the  $SU(8)$ indices of $\phi_i$.  In every case we deal
with below we assume that the 16 $SU(8)$ indices are paired so
that the spin factor is non-vanishing. We noticed the analogy by
asking the question ``What features of the spin factor are
determined only by the $r_i$ and what features require the
explicit assignment of indices?"

Let's begin with the 3-point case in which the derivatives in
\reef{sfcft} give the result (up to a sign): \be \lab{sf3}
\<\phi_1\, \phi_2 \,\phi_3\>\,=\,
\<1\,2\>^{\n_{12}}\<2\,3\>^{\n_{23}}\<3\,1\>^{\n_{31}}\,.
\ee
Since each derivative  $\cd^{(r_i)}_i$ in \reef{sfcft} produces
$r_i$ factors of the spinor $|i\>$, we see that
\bea
\nonumber
\n_{12}\,+\,\n_{31} &=& r_1\, , \\
\lab{cft3}
\n_{12}\,+\,\n_{23} &=& r_2\, , \\
\nonumber
\n_{23}\,+\,\n_{31} &=& r_3\,,
\eea
which uniquely determine the  values
\be
\lab{nuvals}
\n_{ij} = \half (r_i + r_j -r_k)\,,
\ee
where $i,j,k$ is a cyclic permutation of 1,2,3.  At this point,
the reader will undoubtedly recall that the correlation function
of 3 conformal primary operators  $\co_i$ of scale dimension
$(r_i,0)$ is \be \lab{cftt}
\<\co_1(z_1)\,\co_2(z_2)\,\co_3(z_3)\>\,=\,c_{123}\frac{1}{z_{12}^{\n_{12}}z_{23}^{\n_{23}}z_{31}^{\n_{31}}}\,
\ee with {\it the same exponents} $\nu_{ij}$.
Conclusion:  a 3-point spin
factor is completely determined by the $r_i$ just as a CFT 3-point
correlator is completely determined by the 3 scale dimensions. The
forms are strikingly similar. This is not an accident; we can push
further.

The spin factor of any 4-point amplitude obtained from
\reef{sfcft} contains a product of as many as 6 angle brackets, viz.
\be \lab{cft4}
\<\phi_1\, \phi_2 \,\phi_3\,\phi_4\>\,=\,
\<1\,2\>^{\n_{12}}\<1\,3\>^{\n_{13}}\<1\,4\>^{\n_{14}}\<2\,3\>^{\n_{23}}\<2\,4\>^{\n_{24}}\,\<3\,4\>^{\n_{34}} \, .
\ee
The set of 4 equations analogous to \reef{cft3} are not
sufficient to solve for the 6 exponents $\n_{ij}$. What else can
we do to help determine them?  Consider the spin factor for the
MHV amplitude corresponding to the partition $r_1 =7,
\,r_2=5,\,r_3=2,\,r_4=2$, which corresponds to an amplitude with
one gravitino, one graviphotino, and two graviphotons. Let's write
one possible expression which carries the correct scaling weight
$(|i\>)^{r_i}$ for each spinor, namely
\be
\lab{7522}
 \<\phi_1\, \phi_2 \,\phi_3\,\phi_4\>\,
 \sim\, \<1\,2\>^5\<1\,3\>\<1\,4\>\<3\,4\> \, .
\ee There is additional freedom to multiply this by a function
which is {\it invariant} under scaling of all 4 spinors.   It
seems that we can multiply by any function of the variables \be
\lab{vari} \xi \,=\, \frac{\<1\,3\>\<2\,4\>}{\<1\,2\>\<3\,4\>}\, ,
~~~~~ \xi' \,=\, \frac{\<2\,3\>\<4\,1\>}{\<1\,2\>\<3\,4\>}\, , \ee
but they are not independent,  rather $\xi' = 1-\xi$ due to
the Schouten identity.  Similarly  $\xi"
=\<1\,3\>\<2\,4\>/\<2\,3\>\<4\,1\> =\xi/(1-\xi)$.  Thus it appears
that the most general form for our spin factor is \be \lab{7522a}
 \<\phi_1\, \phi_2 \,\phi_3\,\phi_4\>
 \,=\, \<1\,2\>^5\<1\,3\>\<1\,4\>\<3\,4\>\, f(\xi)\, ,
 \ee
where $f(\xi)$  is an arbitrary function of $\xi$.  At this point
the relevance  of conformal field theory is clear.  The properties
of the 4-point spin factor are identical to those of the 4-point
correlator\footnote{See for example section 5.1 of \cite{diFran}.  The
scale dependent product of six factors $(\<i\,j\>)^{r/3 -r_i-r_j}
$ suggested by  (5.28) of  \cite{diFran}  can be used for the
general 4-point spin factor.  However, it involves fractional
exponents, since $r =\sum_{i} r_i\,=\,16$, which is awkward.} of
operators with scale dimension $(r_i,0)$ which involves an
arbitrary function of one ``cross ratio" which may be taken to be
$\Xi  = (z_{13})(z_{24})/(z_{12})(z_{34}).$

One property of spin factors, which is not present in conformal
field theory, is that the exponents $\n_{ij}$ in   \reef{cft4}
must be non-negative integers. This severely restricts the
choice of $f(\xi)$ to $f(\xi)=1,~f(\xi)=\x,$ or $f(\xi)=1-\xi$.
Each choice corresponds to an inequivalent configuration of $SU(
8)$ labels as follows:
\bea
\nonumber
f(\xi)=1 ~~~&\lra&~~~\<\phi_1^{1234567}\, \phi_2^{12345} \,\phi_3^{68}\,\phi_4^{78}\> \, ,\\
\lab{flla}
f(\xi)=\xi ~~~&\lra&~~~\<\phi_1^{1234567}\, \phi_2^{12348} \,\phi_3^{56}\,\phi_4^{78}\> \, ,\\
\nonumber
f(\xi)=1-\xi ~~~&\lra&~~~\<\phi_1^{1234567}\, \phi_2^{12348}
\,\phi_3^{68}\,\phi_4^{57}\>\,.
\eea

A general $n$-point spin factor for a process involving operators
of $\h$-count $r_1,r_2,\dots r_n$ can be expressed as the product
of up to $n(n-1)/2$  independent angle brackets $\<i\,j\>$, each
raised to the non-negtive integer power $\n_{ij}$.  Suppose that
we have obtained one candidate expression which scales as
$\L^{r_i}$ for each spinor.  We would then need to consider
modification of  that expression, involving the possible scale
invariant variables which can be constructed from the spinors.  To
find such variables it is sufficient to scale each
spinor\footnote{We thank Gary Gibbons for this observation.}  to
the form $|i\> \to \tilde{\l}_i^{\da}= (z_i ~1)$. Then each angle bracket
satisfies $\<i\,j\> = z_i-z_j \equiv z_{ij}$ This establishes an
exact correspondence between scale invariant variables constructed
from spinor angle bracket and CFT cross ratios. In the 4-point
case above, we have $\xi = \Xi$! There are $n-3$ independent
variables for an $n$-point function.

Although the CFT analogy is quite perfect, it has been of limited
use for us.  One application concerns the asymptotic behavior of
the spin factors for diagrams in the MHV-vertex expansion of an
NMHV amplitude. 


\setcounter{equation}{0}
\section{Generating Functions for NMHV amplitudes}
\label{nmhvampl}

 We would like to extend our study beyond the MHV sector, but there are
several difficulties. The structure of non-MHV amplitudes with
external gluons or gravitons is far more complicated than MHV, and
the recursion relations they satisfy contain more terms.  It is
also more difficult to extract information\footnote{We solve the $\cn =1$ SUSY \wis\ for NMHV 6-point amplitudes in appendix \ref{app:N1susy}. Some explicit results have also been found in \cite{bdp} and \cite{StStTT6ptdisk}.}  from the  SUSY Ward
identities which relate amplitudes within each non-MHV sector.
Happily it turns out that we can make considerable progress in the
NMHV (next-to-MHV) sector which consists of  all amplitudes
connected by supersymmetry to the $n$-gluon or $n$-graviton
amplitude with 3 negative helicity lines. One needs $n \ge 6$ for a
genuine NMHV amplitude. For $n=5$, the amplitude with helicity
configuration $\langle ---++\rangle $ is the complex conjugate of
the MHV configuration $\langle +++--\rangle$.

In this section we discuss NMHV amplitudes in
$\cn=4$ gauge theory and $\cn = 8$ supergravity.  Our treatment is
based on the  MHV-vertex  expansion developed for gauge theory in \cite{csw} and extended to gravity in \cite{greatdane}.  For external gluon
amplitudes, the method was established before the invention of
modern recursion relations in \cite{bcf,bcfw}, but the version of
recursion relations studied in  \cite{risager} provides the
simplest and most general approach, and clarifies  the validity of
the method.

\subsection{Recursion relations and the MHV-vertex method}
\lab{s:recrel}

Recursion relations express $n$-point tree amplitudes such as $A_n$ as finite sums of  products of two sub-amplitudes $A_{n_1},~ A_{n_2}$ with $n_1,\,n_2 < n$.  They exploit the simple analyticity properties of on-shell tree amplitudes in a variable $z$ which appears through a shift of the spinors used to parametrize the complex momenta. Cauchy's theorem can be used to derive a valid recursion relation provided that the amplitude vanishes as the complex variable $z \to \infty$.  Our applications to $\cn =8$ \sg\ force us to confront this basic fact head on, so it will play an important role in our discussion below (sections~\ref{s:symz} and \ref{s:gravz}). Later we will compare the large $z$ behavior associated with both 3-line and the more common 2-line shift, so we begin with a review of the latter.  See \cite{cabg}, \cite{unexp}
and \cite{nima} for more information on the large $z$ asymptotics.

\subsubsection{2-line shifts}
The simplest recursion relations are based on a complex
continuation of on-shell amplitudes in which the spinors of two
external lines are shifted. Suppose that we are interested in
$n$-point amplitudes of gluons $A_n(1^-,2^-,3,4,\ldots,n)$ or
gravitons $M_n(1^-,2^-,3,4,\ldots,n)$. Particles 1 and 2 have
negative helicity, as indicated, while the helicity of the
remaining particles can be positive or negative. In the method of
\cite{qmc}, the spinors of particles 1 and 2 are shifted as
follows:\footnote{This is usually called a $[-,-\>$ shift. It is known that
$[-,+\>$ and $[+,+\>$ shifts also lead to a valid recursion relation, but a $[+,-\>$ shift does not.}
\bea
  \nonumber
  |1\> &\to& |\hat{1}\> \; =\;  |1\>  -z |2\> \, ,
  \hspace{1cm}
  |2\> ~\to~ |2\>\, ,\\
  |1] &\to& |1]  \, ,
  \hspace{3.17cm} |2] ~\to~ |\hat{2}] \; = \; |2]  +z |1]\,.
  \lab{2shift}
\eea
The shifted momenta are rank 1 products of spinors and
therefore null vectors, and the shift cancels in the sum, so that
momentum is conserved, viz \be
  \lab{pcon}
  (\hat{p}_1+\hat{p}_2)^{\da \b}
  ~=~
  \big( |1\>  -z |2\>\big) [1|\,
  +\,|2\> \big( [2|  +z [1 |\big) ~=~ (p_1+p_2)^{\da \b}\,.
\ee Therefore  the shifted amplitudes
 \bea A_n(z)
=A_n(\hat{1}^-,\hat{2}^-,3,\dots,n)\, , ~~~~~~ M_n(z)
=M_n(\hat{1}^-,\hat{2}^-,3,\dots,n)\,. 
\eea 
are indeed on-shell
analytic continuations of $A_n(0) =A_n({1}^-,{2}^-,3,\dots,n)$ and
$M_n(0) =M_n({1}^-,{2}^-,3,\dots,n)$.

The only singularities  of tree amplitudes are poles where
propagators vanish. Therefore $A(z)$ and $M(z)$ are meromorphic
with simple poles in $z$, a pole for each partition of the
amplitude into a product of two sub-amplitudes connected by a
propagator carrying the $z$-dependent momentum $\hat{P}_I =
\hat{p}_1 +K_1= -(\hat{p}_2 +K_2)$ where $K_1=\sum p_{I_1}$ and
$K_2=\sum p_{I_2}$ are the sums  of unshifted momenta in each
factor. If the important condition that $A(z)\to 0$ as $z \to
\infty$ is satisfied, then Cauchy's theorem can be applied to
derive, see \cite{bcfw,qmc}, the recursion relation 
\be
\lab{recrel1} A_n(1^-,2^-,3,\dots,n) ~=~
\sum_{I}A_{n_1}(\hat{1}^-,-\hat{P}_I,\dots,n)\frac{1}{s_I}A_{n_2}(\hat{P}_I,\hat{2}^-,3,\ldots)\,.
\ee Here $n_1+n_2= n+2$, and $s_I = -(p_1 +K_1)^2$ is the
unshifted Mandelstam invariant associated with the partition $I$.
The sum includes all partitions of the amplitude in which the
shifted lines are on opposite sides. For each such partition, $z$
is evaluated at the pole  $z_I$ determined by
\be \lab{pocon}
  0  ~=~ \hat{P}_I^2 ~=~ (\hat{p}_1 + K_1)^2
  ~=~ -s_I + z_I\<2|P_I|1] \,,
\ee in which $P_I=p_1+K_1$. The internal particle emitted from the
first sub-amplitude can have either helicity. Propagation to the
second  sub-amplitude conserves helicity, but  it is recorded
there with opposite helicity  because of  the outgoing momentum
convention. Graviton amplitudes satisfy  recursion relations
\cite{qmc} of the same form as \reef{recrel1}, but with $A_n$'s
replaced by  $M_n$'s.

\begin{figure}[t!]
  \begin{center}
        \raisebox{-2.9cm}{\includegraphics[width=5.5cm]{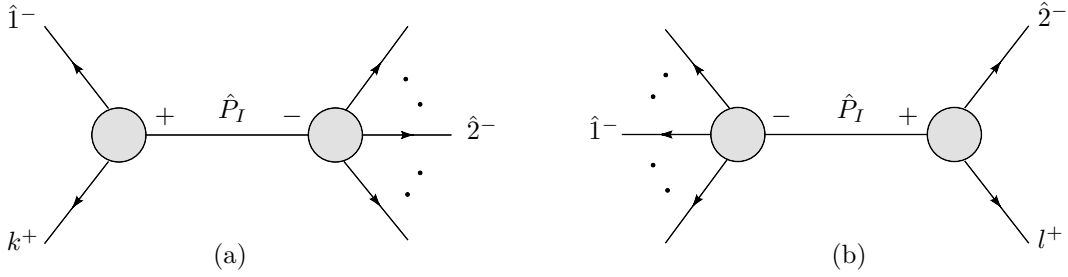}}
        \hspace{2cm}
        \includegraphics[width=5.5cm,angle=180,clip]{MHVrec0.eps}
  \begin{picture}(0,0)(0,0)
      \put(-172,-43){$\hat{1}^-$}
      \put(-2,0){$\hat{2}^-$}
      \put(-2,-84){$l^+$}
      \put(-78,-35){$\hat{P}_I$}
      \put(-103,-37){$-$}
      \put(-55,-37){$+$}
      \put(-218,-43){$\hat{2}^-$}
      \put(-392,-1){$\hat{1}^-$}
      \put(-392,-85){$k^+$}
      \put(-312,-35){$\hat{P}_I$}
      \put(-336,-37){$+$}
      \put(-288,-37){$-$}
      \put(-314,-90){(a)}
      \put(-80,-90){(b)}
    \end{picture}
  \end{center}
\caption{Diagrams from the 2-line shift recursion relations \reef{recrel1} for
MHV amplitudes in both gauge theory and gravity. The 3-vertex in
the right hand diagram vanishes as a consequence of kinematics.}
\lab{fig:MHV}
\end{figure}

It is interesting to examine the types of diagrams that actually
contribute to the recursion relation for various types of
amplitudes. The simplest case is MHV amplitudes in gauge theory in
which color ordering and helicity conservation (see the discussion
below \reef{onea}-\reef{twoa})  imply that only the two diagrams
listed in figure \ref{fig:MHV} can contribute. Both of them
involve the 3-gluon vertex with two positive helicity lines.  But
the two situations are rather different. The 3-vertex in
figure~\ref{fig:MHV}(b) vanishes at the pole by ``special
kinematics'': the null condition $\hat{P}_I^2= \<2 \, l \> [
\hat{2} \, l] = 0$ for the internal line requires $[\hat{2}\,l]
=0$, and therefore $A_3(\hat{2}^-,\hat{P}_I^+,l^+) \propto
[\hat{P}_I\,l]^3 \propto [\hat{2}\,l]^3 = 0$. The 3-vertex in
figure \ref{fig:MHV}(a) does not vanish because $|\hat{1}] =|1]$.
Thus there is only one contributing diagram, and it is not
difficult to show by iteration \cite{bernreview} that the
Parke-Taylor amplitude \reef{nglu} is the solution of  the
recursion relation.  The diagrams for $n$-graviton MHV amplitudes are
similar.  Helicity conservation restricts the possible diagrams to
those containing a 3-graviton vertex, and special kinematics again
forces the diagram of figure \ref{fig:MHV}(b) to vanish. However
there are now more diagrams of the type in figure
\ref{fig:MHV}(a), namely the $n-2$ diagrams containing cyclic
permutations of the positive helicity lines \cite{qmc}. This is
required by Bose symmetry. The simplicity of  MHV recursion
relations was exploited in \cite{ef} to prove a relationship
between $M_n$ and $(A_n)^2$ for all $n$.

The recursion relation is also valid for non-MHV amplitudes, but
more diagrams contribute. Diagrams for the simplest case of the
NMHV gauge theory amplitudes, such as
$A_n(1^-,2^-,3^-,4^+,\ldots,n^+)$, include both NMHV and MHV
sub-amplitudes. An example with  NMHV vertices is shown in
figure \ref{fig:nMHV0}.  This undesirable feature can be avoided
with 3-line shifts as we discuss in section \ref{s:3line}.

Recursion relations fail if amplitudes shifted as in \reef{2shift}
do not vanish as $z \to \infty.$  It is by now well established
that $n$-gluon amplitudes vanish as $1/z$ and $n$-graviton
amplitudes vanish as $1/z^2$ if two negative helicity lines are
shifted \cite{cabg,unexp,nima}.
Indeed this behavior of MHV amplitudes can be directly
observed in \reef{nglu} and \reef{bbgk}.  But the asymptotic
behavior does depend on particle type, a fact of particular
concern for this paper.
For example consider the
amplitude \reef{GSS} for two scalars and $n-2$ gluons. If lines 1
and 2 are shifted, the spin factor in \reef{GSS} contains a factor
$z^2$ which overwhelms the $1/z$ falloff of the $n$-gluon
amplitude. One can see from \reef{n8wi} that  amplitudes in which
a pair of gravitons of opposite helicity is replaced by a pair of
particles of spin $s$ behave as $z^{(2-2s)}$ at large $z$. These
remarks apply specifically to the shift of \reef{2shift}, and
there are other shifts available in these examples. In particular,
if the spinors of particle 1 and 2 are exchanged in \reef{2shift},
the amplitudes vanish as $z \to \infty$ and recursion relations
can be derived.

\begin{figure}[t!]
  \begin{center}
        \includegraphics[width=5.5cm,angle=180]{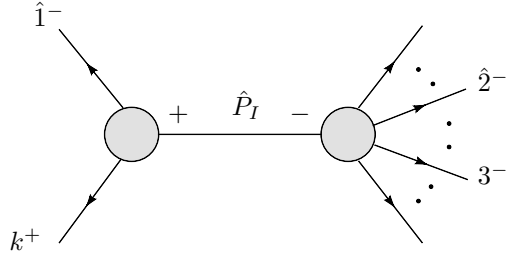}
            \begin{picture}(0,0)(0,0)
      \put(-1,-25){$\hat{2}^-$}
      \put(-1,-60){$3^-$}
      \put(-178,-85){$k^+$}
      \put(-169,0){$\hat{1}^-$}
      \put(-94,-34){$\hat{P}_I$}
      \put(-118,-36){$+$}
      \put(-72,-36){$-$}
    \end{picture}
  \end{center}
\caption{2-line shift recursion relations for NMHV amplitudes
contain non-vanishing diagrams with NMHV vertices. It is an
appealing feature of 3-line shifts that such diagrams vanish due
to special kinematics and the recursion sum consequently contains
MHV subdiagrams only.} \lab{fig:nMHV0}
\end{figure}

\subsubsection{3-line shifts}
\lab{s:3line}

We now discuss the recursion relation which arises from the 3-line
shift first considered in \cite{risager} with further details
 discussed in \cite{greatdane}.
This shift applies to NMHV amplitudes such as
$A_n(1^-,2^-,3^-,4^+,\dots,n^+)$ and
$M_n(1^-,2^-,3^-,4^+,\dots,n^+)$.  In this shift, all $|p\>$  spinors
are unchanged, while three spinors $|1],\,|2]\,\,|3]$ are shifted.  In our applications we will consider the 123-shift and other choices for the 3 shifted lines.
So we define the more general shift
\bea \nonumber
|m_1] &\to& |\hat{m}_1]  = |m_1] + z\, \<m_2\,m_3\>\, |X]\, ,\\
\lab{3line}
|m_2] &\to& |\hat{m}_2]  = |m_2] + z\, \<m_3\,m_1\> \,|X]\, ,\\
|m_3] &\to& |\hat{m}_3]  = |m_3] + z\, \<m_1\,m_2\>\, |X]\,,
\nonumber \eea where $|X]$ is an arbitrary reference spinor, which
will play an important role in our analysis. Shifted momenta
$\hat{p}_i$ remain on shell and total momentum is conserved
because of the Schouten identity.

Let's focus first on 6-point amplitudes. When the $m_i$ are chosen to be the three negative helicity lines, the pure gluon and pure graviton amplitudes vanish for large $z$. Using for example the results for the amplitudes given in the literature, e.g.~\cite{bdp}, we find numerically for large $z$ in gauge theory
\bea
  \nonumber
  \< \hat{-}\hat{-}\hat{-} +++\> &\sim&
  \< \hat{-}\hat{-} +\hat{-}++\> ~\sim~
  \frac{1}{z^4} \, ,\\
   \lab{altgauge}
   \< \hat{-}+\hat{-}+\hat{-} +\> &\sim&
  \frac{1}{z^5} \, ,
\eea
and in gravity (via KLT)
\bea
  \< \hat{-}\hat{-}\hat{-} +++\> &\sim&
  \frac{1}{z^6} \, .
\eea
Higher $n$-point graviton amplitudes are discussed in section \ref{s:npt}.

We will also see later that these exponents change
for amplitudes with other states of the $\cn=4$ or $\cn =8$
theories. However, it is clear that the NMHV amplitudes $A_6$ and
$M_6$ do satisfy recursion relations. We first discuss the gauge
theory case in detail and then the modifications necessary for
gravity.

\subsubsection*{Gauge theory}

The NMHV amplitude $A_n(1^-,2^-,3^-,4^+,\dots,n^+)$ satisfies the
recursion relation \be \lab{rec2}
 A_n(1^-,2^-,3^-,4^+,\dots,n^+) = \sum_{I} A_{n_1}(\hat{m}_1,-\hat{P}_I,\dots,n)\frac{1}{s_I}A_{n_2}(\hat{P}_I,\hat{m}_2,\hat{m}_3,4,\dots)\,,
\ee in which $m_1,m_2,m_3$ are a cyclic permutations of 1,2,3 and
the sum includes all partitions $I$ in which the negative helicity
lines are separated  and all external lines appear in cyclic order
on the right side.\footnote{Formula \reef{rec2} applies specifically to the
case of consecutive ordering of the lines of each helicity.  For
other orderings there are similar relations.}
The contribution of each individual diagram depends on $|X]$. However,
if the amplitude vanishes as $z \to \infty$ for all $|X]$,
Cauchy's theorem ensures that the sum of all diagrams is
independent of $|X]$.

Spinors for the shifted momenta $\hat{1},
\hat{2},\hat{3},\hat{P}_I$ which appear in \reef{rec2} are
evaluated at  the pole of the  variable $z$ in \reef{3line} for
each contributing diagram.  The channel momentum $\hat{P}_I$ can
always be written to include the negative helicity line
$\hat{m}_1$, plus the sum $K$ of the positive helicity lines in
the same sub-amplitude,  i.e.~$\hat{P}_I = \hat{m}_1 + K$. The
pole condition, similar to \reef{pocon}, is
\be \lab{pocon3}
  0 ~=~ \hat{P}_I^2 = - s_I - z_I \<m_2\,m_3\> \<m_1|P_I|X]\,.
\ee
Scalar
products among the shifted (denoted by $m_i$) and unshifted
(denoted by $k$) spinors are required to evaluate sub-amplitudes.
They are given by (see  \cite{greatdane})
\bea
\lab{shsp1}
   \<i\,\hat{P}_I\> &=& \o^{-1}\, \<i\, |P_I |X]\, ,~~~~~~
  \o~=~ [\hat{P}_I\, X]\, ,\\[1mm]
\lab{shsp2}
  [\hat{P}_I \, k] &=& \frac{\o\, \< m_1| P_I |k ]}{\< m_1| P_I |X]}\,,
  ~~~~~~P_I ~=~ m_1 + K\, ,\\[1mm]
\lab{shsp3}
  [\hat{m_2}\,\hat{m_3}]
  &=& [m_2\,m_3] + z_I \, \< m_1| m_2 + m_3|X]\, ,\\[1mm]
\lab{shsp4}
  [\hat{m}_1\,k] &=& [m_1 \,k] - z_I \, \<m_2\,m_3\> [k\,X]\,.
\eea where $z_I$ is determined by \reef{pocon3}.

We now return to the recursion sum \reef{rec2}.  Due to helicity
conservation,   the two  sub-amplitudes must be MHV for partitions
in which both $n_1>3$ and $n_2>3$. The only possibility for
non-MHV subdiagrams in the sum \reef{rec2} arises from diagrams
like that of figure \ref{fig:nMHV0}, but now with $1,2,3$ all
shifted according to \reef{3line}.
As a result of the different shifts, the 3-point anti-MHV
amplitude in this expression now vanishes due to kinematics. This
is easily seen using $A_3(\hat{m}_1^-,k^+,-\hat{P}_I^+) = [k \,
\hat{P}_I]^3/ ([\hat{P}_I\, \hat{m}_1][\hat{m}_1 \, k])$.
Equations \reef{shsp2} then tell us that $[k \, \hat{P}_I] = -
\o\, \< m_1| P_I |k ]/\< m_1| P_I |X] = 0$, because $P_I = m_1 +
k$, while both factors in the denominator are non-vanishing. We
conclude that the sum \reef{rec2} (and its generalizations to
other orderings of the $\pm$ve helicity lines) contains only
diagrams where each subdiagram is MHV. This is the principal
advantage of the 3-line shift. It allows the construction of the
relatively difficult NMHV amplitudes from simpler and familiar MHV
elements.

\begin{figure}[t!]
  \begin{center}
        \includegraphics[width=5.5cm,angle=180]{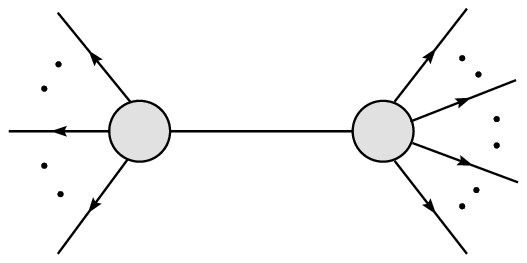}
            \begin{picture}(0,0)(0,0)
      \put(-225,-40){$\mathcal{A}_{n,I} ~~~=~$}
      \put(-16,-77){$l^+$}
      \put(-18,0){$(k+1)^+$}
      \put(-2,-22){$\hat{m}_2^-$}
      \put(-2,-54.5){$\hat{m}_3^-$}
      \put(-181,-77){$(l+1)^+$}
      \put(-154,0){$k^+$}
      \put(-177,-39){$\hat{m}_1^-$}
      \put(-87,-32){$\hat{P}_I$}
      \put(-108,-33){$-$}
      \put(-66,-33){$+$}
    \end{picture}
  \end{center}
\caption{Generic MHV-vertex diagram from the 3-line shift
recursion relations for NMHV amplitudes in both gauge theory and
gravity.}
\lab{fig:nMHV}
\end{figure}

Since the sub-amplitudes are MHV, there is only one choice of helicities
for the internal line, and hence each diagram in the recursion
expansion is uniquely characterized by its pole momentum $P_I$.
Figure \ref{fig:nMHV} shows a typical diagram  $\mathcal{A}_{n,I}$
that contributes to the sum in \reef{rec2} for
$A_n(1,\dots,m_1^-,\dots,m_2^-,\dots,m_3^-,\dots,n)$. Each vertex
amplitude is MHV, so we use the Parke-Taylor formula  \cite{pt} to
write
 \bea
  \nonumber
  \mathcal{A}_{n,I} &=&
  A_{n_1}(\hat{m}_1^-,\dots,k^+,-\hat{P}_I^-,(l+1)^+,\dots)
  \, \frac{1}{s_I} \,
  A_{n_2}(\hat{m}_2^-,\dots,\hat{m}_3^-,\dots,l^+,\hat{P}_I^+,(k+1)^+,\dots) \\[2mm]
  &=&
  \frac{\< m_1\, \hat{P}_I \>^4}
  {\< \hat{P}_I , l+1 \> \cdots \< k \, \hat{P}_I \>}
  \, \frac{1}{s_I} \,
  \frac{\< m_2\, m_3 \>^4}
  { \< \hat{P}_I , k+1 \> \cdots \< l \, \hat{P}_I \>} \, .
  \lab{DI1}
\eea

Each angle bracket with $\hat{P}_I$ can be rewritten using
\reef{shsp1}, giving $|\hat{P}_I \> = \o^{-1} P_I \,|X]$.  Since
the $\o$-factors cancel  in \reef{DI1}, we can ignore them from
the beginning. Thus we will use the simpler rule
\be  \lab{mhvsp}
  |P_I\> = P_I|X] = (p_{m_1}+K) | X]\,.
\ee
This is the CSW spinor prescription for an internal line, and the 3-line
recursion relations thus reproduce the MHV-vertex expansion of
\cite{csw}.

A useful alternate form \cite{ggk} of \reef{DI1} is
\be
  \lab{DI2}
  \mathcal{A}_{n,I} ~=~ \Big(\prod_1^n \<i , i+1 \>^{-1}\Big)
  \, \frac{1}{V_I} \, \< m_2\, m_3 \>^4\, \< m_1\, P_I \>^4 \, ,
\ee where \bea
  \lab{VI}
  \frac{1}{V_I} &=&
  \frac{\<l, l+1\>\<k, k+1\>}
  {s_I\, \< P_I , l+1 \> \< k \, P_I \> \< P_I , k+1 \> \< l \, P_I \> } \,.
\eea The cyclic invariant product is common to all diagrams which
contribute to the full amplitude.

\begin{quote}
{\bf Example: the 6-point gluon NMHV amplitudes}\\
Shifting the negative helicity lines $1,2,3$,  the recursion
relations for the gluon amplitude $A_6(1^-, 2^-,3^-,4^+,5^+,6^+)$
contains 6 diagrams which we label by their poles, namely 12, 23,
34, 61, 612, and 234; the diagrams are shown in figure
\ref{fig:6ptNMHV}. While each diagram depends on $|X]$, their sum
is $|X]$-independent because $A_6(z) \to 0$ for all $|X]$ as $z
\to \infty$.

The two other 6-point gluon NMHV amplitudes  $A_6(1^-,
2^-,3^+,4^-,5^+,6^+)$ and $A_6(1^-, 2^+,3^-,4^+,5^-,6^+)$ can
likewise be computed from recursion relations obtained from
shifting the three negative helicity lines. They contain 8 and 9
diagrams, respectively, and again the sums of diagrams are
independent of $|X]$.
\end{quote}

\begin{figure}[t!]
  \begin{center}
  \begin{tabular}{ccccc}
    \includegraphics[height=2cm]{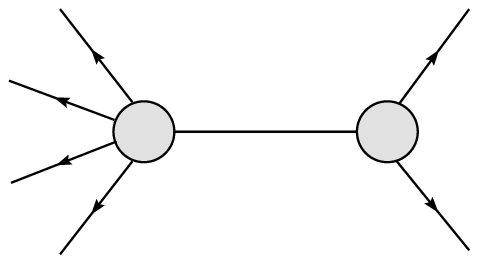}&~~&
    \includegraphics[height=2cm]{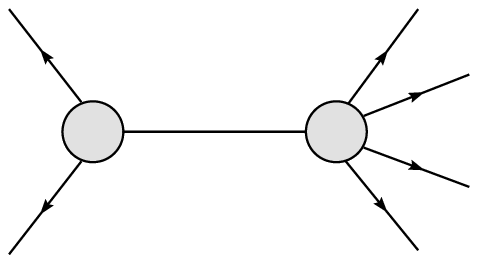}&~~&
    \includegraphics[height=2cm]{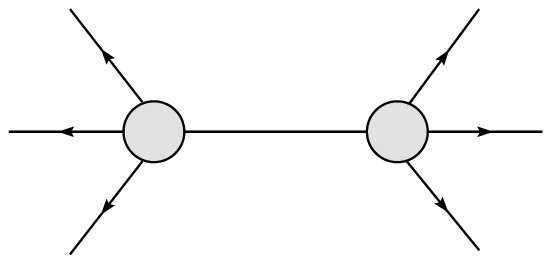} \\[5mm]
        \includegraphics[height=2cm]{nMHVrec6ptA.eps}&~~&
        \includegraphics[height=2cm]{nMHVrec6ptB.eps}&~~&
    \includegraphics[height=2cm]{nMHVrec6ptC.eps}
  \end{tabular}
  \begin{picture}(0,0)(0,0)
      \put(-291,68){$\hat{1}$}
      \put(-291,11){$\hat{2}$}
      \put(-394,11){$\hat{3}$}
      \put(-405,26){$4$}
      \put(-405,50){$5$}
      \put(-394,68){$6$}
      \put(-271,68){$\hat{1}$}
      \put(-271,11){$6$}
      \put(-169,11){$5$}
      \put(-157,26){$4$}
      \put(-157,52){$\hat{3}$}
      \put(-169,68){$\hat{2}$}
      \put(-123,68){$\hat{1}$}
      \put(-138,38.5){$6$}
      \put(-124,11){$5$}
      \put(-21,13){$4$}
      \put(-7,38.5){$\hat{3}$}
      \put(-22,68){$\hat{2}$}
      \put(-291,-8){$\hat{2}$}
      \put(-291,-64){$\hat{3}$}
      \put(-394,-64){$4$}
      \put(-405,-50){$5$}
      \put(-405,-25){$6$}
      \put(-394,-8){$\hat{1}$}
      \put(-271,-8){4}
      \put(-271,-64){$\hat{3}$}
      \put(-169,-64){$\hat{2}$}
      \put(-157,-50){$\hat{1}$}
      \put(-157,-25){$6$}
      \put(-169,-8){$5$}
      \put(-123,-8){$5$}
      \put(-138,-36){$4$}
      \put(-124,-64){$\hat{3}$}
      \put(-21,-64){$\hat{2}$}
      \put(-7,-36){$\hat{1}$}
      \put(-22,-8){$6$}
    \end{picture}
  \end{center}
\caption{The six MHV-vertex diagrams needed for the 3-line
recursion relation for the gluon NMHV amplitude
$A_6(1^-,2^-,3^-,4^+,5^+,6^+)$.
Amplitudes for other external particles of $\cn =4$ theory are
obtained by multiplying each gluon diagram by the appropriate spin factor.}
\lab{fig:6ptNMHV}
\end{figure}

\noindent
Let us now consider the analogous approach to
graviton amplitudes.

\subsubsection*{Gravity}
The 3-line shift gives the recursion relation \bea
 \lab{rec3}
 M_n(1^-,2^-,3^-,4^+,\dots,n^+)
 &=&
 \sum_{I} M_{n_1}(\hat{m}_1^-,-\hat{P}_I^-,\dots)
 \, \frac{1}{s_I} \,
 M_{n_2}(\hat{m}_2^-,\hat{m}_3^-,\hat{P}_I^+,\dots) \,.
\eea For each value of $n_1$, the sum includes all cyclic
orderings of the negative helicity lines and of all distinct arrangements
of the positive helicity lines.
Overall Bose symmetry is then maintained. The sum in \reef{rec3}
only contains MHV-vertex diagrams. This can be shown as in gauge
theory using $M_3 = A_3^2$.

The form of the BGK formula presented in \cite{ef} can be used to
express the two MHV sub-amplitudes and show that the $\o$-factor
in \reef{shsp1} drops out. In more detail: \bea
  \nonumber
  &&
  M_{n_1}(\hat{m}_1^-,-\hat{P}_I^-,i_3^+,\dots,i_{n_1}^+)\\
  \lab{mone}
  && \hspace{1cm}
  ~=~\o^{-4} \sum_{\mathcal{P}(i_4,\dots, i_{n_1})}
  \frac{\<m_1 \, P_I \> \< i_3 \, i_4 \>}{\<m_1 \, i_3 \> \<P_I\, i_4 \>} \,
  s_{\hat{m}_1 i_{n_1}} \left( \prod_{s=4}^{n_1-1} \beta_s  \right)
  A_{n_1}(\hat{m}_1^-,-P_I^-, i_3^+,\dots ,i_{n_1}^+)^2 \, ,~~~~~
\eea with \bea
  \lab{bone}
  \beta_s = -\frac{\< i_s\, i_{s+1} \>}{\< P_I\, i_{s+1} \>}
    \; \< P_I | \, i_3 + i_4 + \dots + i_{s-1} | i_s ] \, .
\eea The $\o^{-4}$ factor comes from setting $|\hat{P}_I \> =
\o^{-1} P_I \,|X]$ in $A_{n_1}^2$.

Likewise,
\bea
  \nonumber
  &&
  M_{n_2}(\hat{m}_2^-,\hat{m}_3^-,\hat{P}_I^+,j_4^+,\dots,j_{n_2}^+)\\
  \lab{mtwo}
  && \hspace{1cm}
  ~=~
  \o^4 \sum_{\mathcal{P}(j_4,\dots, j_{n_2})}
  \frac{\<m_2 \, m_3 \> \< P_I \, j_4 \>}{\<m_2 \,P_I \> \<m_3\, j_4 \>} \,
  s_{\hat{m}_2 j_{n_2}} \left( \prod_{s=4}^{n_2-1} \beta_s  \right)
  A_{n_2}(\hat{m}_2^-,\hat{m}_3^-,P_I^+,j_4^+,\dots,j_{n_2}^+)^2\, , ~~~~~~~~~~
\eea with \bea
  \lab{btwo}
  \beta_s = -\frac{\< j_s\, j_{s+1} \>}{\< m_3\, j_{s+1} \>}
    \; \< m_3 | \, \hat{P}_I + j_4 + \dots + j_{s-1} | j_s ] \, .
\eea
This latter expression contains $\hat{P}_I$ only in the
$\o$-independent combination $\< m_3  \, \hat{P}_I \> [ \hat{P}_I
\, j_s ] = \< m_3 \, P_I \> \< m_1 | P | j_s ] / \< m_1 \, P_I
\>$. (See (\ref{shsp1}-\ref{shsp2}).) The two results for $\o$-factors are only valid for $n_1,n_2
\ge 4$.  For $n_1$ or $n_2=3$ one can simply use $M_3 = (A_3)^2$ to
deduce  the same results. It is obvious now that the $\o$-factors
cancels in $M_{n_1}\, s_I^{-1}M_{n_2}$ yielding diagrams which are
independent of $\o$.
Note that the effect of the shift appears in $|P_I\>$, given in \reef{mhvsp},
and in  $s_{\hat{m}_i j} = \<m_i \, j\>[j\,\hat{m}_i]$.

\begin{quote}
{\bf Example: the 6-point graviton NMHV amplitude}\\
Shifting the negative helicity lines $1,2,3$, the recursion
relations for the graviton amplitude $M_6(1^-,
2^-,3^-,4^+,5^+,6^+)$ contains 21 diagrams which fall in three
classes: three 2-particle ``$--$'' poles ($I=12,13,23$), nine
2-particle ``$-+$'' poles ($I=m_i 4,m_i 5,m_i 6$) and nine
3-particle poles ($I=m_i 45,m_i 46,m_i 56$). One only needs to
compute one amplitude from each class; the rest can be obtained by
momentum relabelling. Our numerical check shows that the sum of
the 21 diagrams is independent of $|X]$.
\end{quote}

The diagrammatic expansions associated with the recursion
relations \reef{rec2} for gauge theory and \reef{rec3} for gravity
are the basis for our treatment of NMHV amplitudes.  We apply them
to amplitudes for general external states of $\cn = 4$ SYM and
$\cn = 8$ \sg\ using the generating functions discussed below to
determine the spin factors for each diagram. It is important that
the amplitudes vanish as $z \to \infty$, and this condition  will
play a crucial role in the application to $\cn = 8$ \sg .


\subsection{NMHV Generating Function for $\cn = 4$ SYM}
\lab{s:GFsym}

In this section we derive the generating function of \cite{ggk}
and discuss several properties that are important for our
application.  The goal is to obtain a correct and efficient
construction of the entire NMHV sector of the  $\cn = 4$ theory.  The
NMHV sector consists of the top $n$-gluon amplitudes  $A_n$ (for
various orderings of the three  negative helicity lines)
together with all other amplitudes related to those by SUSY Ward
identities. The practical definition of this sector is that it
contains all sets of external particles for which the
corresponding differential operator formed from products of $n$
factors from the list \reef{n4ders} is of total order 12, rather
than order 8 which characterizes the MHV sector.  There is another
significant difference between the two sectors. In the MHV sector
there is a single generating function $F_n$, given in
\reef{n4gen}, from which all $n$-point amplitudes are obtained. In
the NMHV sector there is a different generating function
 for each diagram $\mathcal{A}_{n,I}$ in the MHV-vertex
decomposition. After applying the appropriate 12th order
differential operator, the full amplitude is obtained by adding
the results for all diagrams contributing to the recursion
relation \reef{rec2}.

\begin{figure}[t!]
  \begin{center}
        \includegraphics[width=5.5cm,angle=180]{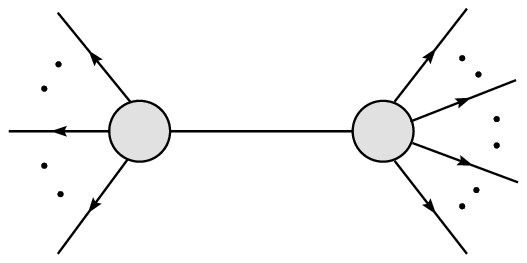}
            \begin{picture}(0,0)(0,0)
      \put(-225,-40){$\mathcal{A}_{n,I} ~~~=~$}
      \put(-16,-77){$l$}
      \put(-18,0){$(k+1)$}
      \put(-2,-22){$\hat{m}_2$}
      \put(-2,-54.5){$\hat{m}_3$}
      \put(-177,-77){$(l+1)$}
      \put(-154,0){$k$}
      \put(-177,-39){$\hat{m}_1$}
      \put(-87,-32){$\hat{P}_I$}
      \put(-108,-33){$-$}
      \put(-66,-33){$+$}
    \end{picture}
  \end{center}
\caption{MHV-vertex diagram from the 3-line shift recursion
relations for NMHV amplitudes with external states of the $\cn =
4$ theory.} \lab{fig:nMHVgenf}
\end{figure}

Consider an NMHV $n$-point amplitude for a general set of external
states, and choose 3 lines, $m_1,\,m_2,\,m_3$, to shift, as in
\reef{3line},  such that the amplitude  vanishes as $z \to
\infty$. Not all shifts produce an amplitude with the required
falloff at large $z$. The issue of the choice of a valid shift is
discussed in section \ref{s:symz}.

Given a valid shift,
the amplitude can  be expressed as the sum of diagrams  in the
recursion  relation \reef{rec2}.
Each diagram contains the product of two MHV sub-amplitudes, as
shown in figure \ref{fig:nMHVgenf}, and each of these can be
expressed as the appropriate eighth order product of derivatives
from the correspondence \reef{n4ders} acting on the MHV generating
function \reef{n4gen}. Thus we can write the generalization of the
amplitude \reef{DI1} of figure \ref{fig:nMHVgenf} to an arbitrary
set of external states as
\bea
  \nonumber
 \mathcal{A}_{n,I} &\equiv&
  \mathrm{sign}(I) \,
  \frac{A_{n_1}(l+1,..,m_1,..,k,-P_I)}{\<m_1\,P_I\>^4}\frac{1}{s_I}
  \frac{A_{n_2}(P_{I'},k+1,..,m_2,..,m_3,..,l)}{\<m_2m_3\>^4}\\
&&~~~~~~~~~~~~~~~ \times\Big( D_{l+1}\ldots D_k\,D_I \,
  \d^{(8)}\big( L \big) \Big)
  \,\Big( D_{I'}D_{k+1}\dots
D_l\,\d^{(8)}\big( R \big) \Big) \,,
\lab{diadan}
\eea
where \bea \lab{lr}
  L ~=~ |P_I\>\, \h_{Ia}+\sum_{i=l+1}^k |i\>\, \h_{ia} \, ,
  \hspace{8mm}
  R ~=~ |P'_{I}\>\, \h_{Ib}+\sum_{i=k+1}^l |i\>\, \h_{ib} \, .
\eea The delta functions $\d^{(8)}$ are defined in \reef{del8}.
The spinors for the internal lines  are\footnote{All factors of
$\o$ have been removed as discussed in the previous section.} \be
\lab{intspin} |P_I\> = P_I|X] \,  = -|P_{I'}\>\,, \ee where $P_I$
is the sum of the external momenta on the left sub-amplitude of
figure \ref{fig:nMHVgenf}.

The differential operators $D_I$ and $D_{I'}$ represent particles
at the left and right ends of the internal line. Since these
particles are opposite helicity states of the same field, the
orders of the operators are related by $d_I+d_{I'}=4$, and they
carry distinct $SU(4)$ indices. Thus
\bea
  D_I \, D_{I'} = \pm \prod_{a=1}^4\frac{\pa}{\pa\h_{Ia}} \, .
\eea

When the derivative operators of the external lines are applied to
$\d^{(8)}\big( L \big) \d^{(8)}\big( R \big)$, they uniquely
determine the split of the four derivatives
$\prod_{a=1}^4\frac{\pa}{\pa\h_{Ia}}$ into the product $D_I \,
D_{I'} $. Starting with an initial ordering of the differential
operators, $D_1 D_2 \dots D_n$ as dictated by the color ordering,
we can therefore write \bea
  \lab{Dorder}
  \Big(\prod_{a=1}^4\frac{\pa}{\pa\h_{Ia}}\Big)
  D_1 D_2 \dots D_n = \mathrm{sign}(I) \,
  D_{l+1}\ldots D_k\, D_I \, D_{I'} \, D_{k+1}\ldots D_l \, ,
\eea where the sign, $\mathrm{sign}(I)=\pm 1$, which also appeared
in \reef{diadan},  arises from the required interchange of
Grassman derivatives.

We now use three facts:
\begin{enumerate}
\item[i.]
The external state derivatives
can all be moved to the left of the expression \reef{diadan} and reordered according to \reef{Dorder}.
\item[ii.] Integration and differentiation are equivalent for
 functions of Grassmann variables, so the 4 $\h_{Ia}$ derivatives can be written as integrals.
\item[iii.]  The 4-fold integral can be performed using the technique of
section 5 of \cite{ggk}.
\end{enumerate}

Using this we can rewrite \reef{diadan} as  the product
$D_1 D_2 \dots D_n$ of derivatives acting on
\be
  \lab{LR}
  \int \prod_{a=1}^4 d\h_{Ia}\,
  \d^{(8)}\big( L \big) \,
  \d^{(8)}\big( R \big)
  ~=~
  \d^{(8)}\Big(\sum_{i=1}^{n}|i\>\h_{ia}\Big)
  \prod_{b=1}^4\sum_{j=l+1}^{k}\<P_I\,j\>\h_{jb} \, .
\ee We started with a product of 16 derivatives in \reef{diadan}
and eliminated the 4 derivatives on $\h_{Ia}$,  The remaining
product of 12 derivatives corresponds exactly to the external
states of the amplitude.

Since the argument applies to any diagram of the general NMHV
amplitude, we have derived the generating function
\bea
\lab{N4gen}
   \mathcal{F}_{n,I} &=&
  \frac{\mathcal{A}_{n,I}^\mathrm{gluons}}{\<m_1\,P_I\>^4 \<m_2m_3\>^4} \,
    \d^{(8)}\Big(\sum_{i=1}^{n}|i\>\h_{ia}\Big)\,
   \prod_{b=1}^{4}\sum_{j\e I}\<P_I \,j\>\h_{jb} \, ,
\eea
where $\mathcal{A}_{n,I}^\mathrm{gluons}$ is the value of the pure gluon MHV-vertex diagram obtained from the same shift,
\bea
  \mathcal{A}_{n,I}^\mathrm{gluons} &=&
  A_{n_1}(l+1,..,m_1,..,k,-P_I)\,\frac{1}{s_I}
  A_{n_2}(P_{I'},k+1,..,m_2,..,m_3,..,l)\, .
\eea
This prefactor ensures that the pure gluon amplitudes are correctly reproduced by \reef{N4gen}.
Thus, given the values of each gluon MHV-vertex diagram $\mathcal{A}_{n,I}^\mathrm{gluons}$ any NMHV amplitude is simply calculated by applying the ordered string of differential operators associated with the string of external states to the sum over generating functions for each diagram
\bea
  \lab{nmhvA}
  A_n^\mathrm{NMHV}
  &=& D_1 D_2 \dots D_n \, \mathcal{F}_n \, ,
  \hspace{1cm}
   \mathcal{F}_n ~=~ \sum_I \mathcal{F}_{n,I} \,.
\eea
Note that this construction automatically produces the correct relative sign of diagrams in the MHV-vertex decomposition.

The sum $\mathcal{F}_{n}$ of generating functions of the MHV-vertex diagrams is the generating function for the whole NMHV amplitude.
Each term $\mathcal{F}_{n,I}$ is a sum of products of 12 distinct $\h$'s,
so an NMHV $n$-point  amplitude is calculated by
applying the appropriate 12th
order differential operator composed of $n$ factors from the list
\reef{n4ders}.
Each distinct NMHV process corresponds to a
particular 12th order operator and conversely. The number of
distinct NMHV processes is thus the number of
partitions of 12 with $n_{{\rm max}} \le 4$ which is 34.  For each
given set of particles, i.e.~for each process, there are several
independent amplitudes
in which their order is permuted.\footnote{In addition there are usually
 inequivalent assignments of $SU(4)$ indices which give independent amplitudes.}  For processes with $n < 12$
external particles, the total number of NMHV processes is $< 34$
because one must count only partitions of length $\le n$.  For
this reason there are `only' 18 distinct 6-point processes. Each of these may have several inequivalent assigments of $SU(4)$ indices.

To gain further confidence in the use of the generating function
\reef{N4gen} let's examine whether the amplitudes obtained from it
satisfy SUSY Ward identities.  One should bear in mind that these
\wis\ need not be satisfied by individual diagrams, but they must
hold for the full amplitude. The $\tQ_a$ \wi\ reads \be  \lab{wi1}
\tQ_a \tilde{F}_n =0\,, \ee and it is satisfied diagram by
diagram, as in the MHV case,  because one is again multiplying the
$\d^{(8)}$ function by its own argument.  The $Q^a$ \wi\ presents
a more interesting situation.
Using the two identities
\bea
Q^a\,\prod_{b=1}^{4}\sum_{k\e I}\<P_I\,k\>\h_{kb}
  ~=~\sum_{i\e I} [\e \,i]\<i\,P_I\>\prod_{b\ne a}\sum_{j\e I}\<P_I\,j\>\h_{jb}\, , \hspace{1cm}
 \sum_{i\e I} [\e \,i]\<i\,P_I\> ~=~ - [\e\,X] P_I^2 \, ,
\eea
we evaluate
\bea
  \nonumber
  Q^a\tilde{F}_n
  &=& \frac{1}{V_I}\big(\prod_{i=1}^{n}\<i\,(i+1)\>\big)^{-1}\d^{(8)}(\sum_{i=1}^{n}|i\>\eta_{ia})\,Q^a\,\prod_{b=1}^{4}\sum_{j\e I}\<P_I\,j\>\h_{jb}  \\
  &=& -  [\e\,X] \, P_I^2 \, \frac{1}{V_I}\big(\prod_{i=1}^{n}\<i\,(i+1)\>\big)^{-1}\d^{(8)}(\sum_{i=1}^{n}|i\>\eta_{ia})\,
  \prod_{b\ne a}\sum_{j\e I}\<P_I\,j\>\h_{jb}\,  .
\eea
This shows that each diagram contributing to a given amplitude vanishes if we choose $X_\a \sim \e_\a$.
However, provided that the amplitude vanishes for large $z$,
the MHV-vertex formalism
ensures that the sum of these diagrams is independent of  the
reference spinor $X_\a$, so the full amplitude will satisfy the
$Q^a$  \wi .

At first sight we now seem to be in the same position as we were
in the analysis of MHV amplitudes.  The full NMHV sector is
determined by the values of the diagrams for the $n$-gluon
amplitudes, and all amplitudes satisfy SUSY \wis. However, there
is an important difference. In the MHV sector there is a unique set of amplitudes which satisfy the \wis\
and agree with the top $A_n$. In the NMHV sector it is not
sufficient to reproduce only $A_n$; additional input is required.\footnote{Relations from $\cn = 2$ \wis\ were used recently
 in  \cite{StStTT6ptdisk} to simplify the calculation of 6-gluon
amplitudes in open string theory.}
However, the MHV-vertex decomposition
contains additional dynamical input, namely the correct
analyticity and factorization properties, so we can be confident
that it generates the right amplitudes --- provided that there is
no contribution from infinity in the recursion relations.


\subsubsection{NMHV spin factors}
\lab{s:SFsym}

We will illustrate the use of the generating function
\reef{N4gen} and the calculation of spin factors through an
example, which will also be relevant for our examination of the
large $z$ behavior of NMHV amplitudes in $\cn = 4$ SYM theory.

Consider the six scalar\footnote{We choose this particular
configuration of three different ``particles'' and their
``anti-particles" because it is the gauge theory analogue of a
6-scalar amplitude in $\cn = 8$ \sg\ which we will study in detail
in section \ref{s:gravz}.} amplitude
\bea
 \lab{6scYM}
 \big\<B^{12}(1)\, B^{13}(2)\,  B^{14}(3)\,
B^{23}(4)\,  B^{24}(5)\,  B^{34}(6)\big\> \, .
\eea
The recursion relations
following from a shift of lines 1,2,3 consists of a sum over the
six diagrams drawn in figure \ref{fig:6ptNMHV}. Each of these
diagrams is the product of the result for 6-gluons times a
\emph{spin factor} obtained by applying the external state
derivatives to the generating function \reef{N4gen} and dividing
by $\< m_1 P_I \>^4 \< m_2\, m_3\>^4$. The spin factor encodes the
state dependence of the amplitude. It is a ratio of products of angle brackets and is homogeneous in $P_I$.

In practice it is simplest to compute the spin factor
from the product $\d^{(8)}(L)\,\d^{(8)}(R)$ in \reef{diadan} and
define it precisely for a generic diagram as
\be
\lab{sfdef}
\mathcal{S}_I \equiv \mathrm{sign}(I)\, \Big(
D_{l+1}\ldots D_k\,D_I \,
  \d^{(8)}\big( L \big) \Big)
  \,\Big( D_{I'}D_{k+1}\dots
D_l\,\d^{(8)}\big( R \big) \Big)/\big(\<m_i\,P_I\>^4
\<m_2\,m_3\>^4\big)\,. \ee
If the diagram is non-vanishing there is a
unique choice of the operators $D_I$ and $D_{I'}$ for the internal
line which produces the result. The derivative operation is
equivalent to a Wick contraction algorithm based on \reef{cont}.

The example of the 6-scalar amplitude \reef{6scYM} will make things
clear. Consider the 12-pole diagram. Using the notation  $\pa_i^a
= \pa/\pa\h_{ia}$ and $\pa_I^a =\pa/\pa\h_{Ia}$,  the derivatives
applied to $\d^{(8)}(L)$ are
\be
  \pa_1^1\pa_1^2 \; \pa_2^1\pa_2^3 \; \pa_I^1\pa_I^2\pa_I^3\pa_I^4\,.
\ee
The total derivative order must be 8, so we included the unique
internal line derivative of order 4.  There is no way to make 4
non-vanishing Wick contractions among these derivatives so the
12-pole diagram vanishes.  The same is true for the 23-pole
diagram. For the 34 pole diagram we write the string of derivatives
\bea
\lab{34p}
\mathcal{S}_{34}&=&(\pa_3^1\pa_3^4 \; \pa_4^2\pa_4^3 \; \pa_I^1\pa_I^2\pa_I^3\pa_I^4)
\times(\pa_5^2\pa_5^4 \; \pa_6^3\pa_6^4 \; \pa_1^1\pa_1^2 \; \pa_2^3\pa_2^4)/\big(\<3\,P_{34}\>^4\<1\,2\>^4\big)\\[1mm] \nonumber
&=&-\frac{\<4\,P_{34}\>^2}{\<3\,P_{34}\>^2}\frac{\<1\,5\>\<2\,6\>\<5\,6\>}{\<1\,2\>^3}\\[1mm] \nonumber
&=&
-\frac{[3X]^2}{[4X]^2}\frac{\<1\,5\>\<2\,6\>\<5\,6\>}{\<1\,2\>^3}\,.
\eea
In the first line we chose the unique 4th order internal
derivative which describes the emission of a negative helicity gluon
from the left vertex and subsequent absorption as a positive helicity
gluon on the right.  The second line gives the unique
non-vanishing Wick contraction. It exemplifies the general feature
that the spin factor is a homogeneous function of angle brackets
and also homogeneous in $|P_I\>$.  In the last line we have used
$\<4 P_{34}\> =\<4 \,3\>[3X]$ and a similar equality.  We invite
readers to compute the remaining 3 non-vanishing spin factors:
\bea
  \lab{sfs}
  \mathcal{S}_{61} &=&
  -\frac{\< 6 P_{61} \>^2}
      {\< 1 P_{61} \>^2}
   \frac{\< 24 \>\< 35 \>\< 45 \>}
  {\< 23 \>^3} \,~=~\,
  -\frac{[1X]^2}{[6X]^2}  \frac{\< 24 \>\< 35 \>\< 45\>}
      {\< 23 \>^3}     , \\[1mm] \nonumber
  \mathcal{S}_{612} &=&
  \frac{\< 35 \>\< 45 \> \< 4 P_{612} \>}
      {\< 3 P_{612} \>^3}
   \frac{\<26\>\< 1 P_{612} \>\< 6 P_{612}\>}
      {\< 12 \>^3} \,~=~\, \frac{\<4|3+5|X] \<1|6+2|X]\<6|1+2|X]}
      {\<3|4+5|X]}\frac{\<26\>\<35\>\<45\>}{\<12\>^3}\, ,\\[1mm] \nonumber
  \mathcal{S}_{234} &=&
  \frac{\< 15 \>\< 56 \> \< 6 P_{234} \>}
      {\< 1 P_{234} \>^3}
   \frac{\< 24 \>\< 3 P_{234}\>\< 4 P_{234} \>}
      {\< 23 \>^3}~=~
      \frac{\< 6|1+5|X]\< 3 |2+4|X]\< 4| 2+3|X]}
      {\< 1 | 5+6 |X]^3}
      \frac{\< 15 \>\< 24 \>\< 56 \>}
      {\< 23 \>^3}
       \, .
\eea

We have checked numerically that the sum of the four non-vanishing
diagrams is independent of the reference spinor $|X]$, and that
the amplitude vanishes for large $z$ under a subsequent shift of
lines 1,2,3.


\subsubsection{Large $z$ behavior in $\cn = 4$ gauge theory}
\lab{s:symz}

Since the NMHV generating function \reef{nmhvA} is based on the
recursion relations obtained from the 3-line shift \reef{3line},
it requires that shifted amplitudes vanish as $z \to \infty$.
Tree amplitudes always behave as the power law  $z^{-\D}$ with
integer $\D$, so we require $\D > 0$ for all amplitudes in order
to use the NMHV generating function with confidence. The exponent
$\D$ depends on the spin of external states.
We now discuss evidence that $\D>0$
for all $n$-point NMHV amplitudes in $\cn = 4$ SYM theory.

Each diagram in the MHV-vertex expansion can be written as a spin
factor  $\mathcal{S}_I$ times the pure gluon diagram. Under a
shift of the 3 negative helicity lines, the pure gluon diagram
goes to zero for large $z$ at least as fast as $1/z^4$. To see
this recall that the gluon diagram is the product of two
Parke-Taylor amplitudes and the internal propagator, \bea
  \lab{nmhvfig1}
  && \frac{\< m_1 \hat{P}_I\>^4}{\< k \hat{P}_I \> \< \hat{P}_I \,(l+1) \>(\dots)}
  \frac{1}{P_I^2}
  \frac{\< m_2\, m_3\>^4}{\< l \hat{P}_I \> \< \hat{P}_I \,(k+1) \> (\dots)} \\[1mm]
  && \nonumber \hspace{1.5cm}    =
  \frac{\< m_1 | P_I | X]^4}{\< k| P_I | X]  \< l+1 | P_I | X]   (\dots)}
  \frac{1}{P_I^2}
  \frac{\< m_2\, m_3\>^4}{\< l | P_I | X] \< k+1 | P_I | X] (\dots)} \, .
  \eea
The factors in $(\dots)$ do not involve $P_I$ and  are not
relevant to our argument, in which we  perform another 3-line
shift of the lines $m_i$, this time with a new arbitrary reference
spinor $| Y ]$. The only factors that can shift are those that
involve the momentum $P_I$. Specifically $\< i | P_I | X]$ shifts,
except when $i=m_1$, and the propagator denominator $P_I^2$
shifts.  Simple power counting in  \reef{nmhvfig1} then shows that
for large $z$ the diagram goes as $1/z^5$ when $k, (l+1) \ne m_1$
and as $1/z^4$ otherwise.

The spin factor denominator $\< m_1 P_I \>^4 \< m_2\, m_3\>^4$
does not shift.  The numerator contains  a fourth order product
of angle brackets containing $|P_I\>$, so under the new $|Y]$-shift,
$\mathcal{S}_I$  can at most grow as $z^4$. Thus, for any 3-line shift the most
divergent behavior possible for any diagram is order $O(1)$.
However, any NMHV amplitude carries a total of 12 $SU(4)$  indices, and group invariance  requires that each distinct index 1, 2, 3 or 4 must occur exactly 3 times among the $n$ external lines. Thus it is always possible to shift 3 lines which have at least one common index, say the index 1.  In every MHV partition of the amplitude that same index must also appear on the internal line of the sub-amplitude containing the line $m_1$. Thus at least one unshifting factor  $\< m_1 P_I \>$
occurs in the numerator of every spin factor. Hence, if one chooses a
shift with at least one common $SU(4)$ index, every diagram will vanish at least as fast as $1/z$.

This argument strongly supports the conjecture that all NMHV amplitudes of $\cn = 4$ theory are constructible by the MHV-vertex method, but it does not  \emph{prove} it.  To eliminate the possibility of a contribution from infinity which could invalidate the recursion relations, one would need to determine the asymptotic powers of NMHV amplitudes without using the form of the expansion itself. Perhaps the method of \cite{nima} can be generalized to 3-line
shifts and applied for this purpose.

Because the general argument does not quite reach its goal, we looked for additional evidence through a numerical study.
We have written a Mathematica code which for given $SU(4)$ indices
of the external states calculates the MHV-vertex decomposition for
any 6-point NMHV amplitude of the $\cn=4$ SYM theory. With this program we
have calculated many  NMHV amplitudes, and in all cases we have
found that there exists at least one 3-line shift such that the
associated sum of MHV-vertex diagrams is independent of the
reference spinor $|X]$.   This is further evidence  that there exists a
``good'' 3-line shift with associated valid recursion relations
for any 6-point NMHV amplitude of $\cn =4$ SYM theory.


\subsubsection*{Large $z$ for the six scalar amplitude in $\cn = 4$ SYM theory}

In a Feynman diagram analysis, the polarization vectors of the 3
negative helicity gluons  provide the power $z^{-3}$ at large $z$.
When the gluons are replaced by scalars, this asymptotic damping
is lost. This suggests that the least favorable
asymptotic behavior occurs for external scalars which have neither
polarization vectors nor external spinors. With this in mind we
discuss the large $z$ behavior of the gauge theory 6-point
amplitude $\<B^{12}(1)\,B^{13}(2)\,  B^{14}(3)\,  B^{23}(4)\,
B^{24}(5)\,   B^{34}(6)\>$ whose spin factors were calculated in section \ref{s:SFsym}.

Using the explicit results for the four non-vanishing spin
factors given in \reef{34p} and \reef{sfs}, one readily sees that
under a subsequent $|Y]$-shift of lines 123, the large $z$
behavior is \bea
  \mathcal{S}_{34} ~\sim~ z^2 \, ,\hspace{5mm}
  \mathcal{S}_{61} ~\sim~ z^2 \, ,\hspace{5mm}
  \mathcal{S}_{612} ~\sim~ z^3 \, ,\hspace{5mm}
  \mathcal{S}_{234} ~\sim~ z^3 \, .
\eea Each spin factor $\mathcal{S}_I$ must be multiplied by the
value of the corresponding diagram in figure \ref{fig:6ptNMHV} for
the 6-gluon process. Since the leading behavior of each of these
gluon diagrams is $1/z^4$, the  leading contribution to the
6-scalar amplitude comes from the 3-particle diagrams and is
$1/z$. There is no cancellation, so the falloff of the full
amplitude is $1/z$.
We also checked the behavior of the 6-scalar amplitude under
2-line shifts. The large $z$ behavior is $1/z^2$ for a $[1,3\>$
shift, but $O(1)$ for a $[1,2\>$ shift and $O(z)$ for a $[1,6\>$
particle-antiparticle shift.

We have also used the generating function to construct all other 6-scalar amplitudes with different configurations of $SU(4)$ indices. In
every case the sum of MHV-vertex diagrams is independent of
$|X]$.


\subsection{Generating function for NMHV amplitudes in $\cn = 8 $ \sg}

At the formal level, it is not difficult to extend the
construction of the previous section to \sg , but the issue of large $z$
behavior will become acute.
The extension is
based on the MHV-vertex formalism for $n$-graviton amplitudes of
\cite{greatdane} which we have discussed in section \ref{s:recrel}.

The NMHV sector of $\cn = 8$ \sg\ consists of all amplitudes
related to the top $n$-graviton amplitude
$M_n(1^-,2^-,3^-,4^+,\ldots, n^+)$ by SUSY \wis.  In analogy with
the $\cn = 4$ theory, the practical definition of this sector is
that it contains amplitudes for all sets of external particles for
which the associated differential operator constructed from
products of $n$ operators from the correspondence \reef{fops} is
of total order 24. We will justify this definition below.

If an NMHV amplitude vanishes at large $z$ under the 3-line shift,
it also obeys a recursion relation (equivalently, it has a valid
MHV-vertex decomposition) similar to \reef{rec3}. The contribution
of a generic diagram corresponding to figure \ref{fig:nMHVgenf} has the same
structure as the gauge theory formula \reef{diadan} and can be
written as
\bea
  \nonumber
 \mathcal{M}_{n,I} &\equiv&
  \mathrm{sign}(I) \,
  \frac{M_{n_1}(l+1,..,\hat{m}_1,..,k,-\hat{P}_I)}{\<m_1\,\hat{P}_I\>^8}\frac{1}{s_I}
  \frac{M_{n_2}
  (\hat{P}_{I},k+1,..,\hat{m}_2,..,\hat{m}_3,..,l)}{\<m_2\,m_3\>^8}\\
 \lab{diagrav}
&&~~~~~~~~~~~~~~~ \times\Big( \cd_{l+1}\ldots \cd_k\,\cd_I \,
  \d^{(16)}\big( L \big) \Big)
  \,\Big( \cd_{I'}\cd_{k+1}\dots
\cd_l\,\d^{(16)}\big( R \big) \Big) \,,
\eea
The arguments $L$ and $R$
of the $\d$-functions are given in \reef{lr}, and the detailed
form of the shifted sub-amplitudes $M_{n_1}$ and $M_{n_2}$ are given  in \reef{mone} and \reef{mtwo}.
Note that the derivatives acting on each $\d$-function are of
total order 16.

We follow the same steps used in the gauge theory case to obtain
the generating function for this diagram
\bea
\nonumber
\tilde{\O}_{n,I} &=&
\frac{M_{n_1}(l+1,..,\hat{m}_1,..,k,..,-\hat{P}_I)}{\<m_1\,\hat{P}_I\>^8}\frac{1}{s_I}
  \frac{M_{n_2}
  (\hat{P}_{I},k+1,..,\hat{m}_2,..,\hat{m}_3,..,l)}{\<m_2\,m_3\>^8}\\
  \lab{n8gen}
&&~~~~~~~~~~~~~~~ \times
\d^{(16)}\Big(\sum_{i=1}^{n}|i\>\h_{iA}\Big)
  \prod_{B=1}^8\sum_{j=l+1}^{k}\<\hat{P}_I \,j\>\h_{jB} \, .
\eea The generating function for the full amplitude is then
$\sum_I \tilde{\O}_{n,I}$,  where the sum runs over all $3(2^{n-3}-1)$
internal pole channels.

In the process of deriving \reef{n8gen} 8 internal derivatives $\cd_I\cd_{I'}$ were
converted to integrals and eliminated. The remaining external line
derivatives $ \cd_1\cdots \cd_n$ which
one applies to $\tilde{\O}_I$ to (re)obtain the diagram
\reef{diagrav} are of total order $32\,-\,8=24$.  Thus each distinct
$n$-point NMHV process in $\cn = 8$ \sg\ corresponds to a
particular 24th order differential operator composed of $n$
factors from the correspondence \reef{fops}. The number of
distinct NMHV amplitudes is the number of partitions of 24 with
$n_{{\rm max}} \le 8$ which is 919. One needs $n$-point functions
with $n \ge 24$ to realize this maximum number. For $n <  24$, the
partition length must be $\le n$, so there are fewer types of NMHV
amplitudes.  For $n=6$ there are 151 distinct processes.

The structure of the formula \reef{n8gen} is analogous to
\reef{N4gen}  in gauge theory.  In gauge theory the dynamical
function which multiplies the Grassmann factors is the value of the $n$-gluon
diagram divided by $\<\dots\>^4 \<\dots\>^4$ for the negative
helicity lines of the diagram. In \sg\ it is the $n$-graviton
diagram divided by $\<\dots\>^8 \<\dots\>^8$ for the negative
helicity lines.  One significant difference is that one must
insert the correctly shifted anti-holomorphic spinors
$|\hat{m}_i]$ for the 3 distinguished negative helicity gravitons.

The contribution of each diagram to a particular amplitude of
interest is obtained by applying the appropriate order 24 product
of external state derivatives to the generating function. Each
diagram has its own spin factor which is obtained in this way.
Our experience indicates that it is easiest to calculate the spin
factor by applying the derivatives to the product of
$\d$-functions for each sub-amplitude. Thus, in analogy with
\reef{sfdef}, the spin factor for a MHV-vertex diagram in \sg\ is
defined by
\be
\lab{sfsg}
\mathcal{S}_I \equiv   \Big(\cd_{l+1}\ldots
\cd_k\,\cd_I \,
  \d^{(16)}\big( L \big) \Big)
  \,\Big( \cd_{I'}\cd_{k+1}\dots
\cd_l\,\d^{(16)}\big( R \big) \Big)/(\<m_i\,P_I\>^8
\<m_2\,m_3\>^8)\,. \ee
 The products include  a pair of  internal line derivatives $\cd_I$
 and $\cd_{I'}$ which are uniquely determined by $SU(8)$ covariance and
 the fact that the total order of derivatives on each $\d^{(16)}$
must equal 16.  The simplest way to calculate uses the Wick
contractions of \reef{cont}. The spin factors of some diagrams may
vanish, implying that the diagram makes no contribution to the
amplitude.

In the previous section, we argued that amplitudes obtained from
the NMHV generating function in $\cn = 4$ SYM theory satisfy the
SUSY \wis. The same argument applies to \sg, so it is clear that
the amplitudes obtained from \reef{n8gen} satisfy the \wis\ of $\cn
= 8$ \sg\ if the MHV-vertex expansion is valid.

The validity of the generating function relies on the vanishing of
the  shifted amplitudes for large $z$. While there is evidence
that such ``good'' shifts can always be found for NMHV amplitudes
in $\cn = 4$ SYM theory, we have found explicit counter-examples
in $\cn =8$ \sg . We will discuss large $z$ behavior in
section \ref{s:gravz},  including several examples and the lessons
they teach us.

\subsubsection{Factorization}
In section \ref{s:mhvGF}
we discussed the factorization of the MHV generating
function for supergravity.  Factorization ensures that MHV
amplitudes are compatible with the operator map \reef{84map} and
that all symmetries are consistently implemented.  A similar
factorization with similar consequences holds for the generating
function of each MHV-vertex diagram in the NMHV sector.   We
observe that the Grassmann terms in \reef{n8gen} factor into a
product of two factors of the analogous terms for gauge theory in
\reef{n4gen}, one each for the $L$ and $R$ gauge theory factors in the
map \reef{map}.  Thus the \sg\ generating function for each
diagram can be rewritten as
\bea
\lab{n8gf}
\tilde{\O}_I &=&
\frac{M_{n_1}(l+1,..,\hat{m}_1,..,k,..,-\hat{P}_I)}{\<m_1\,\hat{P}_I\>^8}\frac{1}{s_I}
  \frac{M_{n_2}(\hat{P}_{I},k+1,..,\hat{m}_2,..,\hat{m}_3,..,l)}{\<m_2\,m_3\>^8}\\ \nonumber
&&~~~~~~~~~~~ \times
 \bigg(
\d^{(8)}\Big(\sum_{i=1}^{n}|i\>\h_{ia}\Big)
  \prod_{b=1}^4\sum_{j=l+1}^{k}\<\hat{P}_I \,j\>\h_{jb}
   \bigg)\,
\times\, \bigg(\d^{(8)}\Big(\sum_{i=1}^{n}|i\>\h_{ir}\Big)
  \prod_{c=5}^8\sum_{j=l+1}^{k}\<\hat{P}_I \,j\>\h_{jc} \bigg)
  \,.
\eea
We see that the spin factors for  supergravity  amplitudes
are products of spin factors for the appropriate gauge theory
amplitudes. This means that $SU(8)$ and SUSY $\tQ_A$ Ward
identities, which hold separately for each term in the MHV-vertex
expansion, are satisfied on the gauge theory side of the map
\reef{map}.  The same conclusion holds for SUSY $Q^A$ Ward
identities after summation of all contributing diagrams, provided that the sum is independent of  $|X]$.


\subsubsection{Large $z$ behavior of NMHV amplitudes in $\cn =8$ \sg}
\lab{s:gravz}

As in gauge theory, the shifted tree amplitudes of $\cn =8$ \sg\
behave as $z^{-\D}$ for large $z$, and the validity of the
generating function requires $\D>0$. The arguments that gave
evidence for this in gauge theory do not carry over to \sg.
Indeed, we will present
explicit counter-examples, namely NMHV amplitudes of $\cn =8$ \sg\
for which no 3-line shift gives $\D>0$.  As in section \ref{s:symz} we begin the
discussion by determining the large $z$ behavior of  typical diagrams
in the MHV-vertex expansion. Each diagram is the product of the result for $n$ external gravitons times a spin factor.   The general discussion will tell us what to
expect at large $z$, and the actual behavior will then be illustrated in
several examples.

\subsubsection*{General discussion}
Our first task is to ascertain the large $z$ asymptotics of a typical
diagram for the $6$-graviton NMHV amplitude $M_6(1^-,2^-, 3^-,4^+,5^+,6^+)$.\footnote{$n$-point amplitudes with $n>6$ are briefly discussed in sec.~\ref{s:npt}.}
The formulas \reef{mone} and \reef{mtwo} contain most of the
information needed to extract the power of $z$ obtained from a
further scaling of $|1],|2],|3]$ in a generic direction $|Y]$ in
spinor space. However 2-particle pole diagrams, which contain the
factor  $M_3$, must be examined separately.  It is not difficult
to obtain the following information about the large $z$ behavior:
\begin{enumerate}
\item[i.] 2-particle pole
diagrams for external gravitons with $--$ helicity vanish at the
rate $1/z^7$.
\item[ii.] 2-particle poles with
graviton helicity $-+$ vanish more slowly, namely at the rate
$1/z^5$.
\item[iii.] 3-particle pole diagrams, necessarily with
$--+$ and $-++$ helicities  in each sub-amplitude, vanish as $1/z^6$.
\end{enumerate}
These estimates apply to each individual diagram. It is possible that
there are cancellations among diagrams, so that the full amplitudes actually fall off faster.

The MHV-vertex decomposition of $M_6(1^-,2^-, 3^-,4^+,5^+,6^+)$ contains 21 non-vanishing diagrams. Numerical results show that
the sum of these diagrams is independent of $|X]$ and vanishes as
$1/z^6$ upon a further shift of lines 123. One can see analytically that
there is a cancellation among the nine 2-particle pole diagrams with
$-+$ helicities. We have also checked that the MHV-vertex method
and the KLT formula produce the same result.

For general external states, the MHV-vertex expansion expresses
the amplitude as a sum of $n$-graviton diagrams multiplied by spin
factors.  See \reef{diagrav} and \reef{sfsg}.  The spin factors
are readily computed for any given process, but it is useful to
have general estimates of their growth rate at large $z$ as an
indication of the behavior of the full amplitude. We consider the
$z$-dependence of the spin factors for a shift of 3 chosen lines
labelled $m_1,\,m_2,\,m_3$. The key parameter that determines the
large $z$ growth rate is the number of $SU(8)$ indices which
appear on all three shifted lines. We let $n_{\rm com}$ denote the
number of common indices.

For the spin factors of $6$-point amplitudes we can prove the following:
\begin{enumerate}
\item[A.]
For diagrams with a 2-particle pole in the $m_2m_3$ channel,  the
product of spin factors grows no faster  than  $z^{8-n_{\rm
com}}$.
\item[B.] For any $3$-particle pole diagram, the maximum growth
rate of the product of spin factors is
   also $z^{8-n_{\rm com}}$.
\item[C.] For diagrams with a 2-particle pole in a channel with one
shifted line, say $m_1$ and
    one unshifted line $a$, the product of spin factors grows at the rate  $z^{r_a}$, where     $r_a$ is the $\h$-count of particle $a$ (defined as the order of the corresponding Grassmann derivative).
\end{enumerate}

The proof of A and B is quite simple.  The numerator of the
product of spin factors in \reef{sfsg} contains a product of 8
brackets  $\<i\,P_I\>$ where $i$ denotes any of the 6 external
lines.  All products except $\<m_1\,P_I\>$ grow linearly with $z$
after a shift. This bracket occurs as $\<m_1\,P\>^{\n_1}$ in the
product of spin factors, so the growth rate of that product is
$z^{8 - \n_1}$.   We will show that $\n_1 \ge n_{\rm com}$ which
will  prove the bound on growth rate stated above.  To establish
this inequality we refer to figure \ref{fig:nMHVgenf}.  Lines $m_2$
and $m_3$ have at least $n_{\rm com}$ $SU(8)$ indices in common.
Therefore these indices cannot appear at the right end of the
internal line. The reason is that each sub-amplitude must
be an $SU(8)$ singlet, so that the 16 indices either one contains must
comprise 8 {\it distinct matched pairs}.  The common indices must
then appear on the left end of the internal line, and at least
$n_{\rm com}$ of them are shared with line $m_1$.  Thus $\n_1 \ge
n_{\rm com}$ and the proof is finished.   (Note that the maximum growth rate $z^{8-n_{\rm com}}$ is actually valid for the spin factors of all diagrams of any $n$-point NMHV amplitude.)

The proof of C follows from the CFT analogy (see section \reef{s:sfcft}) which gives
$D_{m_1} D_a D_I\, \d^{16}(L) \propto \< m_1\, P_I \>^{8-r_a}$. This leaves
 $8-(8-r_a)=r_a$ powers of $\< \cdot P_I\>$ that shift in the product of spin
 factors for the two sub-amplitudes.  (This argument also applies to $n$-point amplitudes).

 The information on the growth of spin factors can now be combined with
 the estimates of the 6-graviton prefactors to give the asymptotic
 growth rates of each type of diagram.
\begin{enumerate}
\item[i.] 2-particle pole diagrams with two shifted lines have prefactors which fall off as
 $1/z^7$. After multiplying by the (worst case) rate $z^{8-n_\mathrm{com}}$ for the spin
 factor we  see that the maximum growth rate of the diagram is $O(1)$.
 \item[ii.] A 2-particle pole diagrams with one shifted and one unshifted line behaves as
 $z^{r_a-5}$  The growth rate is no worse than $O(1)$ unless $r_a \ge 6$.
  We can always choose to shift three lines with $r_{m_i} \ge r_a$.
 Since the total $\h$-count of an NMHV amplitude
 is 24,  the case  $r_a=7$ is then eliminated.
 The only process with $r_a=6$ we need concern ourselves with has partition $6+6+6+6$.
For this four graviphoton amplitude, it turns out that the potential linear divergences cancel by the same mechanism that the pure graviton amplitude's leading $1/z^5$ terms cancels down to $1/z^6$.

Thus we never encounter worse than $O(1)$ asymptotics from the 2-particle pole diagrams.
\item[iii.] For 3-particle pole diagrams  the maximum growth rate is $z^{2-n_\mathrm{com}}$, so we may encounter linear growth and $O(1)$ behavior.
\end{enumerate}

The bound $z^{8-n_{\rm com}}$ on the growth of spin factors suggests that
the optimal behavior at large $z$ can be obtained by shifting 3 lines with the
largest value of $n_{{\rm com}}$.  Indeed,
for $M_6(1^-2^-3^-4^+5^+6^+)$
the conventional $-\,-\,-$ shift realizes the maximal
value   $n_{{\rm com}}=8$.  In our numerical exploration of large $z$ behavior of non-optimal shifts were also included and were instructive.  Since the total $\h$-count of an NMHV amplitude is 24 and $SU(8)$ symmetry requires that every index $1,2,\dots,8$ must appear exactly 3 times among the 6 lines, it is always possible to choose a shift with $n_{{\rm com}}=1$.  One might suspect that the recursion relation would fail for a shift of lines $m_1$, $m_2$, $m_3$ with $n_{{\rm com}}=0$ because a physical pole in the channel $m_1m_2m_3$ would be omitted. However, in an example below we
find a valid recursion relation for such a shift.

 Let us now turn to the results of our extensive numerical study of NMHV amplitudes in $\cn=8$ \sg .  In turn we will discuss ``good'' amplitudes for which the sum of diagrams
 vanishes as $z \to \infty$,  ``bad'' amplitudes in which $O(1)$ behavior occurs in the sum, and ``very bad'' amplitudes with linear growth.    We will argue that ``bad" amplitudes still have valid recursion relations, but ``very bad" amplitudes do not.

\subsubsection*{``Good" amplitudes.}

There are numerous examples of  NMHV amplitudes in $\cn =8$ \sg\
for which there are 3-line shifts with $\D>0$ and hence valid
MHV-vertex expansion. The generating function works for these examples
just as it did in gauge theory. However we will briefly discuss some examples
which reveal interesting regularities.

Consider the amplitudes
$\<b^-\,b^-\,f_1^-\,f^1_+\,b_+\,b_+\>$,
$\<b^-\,b^-\,b_{12}^-\,b^{12}_+\,b_+\,b_+\>$,
$\<b^-\,b^-\,f_{123}^-\,f^{123}_+\,b_+\,b_+\>$,
and $\<b^-\,b^-\,b_{1234}\,b^{1234}\,b_+\,b_+\>$, in which  a
graviton pair is replaced, in turn, by  a pair of gravitini,
graviphotons, graviphotini, and scalars. All amplitudes have 18
contributing diagrams whose sum is independent of $|X]$. Let's
label each particle by its spin $s$, with $s=2$ for the graviton,
$s=3/2$ for the gravitino, etc.  It is interesting to observe a
simple pattern for the spin factors of each diagram as $s$
decreases. For any given  diagram  $\mathcal{A}_{n,I}$ let us simply
denote by $\mathcal{S}_I$ its spin factor in the gravitino amplitude. For
other cases the same diagram has the spin factor $\mathcal{S}_I^{(4-2s)}$.
Furthermore the amplitude with the spin $s$ pair vanishes as
$1/z^{2s+2}$ as $z \to \infty$ under a further 123 shift. The
pattern in the NMHV sector is similar to what occurred in the
analogous set of MHV amplitudes in \reef{n8wi}.

Another example of a ``good'' amplitude is the
scalar amplitude
\bea
  \lab{good6sc}
  \big\< b^{1234}\, b^{1234}\, b^{1234}\,
    b^{5678}\,  b^{5678}\, b^{5678}\big\>\, ,
\eea
whose external states are three identical scalars $b^{1234}$ and their conjugates $b^{5678}$.
A KLT calculation shows that the amplitude \reef{good6sc} has a $1/z^2$ falloff for large $z$ under shifts that do not involve conjugate scalars,\footnote{If the lines shifted involve a scalar and its  conjugate then the shifted amplitude grows as $z^2$ for large $z$.} such as the 123-shift, and the resulting recursion sum of 18 MHV-vertex diagrams therefore gives the correct amplitude and is indeed, as shown in our numerical work, independent of $|X]$.


\subsubsection*{``Bad" amplitudes.}

It was an unwelcome discovery that by a mere change of the
$SU(8)$ labels in the six scalar amplitude, one finds an amplitude
\bea
  \lab{bad6sc}
  \big\< b^{1234}\, b^{1358}\, b^{1278}\,
    b^{5678}\,  b^{2467}\, b^{3456}\big\> \, ,
\eea
for which the sum of MHV-vertex diagrams depends on $|X]$.  This result appears
to be unacceptable, so we proceed to study it further, specifically by an independent
construction using the KLT formula. (The method is explained in more detail below.)
The KLT result is valid for general complex momenta and we can explore the large $z$
behavior by making various 3-line shifts numerically.
The amplitude consists of three pairs of conjugate scalars
(e.g.~$b^{1234}$ and $b^{5678}$) and has only two types of
3-line shifts, namely shifts involving no conjugate pairs  --- such as a 123-shift --- and shifts that involve a conjugate pair of scalars (e.g.~a 124-shift). The former give $O(1)$ for large $z$ and the latter $O(z^2)$ (since the 3-lines  do not share a common index, $n_\mathrm{com}=0$).
Thus there are no 3-line shifts of the six
scalar amplitude \reef{bad6sc} that give large $z$ falloffs faster then $O(1)$, and we therefore categorize it as a ``bad'' amplitude.

At first sight, a MHV-vertex decomposition based on 3-line recursion relations seems to be impossible for ``bad'' amplitudes. However, 3-line shifts with $O(1)$ asymptotics can still be used to derive a recursion formula if the reference
spinor $|X]$ is suitably chosen.  To explain our approach to ``bad" amplitudes
we start with the  example of the pure
gluon amplitude in gauge theory:

\begin{quote}
{\bf Example: $O(1)$ shifts in gauge theory and the role of $|X]$}\\
The gluon amplitude $A_6(1^-, 2^-,3^-,4^+,5^+,6^+)$ is well known
and  we have already seen that it can be calculated with the
MHV-vertex method associated with a shift of the three negative
helicity lines. But it is illustrative to consider the large $z$
behavior of other 3-line shifts of the amplitude. This can be done
numerically and we find
\bea
  \begin{array}{rcl}
  \<\hat{-}\hat{-}-\hat{+}++\> &\sim& \frac{1}{z}\,  , \\[2mm]
  \<\hat{-}-\hat{-}+\hat{+}+\> &\sim& \frac{1}{z}\,  , \\[2mm]
  \<-\hat{-}- \hat{+}+\hat{+}\>  &\sim& \frac{1}{z} \, ,
  \end{array}
  \hspace{5mm}
  \begin{array}{rcl}
  \<- - - \hat{+}\hat{+}\hat{+}\>  &\sim& O(1) \,  ,\\[2mm]
  \<\hat{-}\hat{-} - + + \hat{+}\>  &\sim& O(1) \,  , \\[2mm]
  \< - - \hat{-} \hat{+} \hat{+} +\>  &\sim& O(1)\,   .
  \end{array}
\eea Cauchy's theorem gives valid recursion relations for the
first three  types of shifts for which the amplitude vanishes as $z \to \infty$. In
all three cases, the sum of MHV-vertex diagrams is independent of $|X]$.

The shifts which give $O(1)$ asymptotics also give
rise  to valid recursion relations, but only for the special
values of $|X]$ for which the $O(1)$-contribution vanishes. This
condition is a polynomial equation in $|X]$, and  for each of the three
cases above, we have found the roots $|X_q]$  numerically.
The asymptotic $O(1)$ term vanishes at each root, so the conditions needed
to use Cauchy's theorem are satisfied.  Indeed we have verified
that precisely for these values of $|X]$ the  recursion sum of
MHV-vertex
diagrams agrees with the result from the $-\,-\,-$ shift and thus gives
the correct 6-gluon amplitude.
For any other values of $|X]$ the recursion sum is invalid;
the contribution from ``infinity'' in Cauchy's theorem is missing.

This result is particularly striking for the last two shifts which each involve 3 lines
with $n_{\rm com} =0.$  Thus there is no diagram which contains the gluon pole
in the channel of the shift. Nevertheless this pole is reproduced by the other
diagrams at the special values $|X_q]$.
\end{quote}

We have found that the situation is similar for all ``bad" NMHV 6-point amplitudes
in $\cn=8$ \sg.  This leads to a modified criterion for the
validity of the MHV-vertex method which we now summarize:
\begin{itemize}
\item[---] If a shifted amplitude goes to zero for any $|X]$, then
the   sum of MHV-vertex diagrams resulting from that 3-line shift
must be independent of $|X]$ and  gives a correct expression for the
amplitude.
\item[---] If a shifted amplitude goes to zero only for
specific values $|X_q]$ of spinors $|X]$, then the  sum of MHV-vertex
diagrams resulting from that 3-line shift generally depends on
$|X]$, but the corresponding MHV-vertex method gives the correct result
for the amplitude for the values $|X_q]$ (and only these values). It requires
an independent evaluation of the amplitude to find the  $|X_q]$.
\end{itemize}

With these rules we can apply the generating function to the large
number of the NMHV amplitudes in the $\cn =8$ theory whose best
shifts give $O(1)$ for large $z$.
We have tested this procedure in several examples, including the
six scalar  amplitude \reef{bad6sc}. We first calculate the
amplitude using the KLT formula,
\bea
 \nonumber
 && \big\<
  a(1) \, a(2)\, a(3) \, a(4) \, a(5)\, a(6)
  \big\>
  ~=~
  \bigg\{
  s_{34}\, s_{16}
  \big\< A(1) \, A(2)\, A(3) \, A(4) \, A(5) \, A(6)
  \big\> \\
  \lab{KLT6}
  && \hspace{2.5cm}
  \times \Big[
    s_{15}\,
    \big\< \tilde{A}(1) \,\tilde{A}(3)\, \tilde{A}(4)
           \, \tilde{A}(2) \, \tilde{A}(6) \, \tilde{A}(5) \big\> \\  \nonumber
 && \hspace{3.1cm} +
   (s_{15}+s_{56}) \,
    \big\< \tilde{A}(1) \,\tilde{A}(3)\, \tilde{A}(4)
           \, \tilde{A}(2) \, \tilde{A}(5) \, \tilde{A}(6) \big\>
  \Big]
  \bigg\}+ \mathcal{P}(4,5,6) \, .
\eea
The $a$'s can be annihilation operators for any states of the $\cn =
8$ theory, and $A$  and $\tilde{A}$ denote the decomposition of the
$a$ operators under the map \reef{84map}. We use the NMHV generating
function of section \ref{s:GFsym} to calculate each gauge theory amplitude.

Different ways to split $SU(8) \to SU(4) \times SU(4)$ result in different decompositions $a = A \otimes \tilde{A}$, but the RHS of \reef{KLT6} must give the same result for the supergravity amplitude. Calculating the supergravity amplitudes from different KLT decompositions  provides a useful check on the correctness of the result.

Next we perform a 3-line shift of the supergravity amplitude \reef{KLT6}
with an arbitrary reference spinor $|X]$. The $O(1)$ term for large $z$
is a function of $|X]$, $f=f(|X])$. Setting $f(|X])=0$ gives a
polynomial equation in $|X]$, and its roots $|X_q]$ make the
recursion relation valid. For the six scalar amplitude
\reef{bad6sc} there are six solutions $|X_q]$. We then compute the MHV
vertex expansion for the same shift, and evaluate it
for $|X]=|X_q]$.  The sum of diagrams always agrees with the
KLT result and thus  confirms the validity of the procedure.

The six scalar amplitude \reef{bad6sc} is not the only amplitude
whose best 3-line shift gives $O(1)$ for large $z$. We list here a selection which illustrates that the $O(1)$ behavior occurs in a variety of different cases, not necessarily involving scalars.
\bea
 \nonumber
 &&
 \big\< b^-_{78} \, b^-_{56} \, b^{1567} \, b^{2578} \, b_+^{36}\,  b_+^{48}\big\> \\[1mm]
  \nonumber
 &&
 \big\< b^- \, b^{1234} \, b^{1567} \, f_+^{258} \, f_+^{368}\,  b_+^{47}\big\> \\[1mm]
  \nonumber
  &&
 \big\< f^-_{678}\, b^{1358}\,  b^{1278}\,
 b^{5678}\,  b^{2467}\, f_+^{346}\big\> \\[1mm]
 \nonumber
 &&
 \big\< f^-_{678} \, f^-_{458} \, f^-_{235} \, f_+^{268} \, f_+^{578}\,  f_+^{345}\big\> \\[1mm]
 &&
 \big\< b^-_{12} \, b^-_{34} \, b^-_{56} \, b^-_{78} \, b_+\,  b_+\big\>
 \lab{1exs}
\eea
We have calculated each of these amplitudes using the KLT formula \reef{KLT6},
determined  $|X^*]$ such that the asymptotic $O(1)$ vanished, and verified numerically that the generating
function gives the correct values for $|X]=|X^*]$.\footnote{The
order of the polynomial $f(|X])$ typically varies in the range $2$-$8$.}

We have attempted to test how many of the 151 partitions of distinct 6-point NMHV processes contain ``bad'' amplitudes.
A preliminary count gives 73; this  based on a scan of different $SU(8)$ index structures and tests of large $z$ asymptotics of the diagrams of the MHV-vertex expansion associated with all possible 3-line shifts.

\subsubsection*{``Very bad" amplitudes.}
We finally turn our attention to the ``very bad'' amplitudes. With the help of Mathematica we have analyzed which 6-point NMHV amplitudes have the property that no 3-line shifts give better behavior than $O(z)$ for large $z$. We find that there are only two such amplitudes, namely
\bea
\lab{vbad}
\big\< f^-_{678}\, b^{2568}\,  b^{3478}\,  b^{4578}\,  b^{1367}\,  f_+^{126} \big\>\, , &~~~&
\big\< f^-_{678} \,  f^-_{458}\,  f^-_{238} \,  b^{2468}\,  b^{3578}\,  f_+^8 \big\>  \, .
\eea
We have computed both amplitudes using the KLT relations and confirmed that they grow linearly in $z$ for large $z$  under \emph{any} one of the twenty possible different 3-line shifts.
We have also checked numerically that there are no solutions $|X]$
that simultaneously eliminate the $O(z)$ and $O(1)$ terms.
This means that the generating
function cannot be used to compute these two amplitudes.
However, the two amplitudes \reef{vbad} can be determined from  supersymmetric \wis\ which relate a ``very bad'' amplitude to others that can be computed from recursion relations.

One may wonder how the \wis\ can accommodate amplitudes with different large $z$
behaviors; this is the subject of the next section.


\subsubsection{Supersymmetric \wis\ and large $z$}

As discussed in section \ref{s:swi}, SUSY  \wis\ in the NMHV sector always relate
sets of four amplitudes.  To see this explicitly in the $\cn = 4$ and
$\cn = 8$ theories, one must choose specific values of the
flavor indices.
The generic form of any NMHV SUSY \wi\ is therefore
\bea
  \lab{Awi}
  0 &=& \< \e\, i_1 \> \, A_1+ \< \e\, i_2 \> \, A_2
  + \< \e\, i_3 \> \, A_3 + \< \e\, i_4 \> \, A_4 \, ,
\eea
where $i_k =  1,2,3,4,5$, or 6. There are a variety of possibilities.
A given \wi\ can involve only ``good'' amplitudes, or both ``good'' and
``bad'', only ``bad'', etc. This terminology refers to asymptotic
behavior under an optimally chosen shift of each amplitude. To investigate
the large $z$ asymptotics of the entire \wi, one must use
the \emph{same}  shift to analytically continue all four amplitudes.

Under any  such common shift, we can assume that for large $z$, the
amplitudes behave as $A_i \sim z^{k_i}$ for large $z$. Without loss of
generality, let us assume that $k_1 \ge k_2 \ge k_3 \ge k_4$. Start by setting
$|\e\> = |x_1\>$ in \reef{Awi}. Then we must have $k_2=k_3$, because
either $k_2=k_3=k_4$, or --- if $k_3>k_4$ --- then $k_2=k_3$, so
that the leading powers $z^{k_2}$ and $z^{k_3}$ cancel down to
$z^{k_4}$. Likewise, we determine from $|\e\> = |i_3\>$ that $k_1 =
k_2$. We conclude that the SUSY \wi\ \reef{Awi} restricts the
powers of the leading $z$-behaviors to be $k_1 = k_2 = k_3 \ge
k_4$. Thus for each shift, the four amplitudes in the \wi\ can at
most involve two different large $z$ powers, and the slowest
falloff must occur at least thrice.

Let's see how this works in practice.
Consider the \wi\ \bea
  \nonumber
  0 &=&
  \big\< [\tQ_6 , \, b^-_{78}\, b^{2568}\,  b^{3478}\,
    b^{4578}\,  b^{1367}\,  f_+^{126} ] \big\>\,  \\[2mm]
    \nonumber
  &=&
  \< \e\, 1 \> \,
  \big\< f^-_{678}\, b^{2568}\,  b^{3478}\,  b^{4578}\,
    b^{1367}\,  f_+^{126} \big\>
  +
  \< \e\, 2 \> \,
  \big\<b^-_{78}\, f_+^{258}\,  b^{3478}\,
    b^{4578}\,  b^{1367}\,  f_+^{126}\big\>\\[1mm]
  && +
  \< \e\, 5 \> \,
  \big\<b^-_{78}\, b^{2568}\,  b^{3478}\,
    b^{4578}\,  f_+^{137}\,  f_+^{126}\big\>
   +
   \< \e\, 6 \> \,
  \big\<b^-_{78}\, b^{2568}\,  b^{3478}\,
    b^{4578}\,  b^{1367}\,  b_+^{12}\big\>
    \, .
    \lab{zWI}
\eea We recognize the first amplitude of \reef{zWI} as one of the
``very bad'' amplitudes \reef{vbad}. The three other amplitudes in
the \wi\ \reef{zWI} turn out to be just ``bad''. Under any 3-line
shift, the first amplitude give $O(z)$ for large $z$. Depending on
the choice of which three lines are shifted, the \wi\ \reef{zWI}
accommodates three different combinations of large $z$ behaviors:
\begin{itemize}
\item All four amplitudes grow as $O(z)$ (e.g.~134-shift).
\item One
amplitude gives $O(1)$ and the three others give $O(z)$. This
happens  only in 5 cases: the 156-shift gives $O(1)$ for the
second amplitude, the 126-shift gives  $O(1)$ for the third
amplitude and the 124-, 135- and 125-shifts give $O(1)$ for the
fourth amplitude.
\item The three ``bad'' amplitudes grow as $z^2$
for large $z$ while (as always) the ``very bad'' amplitude grows as $z$. This occurs when the three shifted lines involve three states in the ``bad'' amplitudes which do not share a common index.
\end{itemize}
We have verified this in explicit numerical calculations, with
amplitudes computed by the KLT formula.  Numerical tests included \wis\ for both the $\tQ_A$ and $Q^A$ operators,
The pattern of large of  $z$ asymptotics found here is in complete agreement with  the general analysis.


\subsubsection{2-line shifts vs.~3-line shifts}

We would like to point out some differences --- and relationships
--- between  the 2- and 3-line shifts. First of all, the 3-line
shifts involve the arbitrary reference spinor $|X]$. The fact that
Cauchy's theorem only requires $M_n(z) \to 0$ for $z \to \infty$
for \emph{some} $|X]$, allow us to use the generating function and
the MHV-vertex expansion even for amplitudes whose best shifts go
as $O(1)$ for large $z$; the $|X]$ must be chosen such that the
$O(1)$ term vanishes. This freedom is clearly not available in the
2-line recursion relations.

An example illustrating the differences between the 2- and 3-line shifts
is the ``bad'' six scalar amplitude \reef{bad6sc}. There are \emph{no} valid
2-line shifts for this amplitude; if a pair of conjugate scalars
is shifted, the amplitude grows as $z^2$ for large $z$, while if a
pair of non-conjugate scalars are shifted, then the large $z$
behavior is $O(1)$. On the other hand, the 123-shift recursion
relations give a valid MHV-vertex decomposition of the amplitude
for the six special values of $|X]$ for which the $O(1)$-term of
the large 123-shift vanishes.

It was pointed out in \cite{greatdane} that the 3-line
shifts  can be built from three successive 2-line shifts, viz. \bea
  \nonumber
  |\hat{1} ]  = |1] + z \, \< 2 3 \> | X ] \, ,
  &&
  |\hat{X} \>  = |X\> - z \, \< 2 3 \> | 1\> \, , \\
  \lab{3from2}
  |\hat{2} ]  = |2] + z \, \< 3 1 \> | X ] \, ,
  &&
  |\hat{X} \>  = |X\> - z \, \< 3 1 \> | 2\> \, , \\
  \nonumber
  |\hat{3} ]  = |3] + z \, \< 1 2 \> | X ] \, ,
  &&
  |\hat{X} \>  = |X\> - z \, \< 1 2 \> | 3\> \, .
\eea
The spinor $|X]$ can be chosen as the holomorphic spinor of lines
  4, 5, or 6, since the cumulative shift of
$|X\>$ cancels by the Schouten identity.

The two amplitudes in \reef{vbad} are problematic because they
cannot be  computed with the MHV-vertex method for any $|X]$.
However, there do exist good 2-line shifts for both amplitudes,
e.g.~the $[1,6\>$-shift works for both. (The resulting 2-line
recursion relations will involve anti-MHV vertices.)

The existence of (three) valid 2-line shifts  does \emph{not}
imply that  the combined 3-line shift \reef{3from2} is valid. An
explicit example of this is provided by the second amplitude
$\big\< f^-_{678} \,  f^-_{458}\,  f^-_{238} \,  b^{2468}\,
b^{3578}\,  f_+^8 \big\>$  of \reef{vbad}, for which all of the
twenty possible 3-line shifts give $O(z)$ for large $z$. Under
each of the 2-line shifts  $[1,6\>$, $[2,6\>$, $[3,6\>$ this
amplitude goes as $1/z$ for large $z$, but the combined shift is a
3-line shift with $|X] = |6]$, and we know that the amplitude will
not go to zero for large $z$ under such a shift. In fact, for the
123-shift with $|X] = |6]$ the amplitude goes to a constant. The
reason that one finds $O(1)$ rather than $O(z)$ is that $|X]=|6]$
happens to be one of the solutions to setting $O(z)=0$ for the
123-shift.

We will see next that there are problems with 3-line shifts even for pure graviton NMHV amplitudes for sufficiently large $n$.

\subsubsection{$n$-point NMHV graviton amplitudes with $n>6$}
\lab{s:npt}

One can estimate the large $z$ falloff of the $n$-point graviton NMHV amplitude from the large $z$ behavior of the individual diagrams in the MHV-vertex expansion \reef{rec3}. The large $z$ asymptotics of the two MHV-vertices can be extracted from the BGK formula, as presented in \reef{mone} and \reef{mtwo}.
For $n=6$, the slowest falloff comes from the $+-$ 2-particle pole diagrams and is $1/z^5$. However, due to a cancellation among these diagrams, the falloff of the full amplitude is $1/z^6$.

In \reef{mtwo} the $\beta_s$ factors shift, and for each extra external leg, one therefore gets diagrams which falloff slower by one power of $z$. Provided that the leading large $z$ falloff cancels for $n>6$ as it does for $n=6$, one is lead to expect that under a 123-shift, the NMHV graviton amplitudes behaves as
\bea
  M_n(\hat{1}^-,\hat{2}^-,\hat{3}^-,4^+,5^+,\dots,n^+)
   \sim \frac{1}{z^{12-n}}
\eea
for large $z$. We have verified this behavior in explicit numerical work for $n=5,\dots,11$. This is done by calculating the MHV-vertex expansion for each $n$, testing numerically that the sum of $3(2^{n-3}-1)$ diagrams is independent of $|X]$. Then another 123-shift with an arbitrary reference spinor is performed on the result for the amplitude, and the leading $z$ falloff is read off from a series expansion as $z\to\infty$. As an extra check we have also calculated $M_n$ for $n=5,\dots,9$ with the recursion relations associated with the 2-line shift $[2,1\>$ and numerically tested that the result agrees with the MHV-vertex expansion.

This means that we must expect the MHV-vertex decomposition to break down for $n>12$.\footnote{The borderline case of $n=12$ may be handled by solving the $O(1)=0$ condition. Other shifts $--+$, $-++$ and $+++$ give asymptotic $z^{n-4}$ behavior and thus never valid recursion relations.} It also means that as the number of external legs grow, the spin factors arising from external states other than gravitons will come to dominate the gravity prefactors for large $z$, and so there will be more bad amplitudes, more very bad amplitudes and also very very bad amplitudes. This strongly restricts the validity of the NMHV generating function for higher-point amplitudes of the $\cn =8$ theory.

As a final point, it is worth noting that the $n$-point NMHV graviton amplitudes continue to be calculable from recursion relations based on 2-line shifts. We have indeed checked numerically for $n=5,\dots,10$ that the two line shifts $[-,-\>$, $[-,+\>$, $[+,+\>$-shifts give $1/z^2$ for large $z$, while a $[+,-\>$-shift gives $z^6$. This is expected from the general analysis of \cite{nima}.

Cachazo et al provided in \cite{cabg} the first proof of the validity
of the 2-line recursion relations for graviton amplitudes.
Our numerical work confirms their results for 2-line shifts,
but it disagrees with their statement that the 3-line shift is valid
because it can be obtained by successive 2-line shifts as in \reef{3from2}.

\setcounter{equation}{0}
\section{Discussion and open problems}
\lab{s:disc}

In this paper we have studied $n$-point  tree amplitudes with
general external particles of $\cn =4$ SYM theory and $\cn = 8$
\sg. We have elucidated properties of the generating function
proposed for MHV amplitudes in the gauge theory in \cite{nair} and
extended to the NMHV level in \cite{ggk}, and we have developed
similar generating functions for \sg . The generating function is
a  simple function of auxiliary Grassmann variables. There is a
1:1 correspondence between particles of the $\cn =4$ and $\cn=8$
theories and Grassmann derivatives, and any desired amplitude is
obtained by applying the appropriate product of derivatives to the
generating function.

Any $n$-point MHV amplitude is the
product of the ``top'' $n$-gluon or $n$-graviton amplitude times a
``spin factor'' depending on the external particles.
The spin factor of every MHV process in supergravity is a
homogeneous function of weight 16 (weight 8 in gauge theory) of
the spinors $|i\>$ associated with the external particles. We have
found a curious and rather perfect analogy between the structure
of the spin factors and the structure of holomorphic correlation
functions in conformal field theory on the complex plane.

The MHV generating function neatly encodes the full set of spin factors
and it allows one to count the number of independent MHV
processes. It also clarifies how $\cn = 8$
supersymmetry and $SU(8)$ global symmetry of \sg\ are implemented
in quadratic relations between gauge theory and supergravity
amplitudes such as the KLT formulas and the MHV-level formula of
\cite{ef}. It turns out that, for each permutation in those
formulas, the \sg\ generating function factors into the product of
two gauge theory generating functions.

At the NMHV level the situation is similar, but there
is a different generating function for each diagram in the
MHV-vertex expansion of the amplitude.
Our  application to \sg\ requires clarification of an important feature of past work.
In the MHV-vertex construction, the contribution of each diagram
depends on an arbitrary reference spinor $|X]$. In past work it
has always been assumed that
the full amplitude obtained by summing all diagrams were
independent of $|X]$. It was proven in \cite{csw} that this
property holds for $n$-gluon NMHV amplitudes, but the argument
used does not readily extend to other situations.

The  recursion
relations obtained from the 3-line shift of \cite{risager} provide
a precise framework for the MHV-vertex method, and they clarify
the role and origin of the reference spinor $|X]$ which determines
the shift. The diagrammatic expansion of an amplitude $M$ is valid
if the shifted amplitude vanishes at infinity, $M(z,|X])\,\to\,0$
as $z\to \infty$. If this condition is satisfied for all $|X]$,
then the derivation of the recursion relation from Cauchy's
theorem ensures that the sum of MHV-vertex diagrams will be
independent of $|X]$. If it is not satisfied, the expansion will
not produce the right amplitude because the contribution from
infinity required by Cauchy's theorem is neglected. Then the sum
of diagrams may well depend on $|X]$.

The situation can be made sharper. The 3-line recursion relation
is  valid if $M(z,|X])\,\to\,0$ as $z\to \infty$ for {\it a finite
set of values} of $|X]$, not necessarily {\it all values}. The sum
of diagrams will produce the physical amplitude at precisely those
values.

A study of the large  $z$ behavior of the individual MHV-vertex diagrams as well as explicit MHV-vertex constructions of many 6-point NMHV amplitudes, indicate
that the property $M(z,|X])\,\to\,0$ as $z\to \infty$ holds for all $|X]$ for general external states in $\cn  =4$ SYM theory. But our results also show that it fails for many \sg\ amplitudes: we have found explicit examples of 6-point NMHV amplitudes in $\cn = 8$ \sg\ whose best 3-line shifts behave as either $O(1)$ or $O(z)$ for large $z$. We refer to amplitudes with this property as respectively ``bad'' and ``very bad''.

The NMHV generating function is still valid for ``bad'' amplitudes.
The appropriate values of $|X]$ for which the MHV-vertex expansion is justified can be determined  from a supplementary calculation of the ``bad'' amplitude using the KLT formula.
The special values of $|X]$ are roots of a polynomial which precisely
expresses the condition that the large $z$ portion of the Cauchy
contour integral vanishes. The sum of the diagrammatic expansion
is the same at each of these roots and agrees with the value from
KLT. This is a pragmatic test of the validity of our approach.
Because of the algebraic complexity of NMHV amplitudes in \sg ,
our computations are done primarily after input of numerical values
for the spinors $|i\>,\,|i]$ which contain the information on
particle momenta. However, it is clear that the large $z$
polynomial involves only Lorentz invariant spinor brackets
$\<i\,j\>,\,\,[i\,j]$, so that the procedure does not violate
Lorentz invariance. Furthermore the key role of the generating
function, namely that it encodes the spin factors of all NMHV
processes, is preserved.

It appears that the generating function  approach can be used to
simplify the intermediate state helicity sums needed to obtain the
integrands of Feynman loop diagrams from products of tree
amplitudes by the general cutting techniques widely applied in the
work of Bern, Dixon, and Kosower and their collaborators. The MHV
level sums carried out in section \ref{inthelsum} are really simple, and
preliminary calculations of NMHV level sums in the gauge theory
are promising. It may be difficult to implement NMHV sums in
\sg\ because of the problematic large $z$ behavior we have
discussed above. We plan to look at this question in the near
future.

We also found results which pose difficulties for our
approach. Among 6-point NMHV amplitudes in \sg\ there are
two ``very bad'' amplitudes which grow linearly as $z \to \infty$.
Construction from the generating function is invalid for these.
Also, for $n$-graviton amplitudes with $n \ge 5$, an analytic study
indicates that there are individual MHV-vertex diagrams which grow
at the rate $z^{n-11}$.  Subsequent numerical evaluation of the
sum of diagrams shows that there is leading order cancellation and that
full amplitudes grow at the rate $z^{n-12}$.  Thus it appears that
the MHV-vertex expansion will be invalid for $n \ge 12$. It also means that $n$-point amplitudes with general external states can be expected to have worse large $z$ behavior for $n>6$. Since
the large $z$ behavior of tree amplitudes is related,  see
\cite{unexp}, to UV behavior at the loop level, this result may be
an omen of future problems.

An aspect of $\cn =8$ \sg\ we have not yet
discussed is the nonlinearly realized $E(7,7)$ symmetry.
Details of the action of $E(7,7)$
are nicely described in the recent paper \cite{kallosh}. The 70
scalar fields of the theory are Goldstone bosons of the
spontaneous breaking $E(7,7)\,\to\,SU(8)$. We would like to find
the footprint of $E(7,7)$ symmetry in the set of $n$-point tree
amplitudes we have studied. We expected
that $E(7,7)$ would reveal
itself in the limit of vanishing boson 4-momentum, as in the low
energy theorems for soft pion emission obtained long ago by Adler
\cite{adler} and discussed by Coleman \cite{coleman}. In pion
physics, the low energy limit of a single soft pion is generally
non-vanishing and obtained from the sum of Feynman diagrams in
which the soft pion is attached to other external lines. Graphs
with internal attachment vanish at low energy because of the
coupling to the axial current. The soft pion limit is non-zero in
tree approximation for the process $\pi + N \to 2\pi + N$ even in
a version of the linear $\s$-model with gradient coupling
$\bar{N}\gamma^\mu [ \partial_\mu \s + i
\g_5\vec{\tau}\cdot\partial_\mu\vec{\pi}] N$  so that both
$\pi$ and $N$ are massless.

We examined the one-soft-boson limits of our tree amplitudes and
found that the limit always vanishes. This was puzzling because
the Lagrangian has \cite{dwf} cubic vertices in which the
Goldstone bosons couple to two graviphotons and to a
gravitino-graviphotino pair. This leads to diagrams with external
line insertions, but  their soft limit vanishes when all external
particles are on shell. So our results are consistent, but it is still
puzzling why soft boson limits are
trivial in \sg\ but not in pion physics.\footnote{It appears that
the limit of two soft bosons is closer to the situation in pion
physics. There are low energy theorems which reflect the fact that
the equal-time commutator of two $E(7,7)$ coset currents lies in
the compact $SU(8)$ subalgebra.}

Another aspect of our construction, which was only noted in
section \ref{s:gt},  is the tantalizing 1/2 BPS structure of the
generating functions for tree level MHV amplitudes. Introducing
the `standard' Grassman variables of $\cN = 4$ on-shell superspace
$\theta^a_\da$ and $\bar\theta_a^\a$, one can rewrite the MHV
generating function as an integral over only a chiral half of
superspace since \be \delta^{(8)}(\sum_i \eta_a^i
\lambda_{i}^{\da}) = \int d^8\theta \exp(\theta^a_\da\sum_i
\eta_a^i \lambda_{i}^{\da}). \ee One can go a step further and
observe that the ladder of differential operators acting on the
auxiliary $\eta$'s is strikingly reminiscent of the ladder of
`classical' fields generated by the action of (broken)
supersymmetry on a self-dual (instanton) configuration. The 8
spinors $\theta^a_\da$ are the bookkeeper  for the fermionic
zero-modes.  As in supersymmetric instanton calculus, see \eg
\cite{BKR} for a recent review,  for a non-zero result one must to
`soak up' the 8 fermionic zero modes of chiral $\cN = 4$
superspace. This selects exactly the set of 15 types of MHV
related amplitudes. It is tempting to conjecture that the
functional integral representation of MHV amplitudes is
`dominated' by classical self-dual configurations of the gauge
field whose precise form depends on the boundary conditions
dictated by the choice of external momenta \cite{Rosly,Poppitz}.
This  is further supported by the recent results of
\cite{Brink,Kovacs} where the $\cN = 4$ multiplet is packaged into
a scalar light-cone (LC) superfield that only depends on half the
LC superspace $\theta_{LC}$'s. 
For related work in the supertwistor formulation, see for example \cite{skinner}.
Similar considerations may apply to
$\cN =8$ supergravity, where the recent
 LC $\cN = 8$ superspace approach of \cite{BrinkE7} supports the
1/2 BPS structure of MHV amplitudes and sheds some light on the
role of $E(7,7)$.

\section*{Acknowledgments }

We thank Z.~Bern and L.~Dixon for much advice and encouragement, and we acknowledge illuminating conversations  with S.~Adler, 
I.~Antoniadis, E.~Bjerrum-Bohr, L.~Brink, S.~Ferrara,
G.~Gibbons, M.~B.~Green, D.~Kosower, S.~Kovacs, R.~Kallosh, M.~Mangano, E.~Sokatchev,
S.~Stieberger, and P.~Vanhove.

DZF is grateful for financial support from TH-Division  at CERN,
the Stanford Institute for Theoretical Physics, and the Centre
for Theoretical Cosmology of Cambridge University during his sabbatical in 2007-2008.
The research of DZF is also supported by NSF grant PHY-0600465.

The work of MB has been supported in part by the European
Community Human Potential Program under contract
MRTN-CT-2004-512194, by the INTAS grant 03-516346, by MIUR-COFIN
2003-023852, and by NATO PST.CLG.978785. Preliminary results of
this work were presented by MB at the conference ``String
Phenomenology and Dynamical Vacuum Selection'', Liverpool, UK,
March 27-29 2008. MB would like to thank the organizers for the
kind invitation and the stimulating atmosphere.

HE is supported by a Pappalardo Fellowship in Physics at MIT and
in part by the US  Department of Energy through cooperative
research agreement DE-FG0205ER41360


\appendix

\setcounter{equation}{0}
\section{Conventions}
\lab{app:conv}

Our notation for the spinor helicity formalism is largely inspired
by \cite{dixon,sredder}, but we use different conventions which
are summarized in the following.

\subsection{Spinor helicity formalism}
We work with a mostly-plus metric, $\eta_{\m\n} =
\diag(-1,+1,+1,+1)$.  Gamma-matrices are defined as \bea
  \g^\mu =
  \left(
     \begin{array}{cc}
        0 & \sigma^\mu \\
        \bar{\sigma}^\mu & 0
      \end{array}
   \right) \, ,
  ~~~~~
  \{\g^\m , \g^\n \} = 2 \eta^{\m\n} \, ,
  ~~~~~
    \g_5 \equiv i \g^0 \g^1 \g^2 \g^3
  =
  \left(
     \begin{array}{cc}
        -1 & 0 \\
        0 & 1
      \end{array}
   \right) \, ,
   \lab{gammas}
\eea
with $\sigma^\m = (1,\sigma^i)$, $\bar{\sigma}^\m = (-1,\sigma^i)$
and $\sigma^i$ the standard Pauli matrices.\footnote{Note that
$(\bar{\sigma}^\mu)^{\da \b}
  = - \eps^{\b \g} \eps^{\da \dd}
    (\sigma^\mu)_{\g \dd}$. We use $\eps^{12} = \eps_{12} = 1$.}

Positive and negative helicity solutions of the massless Dirac equation,
$\gamma \cdot p\,  u_s(p) = 0$, are
written in terms of commuting 2-component spinors $\tl$ and $\l$ defined as
\bea \lab{sps}
  u_-(p) =
  \left(
  \begin{array}{c}
    \lambda_\a \\
    0
  \end{array}
  \right) \, ,~~~~~~~~
  u_+(p) =
  \left(
  \begin{array}{c}
    0 \\
    \tl^{\da}
  \end{array}
  \right) \, .
\eea
Projectors $P_\pm =\frac{1}{2} (1 \pm \g_5 )$ then
act as $P_\pm u_\pm(p) = u_\pm(p)$ and $P_\pm u_\mp(p) = 0$.
With the adjoint of a Dirac spinor $\Psi$ defined as
\bea
  \bar{\Psi} \equiv - i \Psi^\dagger \g^0\, ,
\eea
we have
\bea
  \overline{u}_-(p) = ( 0 , - i \, \tl_{\da} )\, ,~~~~~~~~
  \overline{u}_+(p) = ( i \, \l^{\a} , 0)\, ,
\eea
where $\l^{\a} = (\tl^{\da})^*$ and $\tl_{\da} = (\l_{\a})^*$. Note that $\l^\a = \eps^{\a\b} \l_\b$ and
$\tl^{\da} = \eps^{\da\db} \tl_{\db}$.

Defining $p_{\a\db} = p_\m\, (\sigma^\mu)_{\a \db}$ and
$p^{\da\b} = p_\m\, (\bar{\sigma}^\mu)^{\da \b}$, the massless Dirac equation can be written
\bea
  p^{\da\b} \l_\b = 0 \, , ~~~~~
  \tl_{\da}\, p^{\da\b} = 0 \, ~~~~~
  p_{\a\db} \tl^{\db} = 0 \, , ~~~~~
  \l^{\a}\, p_{\a\db} = 0 \, .
  \lab{dirac}
\eea
One can show that
\bea
  \lab{llp}
  \l_\a \tl_{\db} = - p_{\a \db} \, ,~~~~~~~
  \l^\a \tl^{\db} = + p^{\db \a} \,.
\eea

We now introduce the bra-ket notation which is used heavily throughout the paper. Define
\bea
  &&
  | p ] ~=~
  u_-(p) ~=~
  \left(
  \begin{array}{c}
    \lambda_\a \\
    0
  \end{array}
  \right)
   \, , \hspace{1cm}
  | p \>
  ~=~ u_+(p)
  ~=~
  \left(
  \begin{array}{c}     0 \\
    \tl^{\da}
  \end{array}
  \right) \, , \\[1mm]
  &&
  \< p | ~=~ i\,  \bar{u}_-(p) ~=~
  \big( 0 , ~\tl_{\da} \big) \, ,
  \hspace{1cm}
  {}[ p |  ~=~ - i\, \bar{u}_+(p) ~=~ \big( \l^{\a} ,~ 0 \big) \, ,
\eea
It then follows from \reef{llp} that
\bea
  - p_\m  \g^\m = | p ] \< p | - | p \> [ p | \, .
\eea
Spinor products are defined as
\bea
  \< p \, q \> = \tl_{p \, \da} \tl_{q}^{\da} \, , \hspace{1.5cm}
  [ p \, q ]  = \l_{p}^{\a} \l_{q\, \a} \, ,
\eea
and they are related to the dot-product of the momenta by
\bea
  \< p \, q \>\, [ p \, q ] = 2 \, p \cdot q = - s_{pq} \, ,
\eea
where the Mandelstam variables are $s_{pq} = - (p+q)^2$.
For real momenta, the spinor products satisfy $[p\, q]^* = \< q\, p \>$,
so that up to phases $[p\, q] \sim \< p\, q \> \sim \sqrt{2 \, p \cdot q}$.
In applications we often use complex momenta in which case angle and square brackets ($\tl$ and $\l$) will not be complex conjugates, but independent.
We remark on the properties of the angle and square spinors under analytic continuation $p \to - p$. In our conventions, $|-p\> = - |p\>$ and $|-p] = +|p]$.

It is convenient to define ``angle-square brackets'' $\<i | P |j]$ as
\bea
  \<i | P |j] = \sum_{k=1}^m \<i \, k \>\, [k \, j] \, ~~~\mathrm{for}~~~
  P = \sum_{k=1}^m p_{i_k} \, .
\eea

In the spinor helicity formalism polarization vectors can be written as
\bea
  \lab{shpolar}
   \eps_+^\m(p;q)
   = - \frac{[q | \g^\m | p \> }{\sqrt{2}\,  [q \, p]} \, ,
   \hspace{1cm}
   \eps_-^\m(p;q)
   = \frac{ \< q | \g^\m | p ] }{\sqrt{2}\,  \<q \, p \>} \, .
   \lab{polsh}
\eea
One can show\footnote{It is useful to note the following properties:
$\< p | P | q ] = P_\m \< p | \g^\m | q ]$,~
$[q | \g^\m | p \> = - \< p | \g^\m | q ]$,~  $[q | \g^\m | p \>^* = [p | \g^\m | q \>$,
  and $[ p_1 | \g^\m | p_2 \> \< p_3 | \g_\m p_4 ]= 2 [p_1 p_4] \< p_3 p_2 \>$.}  that the polarization vectors are related by complex conjugation and satisfy the orthogonality relations
\bea
  \lab{ortho}
  \big(\eps^\mu_\pm(p)\big)^* = -\eps^\mu_\mp(p) \, ,
  \hspace{1cm}
  \eps^\mu_s(p)^*\eps_{\mu\, s'}(p)= \d_{ss'} \, .
\eea


\subsection{Explicit representation}

Take the momentum to be
\bea
  p^\m
  = (E, ~E\, \sin\th \cos\phi , ~E\, \sin\th \sin\phi,~ E\, \cos\th) \, .
   \lab{p}
\eea
Then
\bea
  p_{\a \db}
  =
  - 2E
  \left(
    \begin{array}{cc}
      s^2 & - c\, s\,  e^{-i \phi}  \\
      - c\, s\,  e^{i \phi}  & c^2 \\
    \end{array}
  \right)\,,
  \hspace{8mm}
    p^{\da \b}
  =
  2E
  \left(
    \begin{array}{cc}
      c^2 & c\, s\,  e^{-i \phi}  \\
      c\, s\,  e^{i \phi}  & s^2 \\
    \end{array}
  \right) \, ,
\eea
where we use $s = \sin\frac{\th}{2}$ and $c = \cos\frac{\th}{2}$.
It is straightforward to show that the two-component vectors
\bea \lab{tws}
  \l_\a = \sqrt{2E}
  \left(
  \begin{array}{c}
    -s \, e^{-i \phi/2} \\
    c \, e^{i \phi/2}
  \end{array}
  \right) \, ,~~~~~
  &&
  \tl_{\da} = \sqrt{2E} \,
  \big( - s \, e^{i \phi/2}  ,~ c \, e^{- i \phi/2} \big) \, ,\\[2mm]
  \tl^{\da} = \sqrt{2E}
  \left(
  \begin{array}{c}
    c \, e^{-i \phi/2} \\
    s \, e^{i \phi/2}
  \end{array}
  \right) \, ,~~~~~
  &&
  \l^{\a} = \sqrt{2E} \,
  \big( c \, e^{i \phi/2}  ,~ s \, e^{- i \phi/2} \big)
\eea
solve the massless Dirac equation in the form \reef{dirac}.

With  $p_\m$ given by \reef{p}, we write the positive and negative
helicity vectors
\bea
  \eps^\mu_\pm(p)
   = \mp \frac{1}{\sqrt{2}}
     \Big( 0,~\cos\th \, \cos\phi \mp i \sin\phi, ~
      \pm i \cos\phi + \cos\th \, \sin\phi , ~ - \sin\th \Big) \, .
  \lab{pol}
\eea
These clearly satisfy \reef{ortho}.
It can be shown that the expressions \reef{polsh} reproduce the
polarization vectors \reef{pol} for an appropriate choice of reference momentum $q$.


\subsection{Majorana spinors}

A Majorana spinor satisfies the condition
\bea
  \psi ~=~ B^{-1} \psi^* \, ,~~~~~~~
  B~=~ \g^0\, \g^1 \, \g^3
  ~=~
  \left(
    \begin{array}{cccc}
       0 & 0 & 0 & -1 \\
       0 & 0 & 1 & 0 \\
       0 & 1 & 0 & 0 \\
       -1& 0 & 0 & 0
    \end{array}
  \right) \, .
\eea
In the $\g$-matrix representation \reef{gammas} this means that
\bea
  \psi ~=~
  \left(
    \begin{array}{c}
       \psi_\a \\[1mm]
       \tilde{\psi}^{\da}
    \end{array}
  \right) \,
  ~~~~~\mathrm{with}
  ~~~~~
  \psi_\a ~=~
  \left(
    \begin{array}{c}
       \psi_1 \\[1mm]
       \psi_2
    \end{array}
  \right) \,
  ~~~~\mathrm{and}
  ~~~~
  \tilde{\psi}^{\da}  ~=~
  \left(
    \begin{array}{c}
       \psi_2^* \\[1mm]
       -\psi_1^*
    \end{array}
  \right) \, .
\eea
It follows from this and
$\tilde{\psi}^{\da} = \eps^{\da \db} \tilde{\psi}_{\db}$ that
$\tilde{\psi}_{\da} = (\psi_\a)^*$.

Let $\varepsilon$ and $\mathcal{Q}$ be Majorana spinors. Then
\bea
  \bar{\varepsilon} \, \mathcal{Q}
  ~=~-i \big(\eps_1^* Q_2^* - \eps_2^* Q_1^*
                  - \eps_2 Q_1 + \eps_1 Q_2\big)
  ~=~-i\big( \tilde{\eps}_{\da} \tilde{Q}^{\da} - \eps^\a Q_\a \big)
  ~\equiv~ -i \big( \tilde{Q} + Q \big)
  \, .
\eea

If $\delta_\eps A$ is the supersymmetry transformation of the field $A$, then the susy generators act as
\bea
  \lab{susyQtQ}
  \delta_\eps A ~=~ i \big[ \bar{\varepsilon } \, \mathcal{Q} , A \big]
   ~=~ \big[ Q + \tilde{Q}, A \big] \, .
\eea

Including labels $a,b,\dots= 1,\dots,\cn$, the generators $Q^a$ and $\tilde{Q}_b$ satisfy the extended supersymmetry algebra
\bea
  \big[ [Q^a , \tilde{Q}_b] , A \big]
  ~=~  \d_b^a \< \eps_2 \, p \> [p\, \eps_1] \, A \, ,
  ~~~~~~
  \big[ [Q^a , Q^b] , A \big] ~=~ 0\, ,
  ~~~~~~
  \big[ [\tilde{Q}_a , \tilde{Q}_b] , A \big] ~=~ 0\, ,
\eea
for distinct susy parameters $\eps_{1,2}$ and $\tilde{\eps}_{1,2}$.


\setcounter{equation}{0} \section{Solution of the $\cn =1$ SUSY
Ward identities for 6-point NMHV amplitudes}
\lab{app:N1susy}

We apply spinor-helicity methods to obtain the solution to the SUSY Ward identities for
6-point NMHV amplitudes in an $\cn=1$ theory originally found in \cite{gris}.\footnote{See also \cite{manparke}.} The
solution is valid for
complex momenta in an arbitrary Lorentz frame and reduces to the solution of \cite{gris}
for real momenta in the center-of-mass frame. We hope that the new method will be useful
for solving the NMHV \wis\ of extended supersymmetry.

Let $b^\pm$ and $f^\pm$ denote the annihilators of the bosonic and fermionic states of an
$\cn =1$ supersymmetric theory. There are 20 independent 6-point NMHV amplitudes for
which we introduce the notation: \bea && G ~=~ \sigma_b \, \big\langle
b^{+}b^{+}b^{+}b^{-}b^{-}b^{-}\big\rangle \, , \hspace{1.35cm} G_{i,I} ~=~ (-)^{i+I}
\sigma_b\, \big\langle f^{+}b_i^{+}f^{+}f^{-}b_I^{-}f^{-}\big\rangle \, , \\[1mm] && F
~=~ \sigma_f \, \big\langle f^{+}f^{+}f^{+}f^{-}f^{-}f^{-}\big\rangle \, , \hspace{1cm}
F_{i,I} ~=~ \sigma_f \, \big\langle b^{+}f_i^{+}b^{+}b^{-}f_I^{-}b^{-}\big\rangle \, .
\eea The momentum (and position) labels $i,j,k$ run over $1,2,3$ while $I,J,K=4,5,6$.
Tthe subscript $i$ on $b^+_i$ means that the particle is in position $i$ with momentum
$p_i$, etc. For example, $G_{1,6} = - \sigma_b \big\langle b^+(1)\, f^{+}(2)\, f^{+}(3)\,
f^{-}(4)\, f^{-}(5)\, b^{-}(6)\big\rangle$.

The supersymmetric \wis\ can be solved to express the 18 amplitudes $G_{i,I}$ and
$F_{i,I}$ in terms of the purely bosonic and fermionic amplitudes $G$ and $F$. The result
is \bea \lab{N1resF} &&F_{i,I} = \D^{-1} \big( \eps_{ijk} \langle jk \rangle \,
\eps_{IJK}[JK]\, F + 4 \< I j \> [ ij ] \, G \big) \, ,\\[1.5mm] &&G_{i,I} = \D^{-1}
\big( \eps_{ijk} [jk] \, \eps_{IJK}\langle JK\rangle \, G + 4 \< iJ \> [ I J ] \, F
\big)\, , \lab{N1resG} \eea where \be
 \D~=~ - 2 \langle ij\rangle \, [ij] ~=~ -2 \<IJ\>\, [IJ]\, , \ee and repeated indices are
summed. The remainder of this appendix is devoted to the proof of
\reef{N1resF}-\reef{N1resG}.

The commutator relations of the $\cn=1$ SUSY generator $\tQ$ with the annihilators is
\bea \nonumber && \big[\tQ,b^{+}(p)\big]~=~ \sigma_b \langle\e\, p\rangle \, f^{+}(p) \,
, \hspace{1cm} \big[\tQ,f^{+}(p)\big]~=~0 \, ,\\ &&\big[\tQ,b^{-}(p)\big] ~=~0 \, ,
\hspace{2.8cm} \big[\tQ,f^{-}(p)\big] ~=~ \sigma_f \langle\e\, p\rangle \, b^{-}(p) \, .
\lab{N1susy} \eea The phases $\sigma_{b,f} = \pm 1$ depend on which $\cn=1$ multiplet is
considered. Similar relations exist for the generator $Q$ which raises the helicity by
1/2.

The SUSY Ward identities from the $\tQ$ commutator relations can be written compactly as
\bea \nonumber \langle\e \, i\rangle \, F_{i,I} + \langle\e\, I\rangle \, G &=& 0\, ,
\\[1mm] \lab{tQwisN1} \langle\e \, i\rangle \, F + \langle\e\, I\rangle \, G_{i,I} &=&
0\, , \\[1mm] \nonumber \eps_{ijk} \langle\e\, j\rangle \, G_{k,I} +
\eps_{IJK}\langle\e\, J\rangle \, F_{i,K} &=& 0 \, . \eea

We will also need a subset of the \wis\ obtained from the conjugate $Q$ Ward identities.
In notation that should be obvious, these read \bea
 \label{conjWIs}
 \sigma \, [\e \, i] \, G - [\e \, I] F_{i,I} ~=~0 \,, \hspace{8mm}
 \sigma \, [\e \, I] \, F - [\e \, i] G_{i,I} ~=~0 \,, \eea with $\sigma = \pm 1$
resulting from the choice of phases in the algebra of $Q$ with the annihilators.

We  start with the solution ansatz
\bea
 \label{ansatz}
 &&F_{i,I} ~=~ M_{i,I} \, G + N_{i,I} \, F  \, , \hspace{1cm}
 G_{i,I} ~=~ K_{i,I} \, G + L_{i,I} \, F \, , \eea where $M, N, K, L$ are $3 \times 3$
matrices. The \wis\ \reef{tQwisN1} split into two sets of equations, one set for the
matrices $N$ and $L$, \bea
 \< \e \, i\> N_{i,I} ~=~0 \, , \hspace{8mm}
   \< \e \, I \> L_{i,I} ~=~ - \< \e \, i \> \, , \hspace{8mm}
 \label{LN}
 \epsilon_{ijk}\, \< \e \, j \> L_{k,I}
 + \epsilon_{IJK}\, \< \e \, J \> N_{i,K} ~=~ 0 \, ,~~~~~
\eea
and another for $M$ and $K$ \bea   \< \e \, I\> K_{i,I} ~=~0 \, , \hspace{8mm}
 \< \e \, i \> M_{i,I} ~=~ - \< \e \, I \> \, , \hspace{8mm}
 \epsilon_{ijk}\, \< \e \, j \> K_{k,I}
 + \epsilon_{IJK}\, \< \e \, J \> M_{i,K} ~=~ 0 \, .~~~~ \eea In addition equation
(\ref{conjWIs}) gives (among other relations)
\bea
  [\e \, I ] N_{k,I} ~=~ 0 \, , \hspace{8mm}
  [\e \, k] K_{k,I} ~=~ 0 \, .
\eea
  Due to the separation of the constraints, we will
focus our attention on the equations for $N$ and $L$; the system of $K, M$ equations is
identical and is treated the same way.

The equation $\< \e \, i\> N_{i,I} =0$ is simply solved by
$N_{i,I} = \epsilon_{ijk} \< j\, k \> n_I$ for any vector $n_I$.
This follows from the Schouten identity. Next, $[\e \, I ] N_{k,I} =
0$ is solved by $n_I = \epsilon_{IJK} [J \, K] \Delta^{-1}$ for
some general function $\Delta$ to be determined. Hence \bea
 N_{i,I} ~=~  \Delta^{-1} \epsilon_{ijk} \< j\, k \> \epsilon_{IJK} [J \, K] \, . \eea
Using the standard identity $\eps_{IJK} \, \eps_{KLM} = (\d_{IL} \d_{JM} - \d_{IM}
\d_{JL})$, the third equation of \reef{LN} then gives \bea
 \frac{2}{\Delta} \eps_{ijk} \< j \, k \> \< \e \, J\> [I \, J]
 ~=~
 - \eps_{ijk} \< \e \, j \> L_{k,I} \, .
\eea
Multiplying both sides with $\eps^{ilm}$ and summing over $i$ we then find
\bea
 \label{Ls}
 \frac{4}{\Delta} \< \e \, J \> [IJ] \< l\, m \>
 ~=~
 - \< \e \, l \> L_{m,I} + \< \e \, m \> L_{l,I} \, . \eea Choosing $\< \e | = \< l |$
(no sum on $l$) provides the solution for $L$; it is \bea
 L_{l,I} ~=~ \frac{4}{\Delta} \< l \, J \> [IJ] \, .
\eea

The only task left now is to determine the scalar function $\Delta$. This is easily done
as follows. Multiply \reef{Ls} by $\< \e' \, I \>$ for some arbitrary spinor $\e'$.
Summing over $I$ and using $\< \e' \, I \> L_{i,I} ~=~ - \< \e' \, i \>$ we obtain \bea
 \label{D1}
 \frac{4}{\Delta} \< \e' \, I \> \< \e \, J \> [IJ] \< l \, m \>
 ~=~
 \< \e \, l \> \< \e' \, m \> - \< \e \, m \> \< \e' \, l \>
 ~=~ - \< \e \, \e' \> \< m \, l \> \, .
\eea
Antisymmetrization of $IJ$ on the LHS of \reef{D1} gives (by Schouten)
\bea
 \label{D2}
 \frac{4}{\Delta} \< \e' \, I \> \< \e \, J \> [IJ]  ~=~  \frac{2}{\Delta} \Big( \< \e'
\, I \> \< \e \, J \> -     \< \e' \, J \> \< \e \, I \> \Big) [IJ]  ~=~
-\frac{2}{\Delta} \< \e \, \e' \> \< I J \> [IJ] \, . \eea Thus the factors of $ \< \e \,
\e' \>$ can be eliminated and we conclude from \reef{D1} and \reef{D2} that \bea
 \Delta ~=~ - 2 \< I J \>  [I J]  \, .
\eea
Note that momentum conservation implies that
\bea
 \Delta ~=~ - 2 \< I J \>  [I J]  ~=~ -2  \sum_{I,J} (p_I + p_J)^2
 ~=~ -2  \sum_{i,j} (p_i + p_j)^2  ~=~ - 2 \< i j \>  [i j] \, .
\eea

This completes the solution for $N$ and $L$. Since the system of equations for $K$ and
$M$ is identical, we have proven that \reef{N1resF}-\reef{N1resG} solve the $\cn=1$ SUSY
\wis .

 
 \end{document}